\documentclass[12pt,preprint]{aastex}

\newcommand{\noprint}[1]{}
\newcommand{\figsetstart}{{\bf Fig. Set} }
\newcommand{\figsetend}{}
\newcommand{\figsetgrpstart}{}
\newcommand{\figsetgrpend}{}
\newcommand{\figsetnum}[1]{{\bf #1.}}
\newcommand{\figsettitle}[1]{ {\bf #1} }
\newcommand{\figsetgrpnum}[1]{\noprint{#1}}
\newcommand{\figsetgrptitle}[1]{\noprint{#1}}
\newcommand{\figsetplot}[1]{\noprint{#1}}
\newcommand{\figsetgrpnote}[1]{\noprint{#1}}

\newcommand{\HI}{\ion{H}{1}~}
\newcommand{\HII}{\ion{H}{2}~}
\newcommand{\ebv}{$E(B-V)$~}
\begin{document}
\title{A Search for Extended Ultraviolet Disk (XUV-disk) Galaxies in the Local
Universe}
\author{David A. Thilker\altaffilmark{1}, 
Luciana Bianchi\altaffilmark{1},
Gerhardt Meurer\altaffilmark{1},
Armando Gil de Paz\altaffilmark{2},
Samuel Boissier\altaffilmark{3},
Barry F. Madore\altaffilmark{4},
Alessandro Boselli\altaffilmark{3},
Annette M. N. Ferguson\altaffilmark{5},
Juan Carlos Mu\'noz-Mateos\altaffilmark{2},
Greg J. Madsen\altaffilmark{6},
Salman Hameed\altaffilmark{7},
Roderik A. Overzier\altaffilmark{1},
Karl Forster\altaffilmark{8},
Peter G. Friedman\altaffilmark{8},
D. Christopher Martin\altaffilmark{8},
Patrick Morrissey\altaffilmark{8},
Susan G. Neff\altaffilmark{9},
David Schiminovich\altaffilmark{10},
Mark Seibert\altaffilmark{8},
Todd Small\altaffilmark{8},
Ted K. Wyder\altaffilmark{8},
Jose Donas\altaffilmark{3},
Timothy M. Heckman\altaffilmark{11},
Young-Wook Lee\altaffilmark{12},
Bruno Milliard\altaffilmark{3},
R. Michael Rich\altaffilmark{13},
Alex S. Szalay\altaffilmark{11},
Barry Y. Welsh\altaffilmark{14}, 
Sukyoung K. Yi \altaffilmark{12}}

\altaffiltext{1}{Center for Astrophysical Sciences, The Johns Hopkins
University, 3400 N. Charles St., Baltimore, MD 21218, dthilker@pha.jhu.edu}
\altaffiltext{2}{Departamento de Astrof\'{\i}sica, Universidad Complutense de Madrid, Madrid 28040, Spain}
\altaffiltext{3}{Laboratoire d'Astrophysique de Marseille, BP 8, Traverse
du Siphon, 13376 Marseille Cedex 12, France}
\altaffiltext{4}{Observatories of the Carnegie Institution of Washington,
813 Santa Barbara St., Pasadena, CA 91101}
\altaffiltext{5}{Institute for Astronomy, University of Edinburgh, Royal Observatory Edinburgh, Edinburgh, UK}
\altaffiltext{6}{School of Physics A29, University of Sydney, NSW 2006, Australia}
\altaffiltext{7}{Five College Astronomy Department, Hampshire College, Amherst, MA 01003, USA}
\altaffiltext{8}{California Institute of Technology, MC 405-47, 1200 East
California Boulevard, Pasadena, CA 91125}
\altaffiltext{9}{Laboratory for Astronomy and Solar Physics, NASA Goddard
Space Flight Center, Greenbelt, MD 20771}
\altaffiltext{10}{Department of Astronomy, Columbia University, New York, NY 10027}
\altaffiltext{11}{Department of Physics and Astronomy, The Johns Hopkins
University, Homewood Campus, Baltimore, MD 21218}
\altaffiltext{12}{Center for Space Astrophysics, Yonsei University, Seoul
120-749, Korea}
\altaffiltext{13}{Department of Physics and Astronomy, University of
California, Los Angeles, CA 90095}
\altaffiltext{14}{Space Sciences Laboratory, University of California at
Berkeley, 601 Campbell Hall, Berkeley, CA 94720}

\begin{abstract}

We have initiated a search for extended ultraviolet disk (XUV-disk)
galaxies in the local universe.  Herein, we compare GALEX UV and
visible--NIR images of 189 nearby (D$<$40 Mpc) S0--Sm galaxies
included in the GALEX Atlas of Nearby Galaxies and present the first
catalogue of XUV-disk galaxies.  We find that XUV-disk galaxies are
surprisingly common but have varied relative (UV/optical) extent and
morphology. Type~1 objects ($\ga$20\% incidence) have structured,
UV-bright/optically-faint emission features in the outer disk, beyond
the traditional star formation threshold.  Type~2 XUV-disk galaxies
($\sim$10\% incidence) exhibit an exceptionally large,
UV-bright/optically-low-surface-brightness (LSB) zone having blue
$UV-K_s$ outside the effective extent of the inner, older stellar
population, but not reaching extreme galactocentric distance.  If the
activity occuring in XUV-disks is episodic, a higher fraction of
present-day spirals could be influenced by such outer disk star
formation. Type~1 disks are associated with spirals of all types,
whereas Type~2 XUV-disks are predominantly found in late-type spirals.
Type~2 XUV-disks are forming stars quickly enough to double their
[presently low] stellar mass in the next Gyr (assuming a constant SF
rate).  XUV-disk galaxies of both types are systematically more
gas-rich than the general galaxy population.  Minor external
perturbation may stimulate XUV-disk incidence, at least for Type~1
objects.  XUV-disks are the most actively evolving galaxies growing
via inside-out disk formation in the current epoch, and may constitute
a segment of the galaxy population experiencing significant, continued
gas accretion from the intergalactic medium or neighboring objects.
\end{abstract}



\section{Introduction}

Imprinted within the disk component of any spiral galaxy is a wealth
of information regarding its formation, and clues concerning its
continued evolution.  For instance, radial variations in the average
surface brightness of the stellar disk are not purely exponential with
a sharp outer truncation as once thought (e.g. van der Kruit \& Searle
1981, Florido et al. 2001), but rather best represented as a complex
broken exponential (Pohlen \& Trujillo 2006, Erwin et al. 2005, Pohlen
et al. 2002) -- sometimes even flattening in the outer disk.  The
color of the disk is also a function of galactocentric distance
(e.g. de Jong 1996, MacArthur 2006, MacArthur et al. 2004,
Mu\'noz-Mateos et al. 2007). Such complex surface brightness and color
profiles trace changes in the disk stellar population, and reflect the
assembly history of the galaxy, modulated by the combined effects of
radial motion (of secular or external origin) and extinction.  The
chemical evolution of a galaxy is also driven by the star formation
history (Boissier \& Prantzos 2000, Chiappini et al. 2003).  Abundance
gradients typically reveal a comparatively enriched inner disk
(Zaritsky et al. 1994, Henry \& Worthey 1999), though barred spirals
are less steep in general. The overall picture which emerges is that
the outer disk is often the locale of most significant recent change,
though it remains considerably less evolved -- as expected for
inside-out disk formation (White \& Frenk 1991, Mo et al. 1998,
Boissier \& Prantzos 1999, Brook et al. 2006, Mu\'noz-Mateos et
al. 2007).
Quantifying the recent star formation rate
($SFR$) across the entire disk in a large galaxy sample will provide
an anchor for interpretation of  enrichment and disk building
processes.

In a broader sense, understanding the typical star formation activity
of outer disks is crucial because it has bearing on diverse
topics including absorption line diagnostics of metal abundance in
the circumgalactic medium, efforts to determine how (and at what rate)
gas accretes onto seemingly evolved galaxies, and how star formation
is regulated (or inhibited) in low-density regions perhaps similar to
proto-galactic environments in the early Universe.

It has been known from deep H$\alpha$ imaging that some galaxies
possess very extended star-forming disks (e.g. Ferguson et al. 1998).
However, the incidence and underlying cause of this behavior are still
unknown.  The new ultraviolet imaging surveys from NASA's GALEX ({\it
Galaxy Evolution Explorer}) mission greatly facilitate the study of
low-intensity, outer disk star formation in a statistically
significant number of galaxies (Gil de Paz et al. 2007).  GALEX
imaging benefits from a very low sky background, high sensitivity, and
a large field-of-view (1.25$\arcdeg$ diameter), enabling efficient
detection of otherwise elusive activity.  By observing in the UV,
GALEX probes a population of OB stars (rather than only the higher
temperature O stars predominantly ionizing \HII regions traced by
H$\alpha$ emission), and thus can catch galaxy evolution processes
occurring at very low SFR surface densities
($\Sigma_{SFR}$).

Thilker et al. (2005) and Gil de Paz et al. (2005) discovered extended
UV (XUV) emission in the extreme outer disk environment of M~83 and
NGC~4625. Both these studies established that the amount and spatial
extent of star formation in the outskirts of a galaxy can be
underestimated by looking for \HII regions alone (as traced by
H$\alpha$ emission).  However, a multitude of unanswered questions
follow from the XUV-disk discovery.  For instance, are XUV-disks rare
or relatively common (e.g. Zaritsky \& Christlein 2007)?  Are certain
galaxy types more likely to host spatially extended star formation?
What is the characteristic morphology of XUV emission?  Is the SF
episodic, or continuous? M~83 and NGC~4625 are rather dissimilar in
their extended UV morphology.  Does that suggest that the outer disk
has a different origin, star formation history ($SFH$), or have we
simply caught these particular XUV-disks at different stages in their
evolutionary progression?  Outer disk star formation does have an
impact on the course of chemical evolution. Gil de Paz et al. (2007b)
show that the metallicity of XUV-disks (M~83, NGC~4625) is very low,
though not primordial.

Much attention has been directed toward the evidence in favor of a
star formation threshold in low density environments (e.g. Kennicutt
1989, Martin \& Kennicutt 2001).  Indeed, many galaxies appear to show
a sharp decline in the radially averaged $SFR$ per unit area derived
from sensitive H$\alpha$ observations, occuring at the radius where
the disk is predicted to become dynamically stable.  The discovery of
XUV-disks raised questions about this interpretation.  Boissier et
al. (2007) demonstrate that at least some truncations in H$\alpha$ may
be an artifact of small number statistics.  Specifically, the number
of \HII regions expected at any given moment in the discrete radial
bins used to compute H$\alpha$ surface brightness profiles becomes
vanishingly small for the observed, UV-derived $\Sigma_{SFR}$ at such
locations.

This paper reports on a campaign to better understand the properties
of XUV-disk galaxies and their relevance in a cosmological context.
Herein, we determine the incidence of XUV emission in a large sample
of nearby disk galaxies.  Necessarily, we suggest a classification
scheme for defining the presence of XUV-disks.  The large number
($\sim 200$) of galaxies considered in our study helps to better
constrain the conditions which allow or encourage an XUV-disk to form.
Our analysis includes a look at the importance of galactic environment
(isolation vs. group membership) and the presence of dynamically
perturbing companions.  The present paper also considers the relation
of spatially extended star formation (traced in the UV) with optical
surface brightness profiles of various types (Pohlen \& Trujillo 2006,
Erwin et al. 2005).  In particular, we comment on the possible
connection of ``anti-truncated'' optical disks and
low-surface-brightness (LSB) galaxies with XUV-disks.

Section 2 describes our galaxy sample along with the GALEX and
ground-based visible-NIR observations we used for this study.  Section
3 outlines our data analysis and XUV-disk galaxy classification
scheme.  We also examine global galaxy parameters and environmental
properties of the XUV-disk class in comparison to the entire sample.
Section~4 presents an extensive discussion and our main conclusions
are listed in Section 5.  In Appendix I, we provide a description and
multiwavelength imaging for each galaxy in our XUV-disk catalogue.
Appendix II presents a short summary of the inside-out disk formation
models used to compare with observed galaxy characteristics.

\label{sintro}

\section{Sample selection and observations}

\subsection{The galaxy sample}

Our survey sample (Table 1) is comprised of 189 disk galaxies ($-0.5
\le T \le 9.5$, consequently S0--Sm) within 40 Mpc included in the
GALEX Atlas of Nearby Galaxies (Gil de Paz et al. 2007) and having $i
\le 80\arcdeg$, $D_{25} > 90$\arcsec, in addition to modest Galactic
extinction, $A_V \le 0.5$.  By selecting only objects from the Atlas,
we ensure that all galaxies were observed with homogeneous (deep)
sensitivity.  M~31 and M~33 have been
excluded, as their large angular size significantly complicates our
analysis.  The upper distance limit ensures that spatial resolution
of our GALEX observations is $\sim 1$ kpc or less.
Galaxy distances were assigned following Gil de Paz et al. (2007), who
established a ranking of reliability for various possible indicators
(e.g. Cepheids, Tip of the Red Giant Branch, etc.)  with use of the
Hubble constant ($H_0 = 70$ km s$^{-1}$ Mpc$^{-1}$) adopted only as a fallback method.

LEDA was employed to extract ancillary quantities.  For example, we
obtained rotation velocities, mean surface brightness, \HI mass, and
$M_{HI}$/$L_B$.  LEDA contains M$_{HI}$ for almost all ($>90\%$) of
our targets, even if some of these (single-dish) \HI measurements
could be underestimates in the case of angularly large
galaxies. Further, Gil de Paz et al. (2007) and Dale et al. (2007)
published observed panchromatic spectral energy distributions (SEDs)
for the GALEX Atlas sample and the SINGS sample (Kennicutt et
al. 2003), respectively.  These data were used to compute
galaxy-averaged quantities such as the extinction-free $SFR$ (from the
union of UV and TIR luminosity), extinction-free specific $SFR$
($sSFR$ = $SFR$/$M_{*}$ = star formation rate per unit of stellar
mass), and the ratio of infrared to ultraviolet luminosity ($TIR/UV$).
$SFR$ surface density ($\Sigma_{SFR}$) was estimated very crudely
using $D_{25}$ as the fiducial disk size.  Note that the $SFR$ used
for computing $\Sigma_{SFR}$ sometimes includes SF outside of the
$D_{25}$ radius, as we only have galaxy-integrated values of IR
luminosity to use in the calculation.  For consistency, the UV
measurements used for $\Sigma_{SFR}$ are also galaxy-integrated. The
dynamical and photometric quantities considered in our study are
included in Table 1.

In order to quantify the environment of each galaxy, we
referenced the hierarchical group assignments of the Giuricin et
al. (2000) Nearby Optical Galaxy (NOG) sample.  This study of a
magnitude-limited ($B \le 14$) sample of nearby galaxies ($cz_{LG}
\le$6000 km/s) showed that 40\% of the local galaxy population is
ungrouped and the remaining 60\% is distributed within galaxy pairs
(15\%) and more populous groups (45\%).  In Table 1 for each of our
189 galaxies, we list the associated NOG Group (NOGG) identification
number and common name (when available), in addition to the number of
galaxies in each group.  Being part of a group does not guarantee that
an object is currently (or recently) perturbed via galaxy interaction.
Therefore, we also computed the tidal perturbation parameter for all
sample galaxies (see Section 4 for details).  Finally, note that the
Giuricin et al. study is expected to miss faint companions of our
sample galaxies (especially at the more distant end of our sample),
and the tidal perturbation parameter would not necessarily reflect a merger in
progress, thus we also identified those galaxies for which the UV
morphology or corollary data indicate evidence for interaction.  They
are flagged in Table~2, which pertains only to galaxies classified as
XUV-disks.



\subsection{GALEX FUV and NUV imaging observations}

GALEX is conducting wide-field (1.25\arcdeg) imaging and
spectroscopic surveys of the sky at far-UV (FUV, 1350--1750~\AA) and
near-UV (NUV, 1750--2750~\AA) wavelengths.  Details regarding the
GALEX surveys and on-orbit performance of the satellite may be
found in Martin et al. (2005) and Morrissey et al. (2007).

  GALEX obtained FUV and NUV direct imaging observations of our sample
targets during primary mission surveys, principally the Nearby Galaxy
Survey (NGS; Bianchi et al. 2003, Gil de Paz et al. 2007) and Medium
Imaging Survey (MIS; Martin et al. 2005).  Total exposure time for these
images is typically 1500 sec, leading to 5$\sigma$ point-source limiting
magnitudes of $FUV(AB)$ = 22.6 and $NUV(AB)$ = 22.7.  
All of the images
analyzed herein were a product of the IR1.1 pipeline.  The data are
already public and can be obtained from the GALEX archive
(hosted by MAST at http://galex.stsci.edu/).  

The GALEX PSF is band dependent.  FUV images have a typical PSF FWHM of 4.3$\arcsec$, with 80\% of the point
source flux enclosed within 6$\arcsec$.  For the NUV band, the FWHM
is $\sim5.3\arcsec$ with an 80\% enclosed energy diameter
 of
7$\arcsec$ (Morrissey et al. 2007)  
The astrometric
error is less than 1.1$\arcsec$ for 80\% of sources lying in the central degree of the GALEX
field.
%
The photometric zero point appropriate to the FUV and NUV bands was
determined using observations of the white dwarf LDS749B.  Uncertainty
in the zero points is about 0.07 mag (Morrissey et al. 2007).
Residual scatter in the flat field is approximately 7\%.

\subsection{DSS-II and SDSS imagery}

During the classification process our GALEX imagery was compared to
the visible morphology and extent of each target.  Second-generation Digitial Sky Survey (DSS-II) imaging
data was used for this purpose. With the ESO/ST-ECF DSS program, we retrieved an image centered on each galaxy from the red (IIIaF + RG610)
plates, covering three times
the $D_{25}$ extent of the disk.

Following our classification (see Section 3), we obtained Sloan Digital Sky Survey (SDSS)
DR5 {\it ugriz} imagery of confirmed XUV-disk objects whenever
possible.  SDSS observations were available for $\sim$ 60\% of the
XUV-disks.  It would have been preferable to use these photometrically
calibrated data for classification purposes, however SDSS is limited to 8000
sq.deg of the sky.

\subsection{2MASS $K_s$ imagery}

To localize the bulk of the old stellar population in our
survey galaxies, we extracted 2MASS $K_s$ images from the NASA/IPAC
Infrared Science Archive (IRSA).  Due to the proximity of our sample,
many galaxies are included in the 2MASS Large Galaxy Atlas (LGA;
Jarrett et al. 2003).  This was our preferred source for $K_s$ data, as
the LGA has already mosaiced individual 2MASS fields to produce a
single image covering these galaxies.  100 of the 189 survey targets
analyzed in this paper were either formally included in the
LGA, or located well within the LGA field of view for a neighboring
object on the sky.  
$K_s$ data for the 89 galaxies unavailable through
the LGA were obtained via the Extended Source Catalog (XSC) IRSA interface
for the 2MASS All-sky data release, otherwise via a coordinate based
cutout for the small number of galaxies with no XSC entry.

\section{Data analysis and results}
\label{sfits}

\subsection{XUV-disk classification}

The first discovered XUV-disk galaxies, M~83 and NGC~4625 (Thilker et
al. 2005, Gil de Paz et al. 2005), are characterized by obvious
UV-bright, spiral structures distributed far beyond the
previously recognized outer limit of any recent SF -- hence our
designation ``extended UV-disk'' (XUV-disk) galaxy.

However, not all outer disk SF is conspicuous. To enable understanding
the properties of extended ultraviolet galaxies in the framework of a
large sample, a classification scheme was needed to define what we
consider to be XUV-emission.  Thus, we first selected a variety of
representative galaxies with which to evaluate possible XUV-disk
criteria, including the shape of the radial surface brightness and
color profiles, the disk surface covering factor of UV-bright clumps,
and other indicators such as the H$\alpha$/UV flux ratio at outer
galactocentric radii. We chose the following scheme to
minimize $SFH$-related bias, avoid pragmatic difficulties in measuring very faint surface brightness profiles, and facilitate application by requiring
only commonly available, broad-band optical data and GALEX
observations.

We define two types of XUV-disk morphology in order to ``catch'' all
apparent cases of spatially extended SF. In particular, the second
definition avoids neglecting galaxies which have active SF in outer
locations lacking an underlying older disk of appreciable stellar
surface density and yet have a well defined radial limit to such star
formation.

We classify XUV-disk galaxies as those
objects which either:
\begin{itemize}

\item  show structured, UV-bright emission complexes 
beyond the anticipated location of the star formation threshold ({\em Type 1}, e.g.
M~83 in Fig. 16; NGC~5055 in Figs. 3, 16), or,

\item  have blue color ($FUV-NIR$) within an exceptionally large, outer,
and optically-LSB portion of the disk ({\em Type 2}, e.g. NGC~2090 in Figs. 4 and
16).
\end{itemize}

\noindent Some galaxies satisfy both definitions and are referred to as {\em mixed-type} XUV-disks.

\subsubsection{Type~1 XUV-disks}
The first of our two XUV-disk definitions has been implemented as
follows.  {\em We require more than one structured complex of
UV-bright emission at positions outside of a centralized, contiguous
region contained by a single surface brightness contour corresponding
to} $\Sigma_{SFR}$ = 3$\times10^{-4}$ M$_{\odot}$ yr$^{-1}$ kpc$^{-2}$ {\em evaluated at 1~kpc resolution}.  Galaxies with such morphology are hereafter denoted as
Type 1 XUV-disks, or alternatively
``M~83-like'' objects, adopting a prototype.  Additional examples can be seen in Fig. 16 of Appendix I. 
\begin{itemize}
\item Our choice of the Galactic extinction-corrected UV surface
brightness, $\mu_{UV}$ [or SFR surface density, $\Sigma_{SFR}$], level used to establish the outer clump
searching region is motivated both theoretically and observationally.
Boissier et al. (2007) demonstrate that for the galaxies of Martin \&
Kennicutt (2001) showing a sharp azimuthally drop (in H$\alpha$)
attributed to a SF threshold mechanism, typical $\mu_{UV}$ are near
our fiducial level when evaluated at the H$\alpha$ ``edge'' -- even if
the edge is merely a consequence of ``missing'' the outer,
short-lived, infrequently-formed \HII regions.  Adopting the SFR
calibration of Kennicutt (1998), $\Sigma_{SFR}$ = 3$\times10^{-4}$
M$_{\odot}$ yr$^{-1}$ kpc$^{-2}$ corresponds to $\mu_{FUV} =$ 27.25
ABmag/arcsec$^2$ or $\mu_{NUV} =$ 27.35 ABmag/arcsec$^2$ for galaxies
only having NUV data. FUV and NUV calibrations are slightly different
because we allow for the spectral shape predicted by Starburst99 for a
young population continuously forming stars (e.g. Iglesias-Paramo et
al. 2006). Also, the local star formation
threshold \HI column density as predicted by Schaye (2004) has been
found to correspond to FUV[NUV] surface brightnesses very similar to
those defining our adopted UV threshold contours (e.g. Thilker et
al. 2007).  Thus, {\em our threshold contour aims to coarsely delineate
where a putative change in the SF properties of the disk may be
occuring.}

\item We require ``structured'' emission complexes
(e.g. spiral segments, irregularly grouped clumps)
 to avoid including objects which have faint, but widespread, slowly
varying UV emission in their outskirts.  Such diffuse emission may result
from causes other than recent SF and might be expected in
the case of scattered light from dust (Hoopes et al. 2005), or {\em in situ}
emission from hot core helium-burning stars.  Structured emission complexes suggest a ``disk'' geometry directly if in the form of spiral-like segments.  In terms of detecting
clumps, we are progressively limited by sensitivity
as galaxy distance increases.
At a fiducial distance of 10 Mpc and MIS/NGS depth (5$\sigma$ $\sim$
m$_{FUV}$(AB) = 22.6, m$_{NUV}$(AB) = 22.7), GALEX detects clusters
having an initial stellar mass of [400, 2600, 6600] M$_\odot$ in the
FUV band for ages of [3, 10, 100] Myr, whereas nearing the 40 Mpc
limit of our sample we are only able to detect 6300 M$_\odot$ clusters
if they are $\sim$ 3 Myr old, otherwise the limit rises to
40--100$\times10^{3}$ M$_\odot$ for the more evolved complexes.  Given the typical appearance of nearby XUV-disks in our sample, we expect outer structures to remain resolved for galaxies in the vicinity of our 40 Mpc distance limit. 
\item The ``UV-bright'' requirement
refers to the structured clumps individually, not the surrounding,
inter-clump environment.  The clumps detected (via visual assessment) in the
UV by GALEX must be barely visible or not discernable at R-band (checked
using the DSS2-red plates).  Underlying optical emission can be seen,
but must have fundamentally different morphology.  For instance, a
smooth/diffuse optical background containing clumped UV emission would
be allowed.
\item We relax the requirement of more than one detected
structured complex in the case where a solitary feature is
unquestionably associated with the galaxy (such as a filament composed
of several bright clumps, aligned with a known \HI arm).
\end{itemize}


\subsubsection{Type~2 XUV-disks}
Our second XUV-disk definition selects galaxies which are forming stars at an elevated rate
over a large area compared to the spatial extent of their evolved
stellar population, whether or not they apparently conform to the
notions of a star formation threshold (gauged by the presence of clump
complexes extending to extreme outer radii).  Objects having such morphology do not
resemble the first-known  XUV-disk (M~83) and were discovered
during the present study.  NGC~2090 (see Figs. 4, 16) is a prototypical
Type~2 XUV-disk.

To quantitatively identify Type~2 XUV-disk structures, we defined a ``low surface
brightness (LSB) zone'' as
the region contained {\em within} the anticipated SF
threshold [$\Sigma_{SFR}$] contour yet {\em outside} a $K_s$-band contour
enclosing 80\% of the total $K_s$ luminosity of the galaxy.  The $FUV-K_s$
color of this optically LSB zone can then be used as a metric of the
luminosity weighted age of the associated stellar population.
{\em If the average $UV(AB)-K_s(AB)$ of the LSB zone is
$\le4$, indicating the relative importance of recent star formation with
respect to the underlying disk {\em and} the area of the zone,
$S(LSB)$, is at least 7 times the enclosed area of the $K_s$
contour, $S(K_{80})$, then we classify the galaxy as an
XUV-disk}.

\begin{itemize}
\item In Fig. 1 we show Bruzual \& Charlot (2003) population synthesis
model colors versus age for three metallicities in two SFH scenarios:
(a) an instantaneous starburst and (b) continuous star formation.  All
models assume a Chabrier IMF (Chabrier 2003).  Our adopted $FUV-K_s$
cutoff was determined empirically (see the
discussion of Fig. 2 below), but corresponds to an instantaneous burst
of age less than $\sim$500 Myr in the case of solar metallicity
($\sim$ 1 Gyr for Z=0.004).  A much older epoch of continuous star
formation is also consistent with some of the bluest $FUV-K_s$ colors
measured within LSB zones. However, this scenario generally can be
excluded on the basis of $FUV-NUV$ (with observed values too red
[slightly positive] for the continuous SF model).  $FUV-K_s$ may also
be interpreted as a metric of the specific star formation rate
($sSFR$, Mu\'noz-Mateos et al. 2007), although the precise calibration
of the relation depends on the type of star formation history assumed.
Fig. 1 also shows predicted values of $FUV-R$ which demonstrate the
expected overall faintness of visible emission in XUV-disks, relative
to their UV appearance.

\item The distribution of relative size, $S(LSB)/S(K_{80})$, versus
Galactic extinction-corrected $FUV(AB)-K_s(AB)$ measured in the LSB
zone [$NUV(AB)-K_s(AB)$ in cases for which FUV data are unavailable]
is presented in Fig. 2. Note the tail of objects with large relative
size and blue $FUV-K_s$ for the LSB zone.  Lines segregate the objects
classified as Type~2 XUV-disks (or mixed-type XUV-disks).  The influence of the bulge component on the
$K_{80}$ contour most likely introduces a bias encouraging the
classification of early-type spirals as Type~2 XUV disks. However, the
observed distribution of points in Fig. 2 shows an abrupt dropoff just
below our cutoff of $S(LSB)/S(K_{80})$ = 7, suggesting that the Type~2
disks are extreme in terms of their disk formation history and T-type
dependent bulges influence our classification minimally.  Our Type~2
classification limits were determined empirically, but after the fact
we compared our measured characteristics to model predictions
describing inside-out disk formation (Mu\'noz-Mateos et al. 2007).
The anticipated locus of 'normal' galaxies of varied T is shown by the
shaded regions in Fig. 2.  Though there is minor overlap between the
Type~2 XUV-disk class and the model predictions, this analysis
demonstrates that Type~2 objects have an anomalously large LSB zone
even for disks assumed to be forming inside-out in a smooth continous manner.
Conversely, the measured $S(LSB)/S(K_{80})$ and color of the LSB zone for non-Type-2 galaxies generally agrees well with the model.


\item Our Type~2 XUV-disk definition does not exclude objects in which
the LSB zone is optically visible, even if
UV-bright clumps can been seen in other bands.
\end{itemize}

\subsubsection{Mixed-type disks and a caveat}

As mentioned above, some objects satisfy both Type~1 and Type~2 XUV-disk
definitions.  We refer to these galaxies as ``mixed-type'' XUV-disks.

It is also worth noting that the continuum of properties found in our
sample is such that there are also galaxies which fail the Type~2
XUV-disk definition but yet have a large, blue outer disk.  An example
is NGC~4625 which has an optically-LSB, outer disk which is blue
enough ($FUV-K_s$ = 2.2) but with a relative size $S(LSB)/S(K_{80})$ =
5.17 that does not meet our cutoff.  However, significant XUV-emission
is distributed beyond the UV threshold contour for NGC~4625 -
qualifying it as an XUV-disk according to our Type~1 definition.
Except for the typical $\Sigma_{SFR}$, this outermost SF appears
similar to that occuring within the UV threshold contour.  If these
regions had been included in the LSB zone, NGC~4625 would have also
met the Type~2 definition.  The purpose of making this point is to
emphasize that even objects belonging {\em exclusively} to a
particular XUV-disk type in our analysis may [to a lesser degree] have
some of the characteristics of the other type.  Furthermore, it
suggests to us a possible future approach to unifying the
classification scheme -- if a robust means of extending the limits of
the LSB zone definition (into lower $\Sigma_{SFR}$ areas) could be
devised.

\subsubsection{Application of the classification scheme}

In this section we present figures illustrating the method of XUV-disk
galaxy classification. Note that our identification of Type~1
extended UV emission remains subjective, so this section is also
intended to convey a sense of the decision making process.  Our Type~2
definition is fully automated.

NGC~5055 (M~63) is a SA(rs)bc galaxy which shows obvious XUV emission
of the ``M~83-like'' variety selected by our Type~1 definition. It has
a significant degree of structured, UV-bright emission outside an
$FUV$ ``threshold'' contour.  Figure 3 has panels showing the GALEX
$FUV$ data and a false-color image using the 2MASS-$K_s$, DSS2-red,
and DSS2-blue data, in addition to the $FUV$ and $K_s$ isophotal
contours described above.  Recall that the $FUV$ contour defines a
region outside of which multiple structured, UV-bright emission
complexes are considered as support for XUV-disk classification.  The
area between the $K_s$ and $FUV$ contours is where we measure the LSB
zone color. In the case of NGC~5055 the zone is too small and red to
satisfy our second XUV-disk definition.

NGC~2090, a Type~2 XUV-disk (Fig. 4), is a SA:(rs)b galaxy which has
only minimal UV emission located beyond the FUV threshold contour, yet
exhibits very blue colors over most of the disk inside this limit.  It
has an expanded LSB zone, extending over an area more than an order of
magnitude larger than the effective (80\%) size of the old stellar
disk.  Incidently, NGC~2090 is a target for which our Type~1 XUV-disk
definition is somewhat ambiguous.  UV emission is found beyond the
threshold contour (green line), however the observed structure (on the W
side of the disk) is a smooth continuation of the disk morphology just
inside the contour.  This type of ``fluff'' at the disk edge was
ignored in our classification analysis given that we are only able to
see such structure in relatively nearby galaxies.

Figure 5 shows an example of a galaxy {\em not} classified as an
XUV-disk.  The object, NGC~7418, is a typical SAB(rs)cd galaxy which
has an obvious color gradient as a function of galactocentric
distance, $r_g$ ($FUV-K_s$ decreasing with $r_g$).  The form of this
gradient is such that in the outermost part of the disk $FUV-K_s$
$\le4$).  However, the spatial extent of this blue ``fringe'' is
small, and $S(LSB)/S(K_{80}) < 1$.  Figure 5 also illustrates the
difficulty of distinguishing distant background galaxies from bonafide
outer disk clusters in the case of isolated UV clumps, not belonging
to a structure such as a clump complex or filamentary feature.  A few
examples of isolated UV-bright sources with little or no optical
emission can be seen in Fig. 5.  Recall that our Type~1 definition
requires structured UV emission features and therefore may (lacking
substantially higher resolution UV observations) miss some galaxies
which do support sporadic outer disk SF at such a low level that it
occurs only in isolated, unresolved clumps.  Such missed XUV-disk
galaxies may be represented by the emission-line dot (``EL Dot'')
phenomenon (Ryan-Weber et al. 2004, Werk et al. 2006).

Figure 6 shows a second example of a galaxy {\em not} classified as an
XUV-disk.  NGC~4736, a nearby (R)SA(r)ab;Sy2 LINER galaxy, has an
inner SF ring contained by the $K_s$-band contour (yellow line),
complemented by an extensive one-arm spiral composed of UV-bright
clumps starting at the circumnuclear ring and eventually reaching the
outer disk (beyond the green contour).  However, optical (DSS2-red)
imagery clearly shows a matching structure of moderate surface
brightness emission.  Our Type~1 XUV-disk definition allows for
optical emission observed cospatially with outer disk UV-clumps, but
we require a clearly different form for the evolved stellar population
traced by the longer wavelength data.  NGC~4736 is a borderline case
falling between tidally induced Type~1 XUV-disks such as NGC~1512 with
LSB underlying optical emission and non-XUV-disk ``trainwreck''
objects such as the Antennae (NGC~4038/39) with {\em in situ} SF
formation occuring inside luminous stellar tidal tails (Hibbard et
al. 2005). Because the future disk-like nature of extreme, equal-mass
mergers like NGC~4038/39 is questionable, we sought to keep them out
of the XUV-disk class.  The similar appearance of moderate surface
brightness optical emission having embedded UV clumps found in
NGC~4736 (versus the Antennae), led us to group them together. Even
though there is no direct evidence that NGC~4736 has experienced a
significant recent merger, it could have undergone merging in the
distant past.  A few other galaxies in our sample have UV/optical
morphology similar to NGC~4736.  In particular, NGC~1068 and NGC~1291
show UV-bright SF complexes within optically matched outer regions
reminiscent of merger activity.  Might these galaxies be examples
of objects which have experienced recurrent XUV-disk episodes and
built up an older stellar population at large galactocentric distance?
Assuming that repeat episodes of extended SF are unrelated to each
other, they should eventually form a quasi-smooth outer disk with
minimal resemblance to the most recent SF episode, making such a
hypothesis unlikely.

\subsection{Incidence of XUV-disk star formation}

Based on the classification scheme outlined above, we have identified
a set of 54 XUV-disk galaxies selected from our current sample of 189
targets.  This shows that XUV-disk SF is a common evolutionary process
in the spiral galaxy population. Our catalogue of XUV-disks also
allows looking for trends related to Hubble type and other galaxy
properties.

We found 30 objects satisfying only the Type~1 XUV-disk definition, 17
objects satisfying only the Type~2 definition, and 7 galaxies matching
both definitions (mixed-type).  Considering the union of the first and
last groups, we find that {\em approximately $\ga 20$\% [37/189] of
the nearby spiral galaxy population is supporting massive star
formation in outer disk locales beyond the classically expected star
formation threshold.  We emphasize this estimate of incidence for
M~83-like XUV-disk structure in our sample is a lower limit.}  The
ease of detecting the star-forming background galaxy population with
GALEX makes it difficult to rule out activity ocurring in very
isolated complexes consisting of a single clump (at our $\sim
5\arcsec$ resolution).  [Such background contamination is less of a
problem for H$\alpha$-based surveys which can select line emission at
the target galaxy's velocity.]  Also, note that some outer, low mass
clump complexes probably remain undetected in galaxies near our 40 Mpc
distance limit. Indeed, cataloged Type~1 XUV-disks are not
distributed evenly throughout our survey volume.  Ignoring Virgo
Cluster members (which set the median distance of the sample), we find
that 17 of 71 galaxies closer than 16 Mpc meet our Type~1 definition
whereas only 8 of 70 beyond 20 Mpc are classified as such.  This
implies that we have likely missed a few galaxies having very
inconspicuous outer disk SF complexes (similar to NGC~300) in the more
distant subset of our sample.  {\em Regarding the objects having an
exceptionally large, blue LSB zone (Type~2 and mixed-type XUV-disk
galaxies), we find that they comprise about 13\% [24/189] of our
sample.}  In our sample, distance effects do not seem to influence the
identification of Type~2 structures, given that 8 of 71 galaxies at D
< 16 Mpc satisfy the Type~2 definition, compared to 10 of 70 beyond 20
Mpc.  As noted above, {\em the overall incidence of XUV-disk star
formation is [54/189] $\ga$29\% in the local Universe.}

\subsection{XUV-disk SF in relation to galaxy properties and environment}

\subsubsection{Distribution in Hubble type and $L_K$}

In Figure 7a we show a histogram of T-type for the entire sample and
both groups of XUV disks.  Our survey sample is clearly weighted
toward intermediate type spirals.  {\em Type~1 XUV-disks are
distributed over a very broad range in T}, with galaxies occupying
nearly every bin with the exception of the extrema in T (for which
there are very few objects in our sample).  We refrain from measuring
relative incidence or internal strength of XUV features for Type~1
XUV-disk galaxies as a function of T, due to the limited number of
objects detected.  K-S testing indicates that the T distribution of
Type~1 XUV-disks matches that of the entire sample at the 82\%
confidence level.  {\em The distribution of Type~2 XUV-disk galaxies
is more skewed toward late-type spirals, with the vast majority of
detections having T $>$ 5.}  With K-S testing, we can exclude ($>99\%$
confidence) the possibility that Type~2 XUV-disks are drawn randomly
from the survey sample.  Note that our Type~2 definition for XUV-disk
classification does not highlight the entire population of very
late-type spirals as being peculiar, despite the trend for globally
integrated bluer $FUV-K_s$ with increasing T (Gil de Paz et al. 2007).
In fact, most sample galaxies with T $\ge$ 8 are not XUV-disks, though
this statement is complicated by the disappearance of a bulge
component in Sm galaxies which could influence our LSB zone
implementation.

The nearly ubiquitous observed distribution in T for Type~1
objects may suggest that the presence of a Type~1 XUV-disk has more to
do with the environment or interaction/accretion history of a galaxy
than then the host galaxy itself.

Figure 7b shows the distribution of our entire sample and that of
XUV-disks, color coded according to type, in the T-L$_K$ plane.  L$_K$
is a reliable, nearly extinction-free tracer for the stellar mass of a
galaxy (Aragon-Salamanca et al. 1993, Gavazzi et al. 1996, Bell \& de
Jong 2001).  There is no apparent offset between the XUV-disk galaxies
and the overall population, leaving open the possibility of episodic
XUV-disk star formation (with XUV active objects drawn from the same
parent sample).

There is a rough separation of the XUV-disk varieties in
terms of $L_K$.  This division is near L$_K$ = 10$^9 L_{\odot}$.
Assuming a mass-to-light ratio of 0.8 (Dale et al. 2007), it would
appear that galaxies are unable to (or already have undergone?)
support a [Type~2] galaxy-wide transformational burst of star
formation when their stellar mass becomes on the order of 10$^9
M_{\odot}$ or greater.  Boissier et al. (2001) demonstrated that more
massive galaxies have lower gas fraction, suggesting that Type~2
XUV-disks are formed only when the gas reservoir is large with respect
to the stellar population.

\subsubsection{Correlation with SFR measures and gas content}

Following up on the observation that Type~2 XUV-disks are
preferentially late-type spirals whereas Type~1 XUV-disks occur in
systems of all Hubble type, we sought to check for correlations with
other galaxy parameters (eg. $SFR$, $sSFR$, $SFR/M_{HI}$).  Although these
variables are correlated with T and $L_K$, XUV disks may still deviate
from the trend of the overall galaxy population.

Figure 8 presents a histogram illustrating the distribution of
integrated $SFR$ for the entire sample, in comparison to the
distribution for XUV disks of each type. We note that {\em despite
seemingly prodigious SF (in terms of the distinctive UV-NIR appearance
and generally widespread outer disk SF leading to their selection),
Type~2 XUV-disks appear to have a maximum $SFR$ of a few
M$_\odot$~yr$^{-1}$.}  Although the Type~2 outer disks are effectively
forming now, they are not ``bursting'' in the manner of nuclear
starbursts characterized by remarkably high star formation efficiency.
The integrated SFR values of Type~1 XUV-disks are dominated by SF in
the main disk, rather than the localized cluster complexes throughout
the outer disk, making the distribution of $SFR$ for these objects
less relevant.  It is important to emphasize that in all XUV-disk
galaxies, regardless of type, the fractional contribution to the total
SFR from SF outside of our estimated threshold contour is low.  The
widespread LSB zone SF in Type~2 XUV-disks is occurring {\em inside} the
threshold, and thus should not be compared directly to the level of
XUV-disk SF in Type~1 objects.

When cast into specific star formation rate, the remarkable short term
history of Type~2 XUV-disks becomes apparent.  Fig. 9a plots
($sSFR$,$L_K$) and illustrates that, {\em although the integrated
$SFR$ may be modest, such activity is sufficent to completely form the
stellar mass of a Type~2 XUV disk generally in less than 1 Gyr.}  The
specific star formation rate here refers to the entire galaxy.  Within
the blue LSB zone, the specific SFR is even higher.  Because we have
plotted the galaxy-averaged quantity in Fig. 9a, the Type~1 XUV disks
appear be very similar to the rest of the sample.  However, as noted
for Type~2 XUV disks already, the $sSFR$ of the outermost disk can be
substantially higher.  Of potential interest is the possible drop out
of XUV-disk galaxies in systems having globally-averaged $sSFR$ less
than 2$\times$10$^{-10}$ yr$^{-1}$, marked with a dashed line in the
diagram.  Of the 26 sample galaxies below this limit in the sample,
only 3 (11\%) are XUV-disk detections.  It does not seem this is a
detection threshold issue, as most XUV-disks are identified on the
basis of clumps which are individually significant, even if sparsely
distributed in the outer disk.  Two of the ways to support continued
star formation of any type (besides closed box recycling, e.g. Boselli
et al. 2001), are accretion from the IGM and gas deposition resulting
from the debris of galaxy interaction processes.  It could be that the
very low $sSFR$ galaxies are those which have started to exhaust their
(internal) gas supply and are not being augmented through such
external means.  If this is true, then the mechanisms to supply the
outer disk with gas (Beckman et al. 2004, Kere{\v s} et al. 2005,
Bournaud et al. 2005) would also be missing -- leading to the observed
lack of XUV-disk structures.

The gaseous content of the galaxies in our sample is well
characterized (Section 2.1). In Figure 9b we show the distribution of
$M_{HI}$/$L_B$ vs. $L_K$ for XUV-disks in comparison to ordinary
galaxies.  $M_{HI}/L_B$ is commonly utilized to classify gas-rich and
gas-poor systems.  {\em XUV-disks of all type (but especially Type~1)
are systematically displaced from the overall distribution in the
sense that they are gas-rich (by about a factor two from the survey
median at constant $L_K$).}  This enhancement is what one would expect
for the Type~1 XUV-disks -- if the outer disk gas is not present, we
would not see UV evidence for star formation unless the \HI was very
recently stripped (as might happen in a cluster environment,
e.g. Boselli et al. 2006, Crowl et al. 2006, Crowl \& Kenney 2006,
Vollmer et al. 2006).
Although the increased gas-richness in XUV-disks is not surprising,
our knowledge is sparse concerning the origin of the outer disk gas
and the mechanisms promoting SF therein.  It will be particularly
informative to study galaxies having extended \HI with {\em and}
without XUV-disk structure, looking for changes in their kinematics,
morphology, and CNM/WNM phase balance.

Figure 9c presents a measure of global star formation efficiency ($SFE$ =
$SFR/M_{HI}$) versus $L_K$.  {\em Type~1 XUV-disks tend to occupy lower
$SFR/M_{HI}$ than the median for the sample at all $L_K$, whereas
Type~2 objects are typical of the entire sample at low $L_K$.}  In Type~1 XUV-disks the average $SFE$ in the outer disk is certainly low (driving the global value down), but probably still reaches typical levels (for within an ordinary disk) on localized scales in HI arms and/or clumps.  Apparent gas
consumption timescales (1/$SFE$) are significant (several Gyr) for most of the
XUV-disks, though the issue is complicated due to star-to-gas recycling and the possibility of \HI production via H$_2$ photodissociation as discussed by Gil de Paz et al. (2007b) and originally by Allen et al. (1986).     

Figure 9 suggests that XUV-disk galaxies are generally gas-rich for
their stellar mass and that the SF activity which makes them
remarkable is occuring in a low SFE mode.  In Figure XXX, we
examine whether XUV-disk galaxies are more gas-rich than other
galaxies of similar global SFR, given the correlation of gas mass with
SFR.  Panel (a) shows $M_{HI}$/$L_B$ versus SFR and panel (b) presents
$M_{HI}$ versus SFR.  We conclude (with and without normalization by
$L_B$) XUV-disk galaxies do have enhanced HI content as a class,
relative to spirals of matching global SFR.

Galactic star formation activity is disclosed completely neither by
emission in IR nor UV bands (except in exceptional cases, e.g. ULIRGs
or UVLGs), but rather by the bolometric energy output of young stars.
It has been shown that the relative contributions of TIR and UV
systematically vary as a function of Hubble type (Gil de Paz et
al. 2007, Dale et al. 2007) and NIR luminosity (Cortese et al. 2006),
but with dispersion approaching $\sim$0.5-1.0 dex in TIR/UV.  What is
the origin of such scatter, far in excess of observation errors?  One
possibility is a mix of internal environments, and the relative
balance (by recent SFR) of these locales.  Both varieties of XUV-disks
possess a region in which the bolometric output is overwhelmingly
dominated by UV luminosity in deference to the IR.  Figure 11 shows
that {\em XUV-disks tend to have a relatively low TIR/UV ratio
compared to other galaxies of the same T-type and stellar mass (traced
by $L_K$), preferentially near the bottom locus of the overall
distribution.}  In Type~2 disks this difference in TIR/UV may be
wholly explained by the UV emission coming from the LSB zone (which
has relatively low dust content and low metal abundance). For Type~1
XUV-disks, the observed offset to bluer TIR/UV (by several tenths of a
dex) is likely due to more than just the contribution of UV emission
from the outer disk, which is very small percentage of total UV
luminosity.

\subsubsection{Evidence for interaction?}

Some XUV-disk galaxies described in Appendix~I are clearly
interacting galaxies, have obvious nearby companions, or show
morphological clues suggestive of external perturbation (optically, or
in the UV).  This should not be surprising, given the fact that
60\% of galaxies belong to a pair or group of massive galaxies
(Giuricin et al. 2000) and a higher percentage must have at least some
low luminosity companion.  Nevertheless, rather isolated galaxies do
exist (Galletta et al. 2006, Varela et al. 2004, Pisano \& Wilcots
1999,2003), even in the XUV-disk population. The situation is
complicated further because the mere fact that a galaxy has companions
does not imply that they have actually had any influence on the recent
star formation activity (100+ Myr timescales).  Boselli \& Gavazzi
(2006) review the important role of environmental effects on galaxy
evolution.  Note that in addition to the stimulation of SF, external
perturbation can remove outer disk gas, stopping SF and producing
truncated UV disks (e.g. NGC~4569, Boselli et al. 2006).

In order to quantify the potential effectiveness of
interaction-induced gas motions as a driver of outer disk SF, we
computed the tidal perturbation parameter, $f$ ($Q_D$ in Dahari 1984,
Salo 1991, Byrd \& Howard 1992), with our calculation following Varela
et al. (2000).  This measure specifies the ratio of the tidal force
imposed on a galaxy ($G$) of size $R$ by a possible perturber ($P$)
and the internal force per unit mass in the outer part of galaxy $G$.
Varela et al. (2000) demonstrated that $f$ can be successfully
computed using relative apparent magnitudes as a proxy for the mass
ratio of galaxies $P$ and $G$, as called for in the seminal papers
referenced above, and the projected separation rather than a true
three-dimensional impact (pericenter) parameter, $b$ (Icke 1985).  For
all galaxies in our sample, we computed the value of $f$ appropriate
for each possible perturber culled from the LEDA galaxy catalog.  Note
that the LEDA catalog is not homogeneous. In this sense, the
computed $f$ might depend strongly on whether or not SDSS catalogs
with coverage for any galaxy of interest have been incorporated into
LEDA. We eliminated clearly unassociated galaxies for each primary in
the same manner as Varela, using checks on the Virgo-corrected
relative radial velocity ($V_{vir}$), spatial size ($D \ge 2$ kpc),
and absolute magnitude ($B \le -12$).  We record the maximum value of
such an array of $f$-values for each sample galaxy, and interpret this
as an indicator of strength for the dominant interaction.  Varela et
al. (2000) denoted galaxies with $f < -4.5$ to be ``isolated'' or,
more precisely, tidally unaffected in the recent past.  They also
defined galaxies with $f > -2.0$ to be ``perturbed''.  
In Fig. 12 we show the distribution of $f$ for all the galaxies in our
sample and for the two classes of XUV-disk galaxy.  We find that {\em
XUV-disks with Type~1 structures have a relatively flat $f$
distribution, despite the strongly peaked $f$ distribution (centered
near $f = -4.5$) for the entire galaxy sample}.  However, a K-S test
shows that this difference is probably not significant -- the Type~1
$f$ distribution agrees with the entire sample at 91\% confidence.  If
Type~1 structures are linked to external perturbation, it is of a
level commonly encountered in the spiral galaxy population, and not
of the strength formally defined as perturbed by Varela et al. (2000).
On the other hand, the {\em Type~2 XUV disks are slightly more
isolated than not (only 29\% chance agreement with entire sample)},
though a wide range in $f$ is still seen for these galaxies.  We note
that the $f$-distribution of Type~2 disks matches that of late-type
spirals (91\% confidence), so the isolation may simply be a reflection
of the progenitor galaxy population.

The tidal perturbation parameter is not the only way of quantifying a
galaxy's environment, and indeed has some limitations.  Note that the
tidal perturbation parameter would not classify a merger remnant as
perturbed (since each merging object would no longer be distinct,
e.g. XUV-disk NGC~2782). Also, a galaxy may live in a high density
location such as a group, without a current perturbation.  For
instance, NGC~3198 is the brightest galaxy in a group of 5 objects,
but is labeled isolated according to the tidal perturbation parameter.
In the framework of hierarchical galaxy assembly, one might expect
that the frequency of galaxy interactions leading to accretion would
scale with the number of neighbors constituting a group, or even more
so with group density.  Fig. 13 shows the distribution of $f$ versus
the number of galaxies in a NOGG group. We find the most isolated
galaxies become increasingly rare in richer groups.  Second, perturbed
galaxies occur in groups of any size. {\em XUV-disks can be found in
galaxy groups both rich and poor}.  In fact, some of the dynamically
isolated XUV disks shown in Fig. 12 are actually located in groups
containing 10 or more galaxies (e.g. NGC~4254 in the Virgo cluster and
NGC~4414 in the ComaI group).

Finally, a galaxy could be perturbed by undetected low mass / low
surface brightness objects in their vicinity, or even high velocity
clouds with gas but no stellar component.  Deep, dedicated survey
observations to disclose any such perturbers are underway, but
presently we looked for morphological clues suggestive of interaction.
We have noted (in Table 2) those XUV disk galaxies which have
peculiarities at UV/optical wavelengths, or which are known through
the literature to be interacting in some manner.  Although based on
heterogenous data and analyses, {\em more than 75\% of the Type~1 XUV-disk
objects show morphological or \HI evidence for interaction/merger or
minor external perturbation of some sort} (e.g. HVCs).  {\em About half of the
Type~2 XUV-disks appear to be recently disturbed}.  In at least some of
these cases, the tidal perturbation parameter falsely indicates that
they are isolated.  Note that we did not scrutinize non-XUV galaxies
in our sample at the same level of detail, as this is beyond the scope
of our paper.  For this reason, the possible importance of
interaction/merger or external perturbation in XUV-disks must remain
suggestive only.

\section{Discussion}

\subsection{Efficacy of SF tracers at low SFR}

Our GALEX observations demonstrate that the domain of star formation
activity in some spiral galaxies has been underestimated, both
spatially and in terms of azimuthally averaged gas column density.
Note that outer disk star formation was previously detected in a few
galaxies (e.g. Ferguson et al. 1998) using deep H$\alpha$ images,
which make it possible to detect a single O star (see Gil de Paz et
al. 2007b for M~83) at distances well beyond the Local Volume.
However, GALEX shows such SF is relatively commonplace.

As shown by Boissier et al. (2007), H$\alpha$
observations of the \HII region population become unreliable in
regions of very low $\Sigma_{\rm SFR}$, merely due to the short
timescale (t$_{HII}$) over which any one \HII region remains visible.
As $\Sigma_{\rm SFR}$ is decreased, there comes a point at which the
probability of an \HII region forming during the preceeding t$_{HII}$
in the (finite) area of interest drops substantially below unity.  In
contrast, UV emitting B stars survive more than an order of magnitude
longer than the O stars responsible for ionization -- increasing the
odds of detecting rare, but still present, star-forming regions.  Our
UV observations, and such an interpretation, are consistent with the
limited population of outer disk \HII regions known in a few galaxies
(Ferguson et al. 1998, Ryan-Weber et al. 2004, Gil de Paz et
al. 2007b).  Moreover, we emphasize that the non-detection of line
emission from any given object (even with good H$\alpha$ sensitivity)
should not be taken as evidence that quasi-steady-state SF is ruled
out at very low $\Sigma_{\rm SFR}$.  Even UV imaging has a similar
limitation - though at much lower $\Sigma_{\rm SFR}$ - beyond which
only CMD analysis for resolved stellar populations can effectively
recover knowledge of past SF activity.

The size of the particular region being surveyed has an impact on
practical limits for the SF tracers because it is the integrated SFR
which ultimately drives the counting statistics of \HII regions and
older (non-H$\alpha$-emitting) clusters.  
In the low SFR regime,
each tracer eventually suffers from stochastic incompleteness and
uncertainty, which cannot be overcome by more sensitive observations.

We have conducted simple Monte Carlo simulations to model the
visibility of a continuously star-forming population with UV and
H$\alpha$ tracers, subject to the size-of-sample selection effect
described above.  We assumed a cluster formation rate (CFR, clusters
Myr$^{-1}$), cluster membership function (CMF), and a Salpeter IMF.
See Thilker et al. (2002) for details of the modeling procedure.  The
number of stars per cluster was allowed to reach down to one.  We
tabulated the number of clusters having one or more ionizing O star
and the number of clusters with surviving OB stars.  These metrics
were evaluated for many independent snapshots throughout each model
run.

In Figure 14, we plot the probability distribution function and
cumulative probability distribution function for the expected number
of clusters still containing O stars, given continuous SF with
average log(SFR) equal to -4.5, -4.0, -3.5, and  -3.0 [M$_\odot$
yr$^{-1}$].  The assumed CMF can have an effect on the predicted
population of O star hosting clusters.  Panels (a-h) of Fig. 14
present two models for each SFR, differing only in the treatment of
the upper CMF.  The slope of the CMF in both cases was set to -2.
However, it is plausible that the most massive/populated clusters may
be inhibited from forming in the outer disk due to a lack of massive
cool clouds.  Braun (1997) showed that the CNM disappears rapidly
beyond the $D_{25}$ radius, leaving \HI predominantly in the form of WNM.
Therefore, we conducted alternative model runs with the cluster
membership function truncated such that the maximum cluster mass is
10$^3$ M$_\odot$.  The default models have a CMF allowing up to 10$^6$
M$_\odot$ as the maximum cluster mass.  In Fig. 14, the default models
are shown in the left column (panels a-d) and the truncated CMF models
are shown in the center column (panels e-h).  We note that for the
default model in low $SFR$ environments (panel d), such as
XUV-disks, the probability of {\em not finding} O star hosting clusters in
the region being studied reaches as high as 80\% despite ongoing star
formation. Truncating the CMF has the effect of increasing the typical
predicted \HII region population, because less mass is
locked up in massive clusters which in turn allows for the more
frequent creation of typical clusters.  

The assumed IMF has a more pronounced effect on the expected
population of O stars at any given time in a low $SFR$ environment,
especially if the maximum stellar mass is a function of cluster mass.
Some theoretical and observational studies support such an effect
(Bonnell 2000), but the idea is still under debate.  For instance,
Elmegreen (2006) reported that the average maximum stellar mass in a
cluster population decreases with decreasing cluster mass, but that
the absolute maximum stellar mass is roughly constant.  In panels
(i-l) of Figure 14, we show plots equivalent to those in panels (e-h :
restricted CMF) except that we imposed a lower limit of 100 stars with
M$>$M$_\odot$ for the initial cluster membership necessary to produce
stars with M$\ge$ 10 M$_\odot$.  Figure 14(i) shows that even for
integrated SFR large as $1\times10^{-3}$ M$_\odot$ yr$^{-1}$
approximately 40\% of the time a region of interest (e.g. sized 1
kpc$^2$ for $\Sigma_{\rm SFR}$ = $1\times10^{-3}$, 10 kpc$^2$ for
$\Sigma_{\rm SFR}$ = $1\times10^{-4}$) will lack an O star hosting
cluster (and associated \HII emission).  At lower SFRs (panels j-l)
the probability is 70--95\%.

The simple simulations presented here represent the best-case
scenario for detection of \HII regions in low $\Sigma_{SFR}$
environments, as our current model has no allowance for extinction,
gas density effects, or variation in overall population age (via the
SFH, hence CFR(t)). Further, the simulations do not account for observational
sensitivity.  In reality, [younger] clusters still hosting at least a
single O star probably have non-negligible extinction, and would
certainly be subject to systematically more attenuation than old
clusters which eventually disperse from the dust associated with their
birthplace.  Initial estimates of extinction toward line-emitting
regions in the XUV-disks of M~83 and NGC~4625 show that values of
\ebv at the level of a few tenths of a magnitude are common (Gil de
Paz et al. 2007b).  The outer disk ISM is substantially more tenuous
than typical environments at smaller galactocentric distances.  Such
low density can have the deleterious effect of dimming the surface
brightness of H$\alpha$ emission associated with \HII regions
(Elmegreen \& Hunter 2006), and in principle could encourage leakage
of Lyman continuum photons for porous or otherwise density-bounded
clouds.  However, it is unclear how much these effects will be
actualized, given that the SF occurs within local
overdensities, even if the general outer disk environment is of quite
low average density.  Gil de Paz et al. (2007b) demonstrate that the few \HII regions found in the XUV-disks of M~83 and NGC~4625 are compact, in keeping with this notion of localized SF. Our current Monte Carlo runs assumed a constant SFR,
although cluster formation times are randomized.  Clearly, models
computed for the period following a burst-like SFH (whether
instantenous, Gaussian, or declining over time) will tend to
exacerbate the bias against detecting \HII regions in the outer disk
illustrated by our current models and Boissier et al. (2007).
Elmegreen \&
Scalo (2006) showed that the (apparent) present-day mass function
(PDMF) is influenced by SFH.

Our Monte Carlo models also allow tracking the number of clusters
containing less massive, but nevertheless UV-emitting, B stars.  As we
noted above, {\em even UV emission may eventually falter as a SFR tracer at
quite low SFRs (when the typical probability of having a cluster
containing one or more B star drops below unity).}  Figure 15 shows the
bivariate PDF representing the predicted number of clusters containing
at least one O star and the number of clusters containing one or more
OB stars.  For moderate--large SFRs (not shown), the N(OB)/N(O) ratio
is effectively constant for the continuous SF scenario.  At low SFRs,
N(O) becomes stochastically limited first (near log(SFR) = -3.5)
followed by N(OB), which starts showing a significant fraction of
``misses'' by log(SFR) $\sim$ -5.5.  Does the universe have any
environments for which this stochastic limit to the UV-emitting
cluster population is relevant?  Adopting the Schmidt Law
parameterization of Boissier et al. (2007), the form of which employs
total (HI+H$_2$) gas surface density, we find that $\Sigma_{(HI+H_2)}$
= 0.1 M$_{\odot}$ pc$^{-2}$ corresponds to a predicted log
($\Sigma_{SFR}$) = -6.0 [M$_\odot$ yr$^{-1}$ kpc$^{-2}$].  Obviously
this prediction neglects the possibility of a star formation
threshold.  Such values of $\Sigma_{(HI+H_2)}$ and $\Sigma_{SFR}$ are
a factor of several below the observed limit of each variable in all
studies to our knowledge, even in the XUV-disk environment.  Thus, it
would seem (at face value) that the stochastic limit of the UV SFR
tracer is most likely never encountered in the overall galaxy
population.  It is remotely conceivable though that this could be a
selection effect, associated with the fact that \HI observations of
nearby galaxies almost never probe such low gas surface densities (one
exception is M~31, see Braun \& Thilker 2004).  We know that low
N$_{\rm HI}$ gas does surround galaxy disks, from QSO absorption line
measurements and directly in emission as demonstrated by Fig. 12 of
Braun \& Thilker.  Although we do not consider it likely, it is worth
mentioning that extremely isolated cool clouds could form
quasi-continuously in the typically warm and diffuse low column
density halo gas and subsequently be missed (as \HII regions are) due
to their short lifetime.  In this case, the stochastic limit of UV SFR
measurement could indeed be relevant.

Radial surface brightness profiles for the first-known XUV-disks, M~83
and NGC~4625, originally raised the issue of potential discrepancy
between H$\alpha$ and UV emission.  In hindsight, {\em the relative
dropout of the H$\alpha$ profile may not have signaled a fundamental
change in the star formation properties between inner and outer disk,
but rather spurned a new view regarding efficacy of tracers for the
young-moderate age populations which do exist and for the remarkable
persistence of star formation in any environment which is locally
populated by a cool ISM component.}  Above we noted
reasons for the lack of detection of \HII regions, or for suppresion
of their formation, but we emphasize that \HII regions are
sometimes present (albeit rarely) in the XUV-disk environment and can
be observed if one images deeply enough (e.g. NGC~0628, Ferguson et
al. 1998; M~83 and NGC~4625, Gil de Paz et al. 2007b).

As noted in the description for a few XUV-disk galaxies studied in
this paper, we detect UV emission from young stars in portions of the
galaxy not even reached by traditional surface brightness profile
analysis (based on the same observations, with profile extent
conservatively set by error limits).  Such sources are located in
regions of the XUV-disk having azimuthally averaged $\mu(FUV) > 30$
mag arcsec$^{-2}$.  At such low $\Sigma_{\rm SFR}$ star-forming
clusters are often difficult to distinguish from the population of
background galaxies, except if there are clusters clumped into a
larger structure.  This demonstrates that, even with GALEX, it will
remain difficult to confidently determine the maximum spatial extent
of star formation in disk galaxies.  A combination of GALEX imaging
and higher resolution follow-up is required for this purpose.  HI maps
can also be used for guidance as to which extended locations are most
likely to host SF.  These considerations suggest that in order to
achieve a robust surface brightness profile in the outermost portions
of a galaxy having sparsely distributed clusters one must resort to
masking of contaminating sources via inspection.  However, we
emphasize that for outer radii (with a negligible source covering
factor) obtaining a azimuthally averaged profile is nearly a moot
point as it is no longer an appropriate description of the
environment.  Rather, {\em the SF characteristics of outer disk, low
$\Sigma_{\rm SFR}$ locations would be better described in a
probabilistic manner based upon the observed PDF for UV emission as a
function of variables such as local $\Sigma_{gas}$, $\Sigma_{stars}$,
and dynamical quantities.}

\subsection{SF thresholds and disk instability}

Given the discussion above concerning SFR tracers in the XUV-disk
environment, how much evidence remains to support the much-debated SF
threshold in all its incarnations (e.g. azimuthally averaged
vs. local, dynamically variant vs. fixed critical column density)?

At the very least, XUV-disks higlight the long-recognized but often
overlooked fact that the Toomre criterion for gravitational stability,
though derived in a one-dimensional manner (radially), should not be
interpreted in the same sense.  It does {\em not} allow one to
delineate an outer zone devoid of all SF.  GALEX shows extensive SF in
subcritical zones for which $Q>1$.  In fact, Kennicutt (1989) and
Martin \& Kennicutt (2001) concluded early that localized peaks in the
gas distribution at otherwise subcritical radii can be gravitationally
unstable and form stars.  In their work on NGC~6822, de Blok \& Walter
(2006) noted that an inherently two-dimensional version of the Toomre
criterion would be ideal, but can only be partially attained given
that the disk rotation velocity cannot be determined on a
point-by-point basis for position angles removed substantially from a
galaxy's major axis. Nevertheless, de Blok \& Walter showed the
improvement in identifying SF-susceptible sites achieved just by
comparing the local $\Sigma_{gas}$ to an azimuthally averaged
$\Sigma_{crit}$, from the traditional Toomre formalism.  To summarize,
it is not surprising that SF persists in outer clumps of gas having
local $\Sigma_{gas}$ $\ge$ $\Sigma_{crit}$ embedded in an
otherwise subcritical environment.

The detailed ISM structure in the outer disk zone is fundamentally
what needs to be considered when judging if SF is likely or unlikely.
  Elmegreen \& Parravano (1994) first suggested
that the {\em local} phase balance of the ISM is directly tied to the
presence or absence of star formation.  Specifically, they showed that
a two phase \HI medium (allowed under conditions of sufficient thermal
pressure) which contains a cool component is a prerequisite to SF.
This view has been recently revived by the modeling of Schaye (2004),
who derived a local column density required for the formation of CNM.
The precipitous decline of the fractional amount of \HI contained in
the cool phase at the edge of the optical disk ($\sim$$D_{25}$ radius) was
first reported observationally by Braun (1997).  {\em The notion that local
conditions are most crucial in determining the observed $\Sigma_{SFR}$
meshes very well with our GALEX observations.}

How then should one best model the anticipated radial distribution of
star formation rate, given the apparent importance of local
conditions?  Elmegreen \& Hunter (2006) presented a multicomponent
model of star formation in disk galaxies which included not only
fundamentally gravitational (Toomre) instability but also allowance
for secondary pathways to cloud collapse, such as spiral wave shocks,
turbulence compression, and localized triggering -- all of which can
persist into the outer disk.  They assumed a lognormal $\Sigma_{gas}$
distribution matched to the average $\Sigma_{gas}$ at each radius,
which naturally allows for a limited fraction of high density regions
even at subcritical average $\Sigma_{gas}$.  Such a lognormal
distribution is appropriate for turbulent media.  The Toomre criterion
was then applied in a local manner to modeled gas surface densities,
to see if the sites of secondary cloud formation became
gravitationally unstable.  A standard Schmidt law paramerization of
the SFR density appropriate for unstable clouds was assumed
thereafter.  {\em The Elmegreen \& Hunter (2006) models appear largely
capable of reproducing diverse SFR profiles such as seen by GALEX for
XUV-disks and by Hunter et al. (2006) for irregular galaxies.}

Even if GALEX UV and earlier H$\alpha$ observations (e.g. Ferguson et
al. 1998) together show that SF frequently occurs well beyond the $D_{25}$
radius and into subcritical radii, {\em we suggest that the Toomre
criterion does rather accurately predict the galactocentric distance
at which ISM properties change from an \HI distribution effectively
dominated by massive cool clouds to one in which the warm diffuse
medium becomes prominent and, as such, has an impact on the SFR (UV
and H$\alpha$) profiles.}  
Braun (1997) showed that this CNM/WNM transition is also associated
with a drop in \HI mass surface density.  The models of Elmegreen \&
Hunter (2006) already indicated that breaks in the SFR profile can be
linked to the general loss of Toomre's gravitationally unstable
clouds, but they assumed a (conservative) smoothly varying gas profile
(exponential or exp. with transition to 1/R) .  In reality, the HI
column density often drops faster than this for a limited range in
galactocentric distance near the $D_{25}$ radius, leading to a
jump/discontinuity in SFR density profiles.  This true discontinuity
worked in concert with the stochastic, size-of-sample effect and other
factors mentioned in Section 4.1 to give the appearance of a genuine
SF threshold (e.g. Martin \& Kennicutt 2001).

In summary, with GALEX we are beginning to probe outer disk
environments with average $\Sigma_{gas}$ low enough (e.g. at $\Sigma_{gas} \le 0.1$ M$_\odot$
pc$^{-2}$ levels near the current limits of \HI mapping) to effectively
exclude star formation  except within
anomalously high density cloud complexes.  {\em Previous claims of a star
formation threshold near the $D_{25}$ radius appear to have marked a
transition in the typical mode of cloud collapse leading to star
formation.  The magnitude of this transition was overly accenuated by
the use of H$\alpha$ as a SFR tracer.}

\subsection{Does galaxy type or environment matter most?}

It appears that galaxy type is of little importance with regard to the
genesis [and support, if SF is not episodic] of sparse, spatially
extended star formation, given the detection of Type~1 XUV-disks
associated with spiral systems of all type in the S0-Sdm range
(Section 3.3.1).  Rather, {\em what appears crucial is the means to
supply a rich, widely-distributed reservoir of gas, whether through
(tidal) removal of the interstellar medium from the ordinary disk or
infall/accretion processes}.  Clearly, in many cases (e.g. NGC~1512,
M~81) galaxy interaction is responsible.  For the majority of our
XUV-disk class the source of outer disk gas is not so obvious.  We do
see evidence for some type of external perturbation (HVCs; low mass,
LSB companions) in 75\% of Type~1 XUV disks and in about half of the
Type~2 objects, but it is unclear if this minor level of perturbation
can provide the needed amount of gas into the outer disk.  It seems
more likely that the minor external perturbation could be enough to
stimulate XUV-disk SF if the gas was already present, but quiescent.

Gaseous accretion from the IGM is also expected (e.g. Keres et
al. 2005) in two dissimilar modes, hot and cold.  Cold mode accretion,
in which the gas is not shock heated to the virial temperature of the
system, is thought to dominate for lower mass halos (log
M$_{baryonic}$ $<$ 10.3 M$_\odot$, or M$_{halo}$ $\la$ 11.4
M$_\odot$).  The observed lack of Type~2 XUV-disk systems at L$_K$
$\ga$ 10$^9$ L$_\odot$, and resultant skew toward late-type disks,
hints that perhaps these galaxies are those which are now growing
substantially through efficient cold accretion gas feeding along IGM
filaments.    Looking from a
galaxy-wide integrated perspective, we showed in Section 3.3.2
(Fig. 9b) that XUV-disks are nearly twice as gas-rich as non-XUV-disk
galaxies having similar Hubble type.  This could indicate that
whatever mechanism is supplying the gas needed for XUV-disk formation
has been working for a prolonged duration to achieve this level of
richness, as it seems unlikely that a short term event would generally
be capable of doubling the gas content without leaving signs of a past
major merger.  The elevated $M_{HI}/L_B$ values for XUV-disk
galaxies are not caused by a decrease in $L_B$ at fixed
T-type.

Returning to the issue of galaxy interactions and/or potentially minor
perturbations, a few comments are appropriate.  We mentioned above how
NGC~1512 and M~81 show outer disks which have been tidally structured.
Other systems (e.g. NGC~2782, NGC~4483, NGC~6239, NGC~1140) appear to
have suffered a quite significant, but dissimilar galaxy interaction,
such that the spatially extended star formation we note as XUV-disk
structure comes only from tidal or peculiar features rather than a
true disk.  Will such objects return to being otherwise normal
galaxies after the current burst of XUV-disk SF ceases?  Many ordinary
galaxies having peculiar \HI arms are found in the literature -
objects which could have already experienced a burst of SF in the
(tidal) \HI features.  Finally, some XUV-disks show evidence for
perturbation of a less encompassing variety.  Bekki \& Chaiba (2006)
numerically modeled the influence of dark subhalos in orbit around a
galaxy, finding that fine structure and holes can be created in the
extended \HI distribution provided that characteristic subhalo is
massive enough.  We consider this hypothesis viable for features of a
disorganized nature, but note that many of the XUV-disks show large
scale spiral morphology in their outskirts.  Dark impact such as that
modeled by Bekki \& Chaiba (2006) could not produce such coherent
gaseous filaments nor the following star formation.

As it may provide a clue to the specific mechanisms which promote or
trigger outer disk star formation in the (azimuthally-averaged)
subcritical regime once gas is in place, we note that {\em some extended HI
disks are known which do not exhibit any sign of recent star
formation.}  Examples are NGC 2915 and NGC925 (as noted in Boissier et
al. 2007).  Given that galaxies such as these exist, will they remain
stable?  Have they formed extreme outer disk stars before, then
transitioned to a quiescent phase?  Or did they recently acquire their
gas, on a timescale too short for SF to have become organized?  We do
not yet have a good handle on which of these scenarios is correct, or
even if a single explanation suffices.  
{\em Even if we assume that XUV-disk SF is episodic, we cannot
yet constrain the duty cycle except to say that some bursts seemingly
last for a minimum of a few times 10$^8$ yr (considering the detection
of both H$\alpha$ and aging UV sources), and that the relative amount
of time spent in the quiescent phase is probably small for galaxies
with extended HI.}  This last assertion is based on the fact that only
a handful of the extended \HI galaxies of which we are aware in the
present sample (Table 1) do not show XUV-disk structure.  Note that
there could also be galaxies in which XUV-disk SF has largely consumed
the available outer disk gas, leaving only an imprint in the radial
distribution of stellar mass and characteristic age (e.g. Mu\'noz-Mateos
et al. 2007).  If inside-out disk formation models are accurate, most
spiral galaxies may have already experienced XUV-disk SF to at least
some degree.  

\subsection{XUV-disks and LSB galaxies}

The preceeding discussion focused on the possible relation of XUV SF
and ordinary galaxies, but there are also reasons to
consider atypical objects.  The apparent similarity of physical
conditions (low $\Sigma_{SFR}$, low metallicity, sparse
morphology) in Type~1 XUV-disk environments and massive
LSB (MLSB) galaxies makes comparison of these objects compelling.  Do
MLSB galaxies represent an extreme variant of the XUV-disk class?

Barth (2007) recently challenged the current understanding of the
prototype MLSB Malin 1's galactic structure based on HST/WFPC2 I-band
(F814W) imaging, questioning the central surface brightness and
scalelength of the disk.  He concluded that a brighter and more
steeply declining disk component had been missed in ground-based
images, with the bulge actually more compact than thought before.
Comparing his revised estimates for Malin 1's disk parameters with
those of other S0 galaxies (Aguerri et al. 2005), Barth reported the
underlying inner disk was unremarkable, and not faint enough to
qualify Malin 1 as an LSB galaxy.  GALEX observations of Malin 1 show
faint UV emission distributed over the extent of the \HI disk, such
that Malin~1 certainly should be classified as an XUV-disk according
to our criteria.  {\em If it can be shown that other massive LSB
galaxies categorically possess small scalelength evolved stellar disks
like that of Malin 1, their faint outer regions quite likely would
also be well described as XUV-disks.}  NGC~262 is a MLSB
galaxy in the GALEX Atlas (Gil de Paz et al. 2007) which is similar to
Malin~1, but not included in our sample because of distance.  {\em
MLSB objects could merely be the high mass tail of the XUV-disk
population.} No matter what their classification, MLSB galaxies again
demonstrate the pervasive nature of SF -- even in environments which
are characterized by overall low density.

There may also be lessons to be learned from XUV-disks in the context
of LSB galaxies at the low end of the mass spectrum, namely LSB dwarf
galaxies.  Heller et al. (2001) completed an H$\alpha$ survey of LSB
dwarfs in the Virgo Cluster, finding small integrated SFRs (several
$\times$ 10$^{-3}$ to 10$^{-2}$ M$_\odot$ yr$^{-1}$) with the bulk of
their emission coming from a few discrete \HII regions per target (if
at all).  Just as the low $\Sigma_{\rm SFR}$ values characterizing the
XUV-disk zone in giant spirals can lead to a typical probability $<1$
of catching a stellar cluster still young enough to produce an \HII
region, {\em we caution that some fraction of the LSB dwarf population may have recent SF
which has been missed entirely, or a time-averaged SFR which has been
underestimated, due to stochastic effects.}  A comprehensive survey of
LSB dwarf galaxies in the UV is needed to assess this possibility.

\subsection{XUV-SF and the process of disk-building}

What is the evolutionary impact of XUV-disk star formation over
cosmological time periods, and how is this influence reflected in the
non-UV properties of a galaxy?  Clearly the answer will vary depending
on the length of time over which the XUV-disk star formation has been
sustained, and the degree of SF intermittency during this time.
{\em The high specific star formation rate ($sSFR$) of Type~1
XUV-disks beyond the threshold contour and overall for Type~2
XUV-disks suggest that XUV-SF is an important disk building
mechanism.}  This process is much harder to study in the distant universe (e.g. Trujillo \& Pohlen
2004) at an epoch during which the bulk of galaxy disks were
assembled.

There has been much contemporary progress on evolved stellar
populations in the outer disk, including insights possibly linked to
XUV-disk star formation.  The low surface brightness radial extent of
the evolved stellar disk is now known to be much larger than
previously thought and exhibit a variety of structure in terms of
surface brightness profile shape.

Ibata et al. (2005) reported an exponential disk-like
component in M~31 extending to at least 40 kpc radius (about twice the
previously known size) having kinematics consistent with the expected
circular motion at the radii probed.  The stellar population of a
small section within the outer M~31 disk was found to have an intermediate age of 4-8 Gyr by Brown et al. (2006).  A vast, purely
exponential disk of very low surface brightness was also recently
suggested in NGC~300 by Bland-Hawthorn et al. (2005), reaching out to
ten times the optical scalelength.  

Simple exponential profiles, however, are rare (10\% according to
Pohlen \& Trujillo 2006).  Disk galaxies with profiles well described
by a broken exponential are now thought to consitute a majority of the
disk population (Pohlen et al. 2002, Pohlen et al. 2004, Pohlen \&
Trujillo 2006, also Hunter et al. 2006 for Im galaxies) -- perhaps
reflecting a change of star formation and disk assembly processes at
the break radius, and generally emphasizing the complicated nature of
disk-building processes.  Of the broken exponentials, some profiles
are up-turned at the break radius and others are down-turned.
Down-turned broken exponentials are easier to explain in terms of a SF
threshold framework (or, alternatively, the new view of reduced SF
efficiency due to large scale zones with $Q > 1$ and reliance on
secondary, non-gravitational cloud collapse mechanisms therein).  They
also completely replace the sharp radial cutoffs proposed by van der
Kruit (1979). Of perhaps even more relevance to XUV-disks, Erwin et
al. (2005) and Pohlen \& Trujillo (2006) discovered ``anti-truncated''
spiral galaxies, disk systems with surface brightness profiles
best-fit by a double exponential in which the outer disk has a larger
scalelength than the inner disk.

Naively, one would suppose that up-turned, anti-truncated optical
disks might be associated with galaxies which are extended in terms of
their gas distribution, whereas downturned galaxies have a
consistently declining $\Sigma_{gas}$ profile. This originally
suggested to us (Thilker et al. 2005b) that anti-truncated disks might
be the evolved counterparts of XUV-disks.  Tantalizingly, the
Elmegreen \& Hunter (2006) SF model predicts double exponential SFR
profiles with a downturn in the outer disk for the case of a purely
exponential radial decline in average $\Sigma_{gas}$, but also
predicts purely exponential or even broken, up-turned profiles in the
case of galaxies having a more gradual (1/R) drop in the gas
distribution.  They further confirmed the expectation that the stellar
mass surface density profiles built up by SF over many Gyr have a form
resembling the predicted SFR profile.

However, {\em the observed relation between XUV-disk SF and optical
profiles remains unclear at this point}.  In XUV-disk galaxy NGC~300,
the GALEX surface brightness profiles in FUV and NUV exhibit a
down-bending break near the $D_{25}$ radius (Fig. 3 of Gil de Paz et
al. 2007), but Bland-Hawthorn et al. (2005) report a pure exponential
form of the underlying evolved stellar disk out to well beyond this
radius.  Some of our sample galaxies were studied optically by Pohlen
\& Trujillo (2006).  UGC~04393, optically fit with a pure exponential
by PT06, is classified as a mixed-type XUV-disk.  In
the case of Type~1 XUV-disk NGC~3359, categorized as down-bending
optically (their type II-AB) by PT06, we observe well-defined outer
spiral structure in the GALEX UV imaging.  Even more perplexing is
NGC~2541, which we designate as a Type~2 XUV-disk and PT06 call a
down-bending optical exponential (their type II-CT).  Five of the
up-bending profile galaxies of PT06 (their type III and III-d) are
included in our survey sample.  They are NGC~1084, NGCC2976, NGC~4030,
NGC~5713, and NGC~5806. Only one of these targets (NGC~2976) exhibits
clear disk-like properties at large radii in the optical (III-d), but
nevertheless we find it intriguing that none of these galaxies are
classified as XUV-disks.  This could imply that, if PH06 type III
galaxies are evolved XUV-disks, there comes a time when XUV-disks run
out of gas to convert into stars or at least that quiescent periods
last for a few hundred Myr.

In any case, complex outer disk optical profiles demonstrate that SF
in low-density outskirts of a galaxy is commonplace, or more precisely
that it was at some time in the past, and that such outer disk SF is
supported to varying degrees (perhaps determined by availability of
gas through accretion and/or interaction).

\subsection{Morphological structuring of XUV emission and HI}

Typically, the spatial extent of outer disk SF matches
the apparent \HI size of a galaxy evaluated at the sensitivity
attainable with the VLA at GALEX resolution (N$_{\rm HI}$ $\sim$
10$^{20}$ cm$^{-2}$), {\em with UV peaks generally following large scale HI
emission structures but not necessarily corresponding with \HI clumps
individually}.  This is expected, however, as the mechanical luminosity
of the clusters (from SNe and stellar winds) should be sufficient to
create holes or bubbles in the local distribution of gas and
effectively break down the Schmidt Law on small scales.
Again, though an extended \HI distribution appears
to be a XUV-disk prerequisite in all but the most bizarre cases
(e.g. NGC~4254), we reiterate there are extended \HI galaxies
without evidence of recent SF in their outer disk (NGC~925, NGC~2915).

Despite the large area over which XUV-disk star formation proceeds,
established by the \HI disk size and CNM content therein, the total outer disk
SFR in most XUV-disk galaxies is modest.
The low covering factor of the UV-emitting sources explains
the modest total XUV-disk $SFR$s, even if on small scales the star
formation surface density in clump complexes becomes significant.  
Even when
cast into a fractional form ($SFR_{XUV-disk}/SFR_{galaxy}$) {\em the XUV-disk
star formation is generally limited to a few percent of the
galaxy-wide integrated level, with the exception of
Type~2 XUV-disks}.  One could argue that the small total and
fractional SFRs associated with XUV-disk environments means that the
SF occurring therein is unimportant in a cosmological sense.  It is
true that at the present epoch XUV-disk activity is a secondary
mode of baryonic transformation from gas to stars, when averaged over
the galaxy population as a whole.  Nevertheless, it has profound
consequences for the future chemical and photometric evolution of the outer disk and provides a
snapshot of how disk galaxy formation might occur in general -
via building the stellar disk from the inside-out.  

We observe a wide variety of UV morphology amongst the XUV-disks
described in Appendix~I and displayed in Fig. 16.  {\em The overall
morphological distribution is represented by our two types of XUV-disk
structure (Type~1 and Type~2), but there is subtlety -- with some
features observed in both Type~1 and Type~2 galaxies}.  Here we
highlight some of the morphological diversity.  The GALEX
data show: wispy extensions of the inner disk at markedly lower
intensity (NGC~300, NGC~4414), spiral-arm continuation with gradual
fading (NGC~628, NGC~3344, M~101), warped \HI disks with prominent
internal structure in either ringlike or spiral form (NGC~2841,
NGC~5055, M~83), tidal filaments (NGC~772, NGC~1512, NGC~2782, M~81,
NGC~6239) or interaction debris (NGC~4438), ``filled'' envelopes
with/without spiral substructure (NGC~1672, NGC~1042, NGC~3621,
NGC~4625, NGC~6902 - with fading too), very LSB with isolated clumps
(NGC~2146A, NGC~3185), unambiguous well-isolated spiral arms
(NGC~3359), swept/lopsided appearance (NGC~4254),and outer spiral
structure unrelated to optical appearance but also seen at inner radii
(NGC~4383).

Finally, it is interesting to compare the clumpy XUV-disk emission
against compact features dominating inner disk UV morphology.
Preliminary measurements indicate that the distribution of peak
surface brightness and luminosity for outer disk clumps is
substantially fainter than for compact structures occupying the inner
disk.  Indeed, {\em many of the XUV clumps have peak surface brightness
comparable to the interclump emission of the main disk.  Perhaps the
clumped outer disk emission is more similar to inner disk diffuse UV
light -- now thought to be composed of dissolved clusters} (Pellerin et
al. 2007).  The process of ``infant mortality'' is understood to occur
on rather short timescales, conceivably even less than 10--25 Myr.  In
principle, the disruption of clusters may begin with the first SNe in
a star-forming complex.  Our Monte Carlo simulations suggest that the
majority of sources in the outer disk should be older than this very
young age.  The combination of moderate age (with clumps being
dominated by B stars rather than O) and a relatively diffuse ISM,
which might encourage cluster disruption, leads us to believe that
infant mortality of clusters through disruption of the natal star
forming cloud may be quite effective in the XUV-disk environment.  We
have already obtained HST UV--optical imagery of locations in M~83 and
NGC~5055 which will be used to test this hypothesis.

\section{Conclusions / Summary}

We searched for extended ultraviolet disk (XUV-disk) galaxies in the
local universe, following the discovery of extreme outer disk star
formation in M~83 and NGC~4625.  Our initial survey of 189 spiral
galaxies within 40 Mpc observed by GALEX with uniform (surface
brightness) sensitivity showed evidence for two varieties of XUV-disk
structure.  Type~1 XUV-disks similar to our prototypes (e.g. M~83,
NGC~4625), have UV-bright cluster complexes (often arranged in
spiral-like features) at extreme galactocentric distances, where the
mechanisms leading to initial cloud collapse (and eventual
gravitational instability) begin to phase out.  Type~2 XUV-disks are
dominated by a large, outer, blue LSB zone having rather high specific
star formation rate.  In Type~2 disks the SF does not necessarily
reach to extreme disk radii, and can even show a sharp outer edge.
The actual galaxy population exhibits a continuum of properties, even
in the case of XUV-disk star formation.  The main two types may be
viewed as illustrative of the modes of galaxy evolution via inside-out
disk formation (Type~1 gradual, Type~2 extreme/non-linear), but are
not mutually exclusive.


The apparent, present-day incidence of XUV-disk star formation
corresponds to  $\ga$20\% of spiral galaxies having Type~1 (M~83-like)
structures and $\sim$ 10\% classified with Type~2 XUV-disk morphology.
Actual incidence over cosmological periods could be even higher if
XUV-SF is episodic, or if it has already happened in some fraction of
the galaxy population not classified as XUV-disks.  We find that
Type~1 XUV-disks span the entire range of Hubble types considered in
our survey, while Type~2 XUV disks appear to be preferentially
late-type/low-mass spirals.  Global characteristics are similar to
that of the general galaxy population, albeit with mildly increased
gas richness and hints that Type~1 XUV-disks may be stimulated by
minor perturbation.  The lack of other notable changes in integrated
properties between the XUV-disk galaxies and the entire sample
suggests that perhaps XUV-disk building is a part of the
natural process of galaxy growth which is inherently tied to the
environmental influences acting upon on any particular object.
Recent/ongoing interaction events and a high-specific rate of gas
accretion are two possible triggers for XUV-disk formation.

\acknowledgments
GALEX (Galaxy Evolution Explorer) is a NASA Small Explorer, launched
in April 2003. We gratefully acknowledge NASA's support for
construction, operation, and science analysis for the GALEX mission,
developed in cooperation with the Centre National d'Etudes Spatiales
of France and the Korean Ministry of Science and Technology.  AGdP is
financed by the MAGPOP EU Marie Curie Research Training Network and
partially by the Spanish Programma Nacional de Astronom\'{\i}a y
Astrof\'{\i}sica under grants AYA2003-01676 and AYA2006-02358.

This research has made use of the NASA/IPAC Extragalactic Database
(NED).  We acknowledge the usage of the HyperLeda database
(http://leda.univ-lyon1.fr).  The Digitized Sky Surveys were produced
at the Space Telescope Science Institute under U.S. Government grant
NAG W-2166.  Funding for the SDSS and SDSS-II has been provided by the
Alfred P. Sloan Foundation, the Participating Institutions, the
National Science Foundation, the U.S. Department of Energy, the
National Aeronautics and Space Administration, the Japanese
Monbukagakusho, the Max Planck Society, and the Higher Education
Funding Council for England. Some images presented in this paper
were obtained from the Multimission Archive at the Space Telescope
Science Institute (MAST). STScI is operated by the Association of
Universities for Research in Astronomy, Inc., under NASA contract
NAS5-26555. Support for MAST for non-HST data is provided by the NASA
Office of Space Science via grant NAG5-7584 and by other grants and
contracts.  This research has made use of the NASA/ IPAC Infrared
Science Archive.  This publication makes use of data products from the
Two Micron All Sky Survey.

{\it Facilities:} \facility{GALEX}, \facility{CTIO:2MASS}, \facility{FLWO:2MASS}, \facility{Sloan}, \facility{PO:1.2m}, \facility{UKST}

\appendix
\section{Notes on individual XUV-disk galaxies}

We now comment on each XUV-disk galaxy in our GALEX survey, and
provide background information from the literature when relevant.
Figure 16 shows GALEX, visible (DSS2-red or SDSS {\em gri} color
composite), and 2MASS-$K_s$ imaging data for each galaxy described in this
section.  We have grouped the galaxies according to their XUV-type,
first the Type~1 disks, then mixed-type objects, and finally
Type~2 XUV-disks.  In the print edition, we only show two illustrative
galaxies per group as a guide to content.  The electronic version
contains a complete version of Figure 16.

\subsection{Type~1 XUV disks}

{{\em NGC~300:}} GALEX imaging of this SA(s)d galaxy reveals a bright
inner disk filled with SF complexes in addition to an outer disk
occupied by low surface-brightness, wispy arcs of clumpy XUV emission.
Radial surface brightness profiles in FUV and NUV from the GALEX Atlas
(Gil de Paz et al. 2007) indicate a downward break in the slope of the
profile at the $D_{25}$ radius. NGC~300 is known to have a large HI
envelope (D$_{HI}$ $\sim$ 61\arcmin, Puche, Carignan, \& Bosma 1990)
with a warp beginning at radii of 10\arcmin.  The outer parts of this
\HI distribution are only very sparsely populated with UV clumps.  Some
of the XUV disk clusters associated with NGC~300 can be recognized on
the DSS2-red plate, using the UV imagery as a prior.  Bland-Hawthorn
et al. (2005) recently reported a disk-like component of the resolved
stellar population out to ten radial scalelengths (24\arcmin, or
2.2 $R_{25}$) in this galaxy.  A causal relation between the outer disk star
formation evident from our GALEX data and the extended disk of
Bland-Hawthorn remains speculative though.  A member of the Sculptor
group, NGC~300 is frequently described as a twin of M~33.  It dominates
a ``loose quartet'' (Karachentsev et al. 2003) of Sculptor galaxies
including NGC~55, ESO~410-05 and ESO~294-10 (later two are not listed in
the NOG catalog, Giuricin et al. 2000).

{{\em NGC~628:}} This SA(s)c galaxy has received attention (along with
NGC~6946 and NGC~1058) as an example of a spiral with faint \HII regions
populating the outer disk (Ferguson et al. 1998a,b; Lelievre \& Roy
2000).  GALEX imaging shows UV clumps at the positions of each
previously known outer disk \HII region, as well as numerous additional
UV-only complexes located at modestly larger radii (at least to
10.3\arcmin, maybe further to E and NW).  The census of recent star
forming sites (even in areas populated by outer disk \HII regions) is
therefore more complete in our GALEX data.  Spiral structure is
apparent in the XUV disk, although the covering factor of SF detected
complexes is small.  The XUV-disk spiral structure appears to be a
continuation of the inner disk pattern, though perhaps more tightly
wound.  The \HI distribution of NGC~628 is known to be extended
(Kamphuis \& Briggs 1992, Auld et al. 2006) within a diameter
of approximately 40' at a limit of N(HI)=$2.2\times10^{18}$~cm$^{-2}$.
Kamphuis \& Briggs (1992) noted the presence of anomalous high
velocity gas.  Smaller companions are: UGC~01104, UGC~01171,
UGC~01176(DDO13), UGCA~20, and KDG10, plus dw0137+1541 (Briggs 1986).
NGC~628 is paired with the disturbed, polar-ring galaxy NGC~660 which
also has companions.

{{\em NGC~772:}} GALEX imaging of this SA(s)b galaxy shows a complex,
asymmetric spiral structure with an anomalously bright arm extending
to the NW of the nucleus for approximately 3\arcmin.  This UV
morphology (and the visible appearance) is highly suggestive of past
or present galaxy interaction, probably with NGC~770 (3.3\arcmin SW).
Such excitation of one-armed spiral modes has been discussed by Taga
\& Iye (1998) and Lovelace et al. (1999). The inner part of the
NGC~772 disk is surrounded by tightly wrapped, generally continuous
spiral features which attain their greatest extent (7.1\arcmin) to the
S of the galaxy. NGC~772 dominates a NOG group (NOGG 124) of three
bright galaxies. Zaritsky et al. (1997) further associate three lower
mass companions with NGC~772.  Of these, NGC~772b appears optically
disturbed (Gutierrez et al. 2006).  NGC~772 has a comparatively large
\HI disk (Rao \& Briggs 1993, Iyer et al. 2001).

{{\em NGC~1042:}} This SAB(rs)cd NLAGN galaxy has along its perimeter
(predominately to the E and W) an annular zone of reduced SF
intensity, for which both GALEX and H$\alpha$ data (SINGG, Meurer et
al. 2006) indicate a population of distinctly lower luminosity SF
regions.  SDSS data does not show the vast majority of emission
features detected by GALEX in this zone.  NGC~1042 belongs to a group
of 11 galaxies dominated by NGC~988 (Giuricin et al. 2000), but also
containing NGC~991 which we have classified as a Type~2 XUV
disk.  The \HI disk of NGC~1042 has been studied in detail by Kornreich
et al. (2000), who find that the gas distribution is asymmetric and
conclude that this resulted from interaction with NGC~1052 (also
suggested by van Gorkom et al. 1986).  Although it reaches past the
optical disk emission, the relative \HI extent is not particularly
large (D$_{HI}$/$D_{25}$ = 1.7,
Kornreich et al. 2000) in comparison to other spirals.  The outer gas disk also
doesn't appear substantially structured into filaments at 20\arcsec
($\sim$1 kpc) resolution.  The net effect of these two factors is that
the XUV-disk emission in NGC~1042 does not attain truly extreme
galactocentric distances, such as found in many of our other objects.

{{\em NGC~1512:}} This SB(r)ab galaxy is in the midst of a close
interaction with NGC~1510 (located 5\arcmin to the SW, projected inside
of the XUV-disk).  The UV morphology of the pair is dominated at high
surface brightness by an inner SF ring in NGC~1512 and starburst
activity in the nuclear region of each galaxy.  A single spiral arm
begins at the NE edge of the SF ring and wraps continuously around
NGC~1512 for nearly 540 degrees in azimuth, eventually reaching a
galactocentric distance of 11.4\arcmin SW of the galaxy.  \HI synthesis
imaging (M. Thornley, priv. comm.) shows that the spatial distribution
of atomic gas and UV-bright complexes are tightly correlated.  The tip
of the continuous, winding XUV arm coincides with the position where
the \HI arm becomes undetected.  Using the \HI image as a prior we can
discern what may be a continuation of this arm directly to the W of
NGC~1512, where we find a UV clump (at 12.6\arcmin) coincident with a
local maximum in an elongated \HI feature oriented N-S.  Arcs of
UV-bright complexes can be found associated with NGC~1510, just to the
N of this SA0 pec; HII BCDG 
companion.
Disorganized UV filaments can be found scattered NW of NGC~1512, which
cannot be confidently associated with either galaxy in the pair.
Perhaps these features represent SF occuring in tidally disturbed
material.
The
large extent of the outer spiral features in NGC~1512 was noted by both
Hawarden et al. (1979) and Sandage \& Brucato (1979) using optical
observations, though the degree of recent star formation activity in
these locales is a new GALEX result.  The NGC~1512 group also contains
NGC~1487.

{{\em NGC~1672:}} This southern (R'1:)SB(r)bc Sy2 
 galaxy has an XUV-disk with a high surface covering factor, composed
of 
spiral segments continuing outward from
the prominent inner disk arms.  The elongation and position angle of
the outer disk is similar to that of the galaxy's bar, and differs
from the optical isophotal contours at intermediate surface
brightness.  This may be the  signature of SF occuring in a warped
\HI disk.  Only NUV observations are available for this target, making
the task of identifying the most spatially extended complexes more
difficult than usual.  Sandage \& Bedke (1994) previously noted faint
outer disk structure in this galaxy on the basis of deep optical
prints.  NGC~1672 is the main galaxy in a small group containing at least eight
galaxies (from NOG catalog). The NGC~1672 group is part of the yet larger
Fornax Wall and Dorado group.

{{\em NGC~2146A:}} This SAB(s)c: galaxy  
 has a very low surface brightness outer
disk, which is most apparent to the W and NE.   NGC~2146A is isolated,
according to the NOG catalog of galaxy groups.


{{\em NGC~2710:}} 
 This SB(rs)b galaxy has an inner disk accentuated by bright UV
emission from a bar and adjoining spiral arms.  The spiral arms fade
in surface brightness and become indistinguishable from a fainter,
clumpy distribution filling the optical extent of the galaxy.  An
arclike, XUV spiral arm is found beyond the NW limit of the inner disk
about 1.5\arcmin from the galaxy nucleus.  SDSS observations present a
similar morphology as GALEX, but do not show the NW feature.
NGC~2710 is isolated according to the NOG clustering analysis.

{{\em NGC~2782:}} The clearly disturbed ultraviolet and optical
morphology of this SAB(rs)a;Sy1 starburst galaxy has been attributed to a
previous galaxy merger by Smith (1991).  The merger remnant exhibits a
very large, double-sided \HI tail (Smith 1991, 1997; Noordermeer et
al. 2005) which has modest extent directly E of the galaxy (coincident
with low surface bright plume in DSS imaging) but more significantly
reaches 5\arcmin to the NW in the form of a long curving feature.  Our
GALEX images show clumpy UV emission within the NW tail, matching the
morphology of the atomic gas, and a component of more diffuse and redder UV light
originating from the eastern \HI cloud.  Smith (1991) reports that LSB
($\mu_B = 25$ mag~arcsec$^{-2}$) optical emission can be seen within
the NW tail, but the UV clumps seen in our data are more conspicuous
and demonstrate that SF is occuring in the structure.  SDSS image data is
not sensitive to the young clusters in the NW tail.  NGC~2782 is the
brightest galaxy in a small group of four objects.

{{\em NGC~2841:}} GALEX imaging of this SA(r)b:;LINER Sy1 galaxy shows
nearly continuous spiral arms wrapping the outside of a bright,
flocculent inner disk.  The \HI disk is also quite extended (Bosma 1981
and Walter et al. (THINGS)).  XUV complexes can be identified out to
at least 4 times the $D_{25}$ radius at both ends of the galaxy major axis.
These particular features are corroborated by the \HI morphology in the
same area (Walter et al. in prep).  Star formation activity is ongoing throughout the entire HI
distribution, which exhibits a notable ring-like appearance just
outside the optical disk.  NGC~2841 is dramatic in terms of the
difference in mean UV surface brightness between the inner
(optically-prominent) disk and the XUV-disk.    Some of the
XUV-disk complexes in NGC~2841 are luminous enough to be detected
individually in SDSS observations.  A forthcoming paper by Bianchi et
al. (in prep., also see Bianchi et al. 2006-IAUS235) will discuss population synthesis SED-fitting of these
sources.  NGC~2841 is a relatively isolated galaxy, as it is only
paired with one other object (UGC~04879).

{{\em NGC~3031:}} Primary member amongst one of the most well-studied
interacting galaxy groups, this SA(s)ab;LINER Sy1.8 object has
extensive tidal arms apparent in the \HI distribution (Yun, Ho, \& Lo
1994, 2000).  GALEX observations show that the tidal arms host SF at a
level greater than appreciated via the H$\alpha$ tracer (comparing to
the emission line image of Greenawalt 1998).  SDSS imaging reveals
some of the XUV-disk cluster complexes in NGC~3031.  Within the large
scale tidal features, molecular gas has been found (Brouillet et
al. 1992, Braine et al. 2001).  Tidal dwarf galaxies (van Driel et
al. 1998, Boyce et al. 2001, Makarova et al. 2002), isolated \HII
regions (Flynn et al. 1998), and a diffuse stellar population (Sun et
al. 2005) were previously known to exist within the \HI streamers.  MUV
and FUV imagery of NGC~3031 from UIT (Hill et al. 1992, Marcum et
al. 2001) illustrated a red bulge and blue SF arms, but not the
XUV-disk complexes.

{{\em NGC~3185:}} The ultraviolet morphology of this (R)SB(r)a Sy2
galaxy is dominated by a UV-bright, inner ring and a luminous nuclear
source.  Outside of this and lying toward the SE major axis, our GALEX
data reveal a string of blue ($FUV-NUV$) clumps qualifying the galaxy
as an XUV-disk -- though quite limited in nature.  Background galaxies
may be confused with some of this emission.  NGC~3185 is a member of
the group dominated by NGC~3190. SDSS imaging also shows the SF disk in
the main disk, but doesn't detect the XUV clumps.

{{\em NGC~3198:}} This inclined SB(rs)c galaxy was studied in some
detail following the discovery of an apparent sharp outer edge to the
\HI distribution (Maloney 1993, van Gorkom et al. 1993, van Gorkom
1991).  \HI extends to at least 634'' radius, about 2.5x $D_{25}$ radius
(Sicking 1997). The \HI distribution is asymmetric with a filament in the
NE portion of the galaxy (Bosma 1981, Sicking 1997, THINGS).  GALEX
shows that SF is occuring in spiral waves throughout the known extent
of HI, with lower average UV surface brightness in the XUV-disk.
Nevertheless, localized UV-bright clumps are obvious even in the most
extended parts of the \HI disk.  As expected, patterns in the
distribution of XUV complexes are tied to arm features the HI
morphology of NGC~3198.  We can discern SDSS detections for a limited
number of the brightest inner XUV-disk UV complexes.  NGC~3198 is the
brightest member of group containing five galaxies, including NGC~3319
(a Type~2 XUV-disk).  However, computation of a tidal
perturbation parameter (see Section 3.3.3) indicates that NGC~3198 is
effectively isolated in terms of tidal interaction.

{{\em NGC~3344:}} This (R)SAB(r)bc \HII galaxy has a large UV disk
which gradually fades with radius (similar to NGC~628, but less
obvious).  Both Verdes-Montenegro et al. 2000 and Corbelli et al. 1989
present \HI maps of NGC~3344.  UV emission is smoothly extended but bright,
even if not so in the optical.  NGC~3344 has two companions (UGC~05672 and
NGC~3274, altogether constituting NOGG 468), though is not presently
tidally interacting (Section 3.3.3 again).  SDSS imaging 
helps to differentiate the outermost clumps from background galaxies.

{{\em NGC~3359:}} A two-arm SB(rs)c \HII galaxy with XUV-disk complexes
distributed in long, wrapping spiral features beginning at the ends of
the major axis where the optical spiral arms cease.  The eastern XUV
arm is of higher surface brightness near the inner disk, whereas the
western XUV arm is more prominent at large galactocentric distances.
Nevertheless, both UV-bright complexes in both arms can be seen out to
nearly three times $R_{25}$.  NGC~3359 has been categorized as an
``isolated'' galaxy in several studies (Karachentseva 1973, Verela et
al. 2004, Giuricin et al. 2000, Verdes-Montenegro et al. 2005) though
it is known to have a low mass, LSB object detected in \HI (Ball 1986)
and Palomar plates.  This object is 15\arcmin~SW of the galaxy, beyond
the limits of the multiwavelength images we present.  However, our
GALEX data do recover the Ball et al. LSB object as a resolved
FUV-bright clump.  The \HI structure of NGC~3359 and this LSB companion
has also been studied by Broeils 1992 and
with new WSRT data in van der Hulst \& Sancisi 2005 (who measure an HI
mass of 1.8$\times10^8$~M$_{\odot}$ for the companion).  The van der
Hulst \& Sancisi observations are sensitive enough to discern a
connection between the outer LSB cloud and NGC~3359's spiral structure,
meaning our UV detection of the object makes NGC~3359 amongst the most
extended UV disks known (relative to optical size).  SDSS observations
are sensitive to only the brightest of the XUV complexes.

{{\em NGC~3621:}} This SA(s)d galaxy is of very large angular size in
the UV, at least partially owing to its proximity (8.3 Mpc).  GALEX
imaging reveals SF occuring throughout almost all of the known HI
extent (32 x 11', Koribalski et al. 2004; THINGS).  A linear feature running along the major axis in HI
maps could be the remnants of a previous interaction, or even a dwarf
currently being accreted (see LVHIS map of W. Walsh, esp. velocity
dispersion).  The XUV-disk has a high surface covering factor in the
UV and can be traced as a LSB optical structure as well.  Spiral
structure is evident throughout the XUV-disk.  An abrupt drop in the
radial UV surface brightness profile (Gil de Paz et al. 2007)
indicates the transition to the LSB outer disk.  Incidently, this
transition occurs within the $D_{25}$ extent of the galaxy, despite the
continuation of LSB features to more extreme galactocentric distance.
The NOG catalog 
indicates that NGC~3621 is isolated, though the linear feature in the
\HI maps suggests it has not always been so.  Perhaps the elevated
degree to which SF is now occuring in the outer \HI disk was stimulated
by this interaction.  If so, the age of the interaction could be
inferred following from the mean age of the LSB stellar population.

{{\em NGC~3705:}} Belonging to the LeoI group, this SAB(r)ab;LINER \HII
galaxy has faint discrete XUV complexes E and W of the disk
superimposed on a more diffuse component in our NUV-only GALEX imagery
-- though hard to ``confirm'' even with SDSS images.  Though it lives
in a group environment, NGC~3705 is not presently interacting according
to the tidal perturbation parameter.  This is a somewhat
marginal XUV-disk, as the UV structure is limited in nature.

{{\em NGC~4254 [M~99]:}} This Virgo cluster member, an SA(s)c galaxy, exhibits
peculiar lopsided XUV emission in our deep NUV GALEX imaging.  The UV
clumps are located to the SW of the bright inner disk over an
azimuthal range of about 45 degrees and extending to approximately two
times the $D_{25}$ radius.  The XUV structure, although of a high surface
covering factor, does not appear to be related to the spiral structure
of the inner disk.  The XUV disk of NGC~4254 is also odd because the HI
disk of the galaxy is slightly extended in an opposite direction, to
the N of the disk (VIVA of Kenney et al., Braun et al. 2007).
Furthermore, ALFALFA observations have demonstrated that NGC~4254 has a
very low column density \HI stream (ref: Giovanelli ALFALFA followup
proposal) running $\sim 30$\arcmin NNW of the galaxy and including
VIRGOHI21 (Minchin et al. 2005).  SDSS imaging shows little in the
vicinity of the lopsided XUV-disk.

{{\em NGC~4258 [M~106]:}} GALEX observations of this SAB(s)bc;LINER Sy1.9
galaxy indicate a very large extent ($\sim 25\arcmin$), although the
substantial inclination of the disk (and the orientation of its warped
outer HI) make it difficult to say precisely to what galactocentric
distance XUV features are seen.  The most significant area with
XUV-disk clumps is S of the disk.  SDSS data  detects very few
of these star clusters.  NGC~4258 is the brightest galaxy in a group
having over twenty members, possibly interacting with NGC~4248
(f=-2.2).

{{\em NGC~4383:}} This Sa? pec \HII galaxy found in the outskirts of
the Virgo cluster has a very bright inner disk, centered within an
S-shaped, spiral pattern of intermediate surface brightness UV clumps.
This spiral morphology is unapparent in visible light imagery, with
some authors calling the galaxy amorphous.  UV clumps are distributed
throughout the extent of the \HI mapped by Kenney et al. (VIVA) ---
most notably to the SSE, E and NE of the galaxy.  The H$\alpha$
emission of NGC~4383 shows some \HII regions associated with the
extended UV clumps (Gavazzi et al. 2003, Gavazzi et al. 2006) and an
inner bright, biconical filamentary structure (Koopman \& Kenney
2006).  SDSS imagery reveals only the blue clumps composing the
innermost northern part of the S-shaped spiral pattern.
NGC~4383 has an overall appearance similar to NGC~1140, though they
are in different XUV classes.  This illustrates the difficulty of
classifying objects having such complex structure.

{{\em NGC~4414:}} A member of the ComaI group (dominated by NGC~4631),
our observations of this SA(rs)c? LINER galaxy reveal a bright main
disk surrounded by XUV clumps distributed in disorganized (tightly
wrapped?) spiral features at almost all position angles.  The clumps
are not located at extreme galactocentric distances but rather tend to
hug the main disk.  This galaxy is one of the few to have confirmed
outer disk molecular clouds (Braine \& Herpin 2004) at a limited
number of positions near \HI peaks (Thornley \& Mundy 1997).  Deep
H$\alpha$ imaging reveals sporadic \HII regions coinciding with some of
the XUV clumps (Gavazzi et al. 2003, Gavazzi et al. 2006).  SDSS
imagery allows one to see the brightest XUV-disk stellar
complexes. Despite belonging to the ComaI group, NGC~4414 is
dynamically isolated in the sense that it has no known tidally
perturbing companion.

{{\em NGC~4438:}} This SA(s)0/a pec: LINER galaxy is interacting with
NGC~4435 as they cross the center of the Virgo cluster at high speed
(Vollmer et al. 2005). The high-speed collision is discussed by Kenney
et al. (1995).  GALEX observations analyzed by Boselli et al. (2005)
show the presence of very recent star formation in a tidal tail
extending to the NW of the galaxy -- this feature was undetected at
other wavelengths (including R down to 27.8 mag/sq.arcsec via
Subaru). The UV-visible-NIR SED fitting of Boselli et al. suggests
that the burst of star formation is notably younger than the dynamical
age of the NGC~4438/NGC~4435 interaction.  SDSS imagery highlights a
large dust concentration on the W side of NGC~4438.  It is  interesting to note that the \HI streamer of nearby NGC~4388
(Oosterloo \& van Gorkom 2005) points very near to NGC~4438's NW XUV
feature with similar orientation. 

{{\em NGC~4559:}} GALEX observations of this SA(r)0+;\HII LINER galaxy
show diffuse off-plane FUV emission perhaps associated with
extraplanar \HI of Barbieri et al. (2005).  Discrete XUV-disk complexes
are arranged in spiral arms tightly wrapping the on galaxy periphery
in a manner similar to NGC~4414.  SDSS imaging at blue wavelengths
detects some of the UV-bright clumps.  NGC~4559 is a ComaI group
member.

{{\em NGC~4625:}} One of the first known XUV-disks, GALEX observations
of this SAB(rs)m pec galaxy show that the outer disk is organized into
spiral arms, though the surface covering factor remains quite high
compared to M~83 or NGC~5055, for instance, which both have a more
filamentary appearance. Abundance is low (though not primordial) in
the outer disk, as measured in a population of faint \HII regions (Gil
de Paz et al. 2007b).  Most of the XUV-disk stellar complexes are too
faint for individual detection by SDSS.  NGC~4625 is interacting with
NGC~4618 and possibly NGC~4625A (Gil de Paz et al. 2005).  NGC~4625 is
a member of the NGC~4631 group. \HI data for the NGC~4625/18 pair were
obtained by the WHISP project and show the usual correlation between
\HI subtructure and the outer disk UV clumps.

{{\em NGC~5055:}} M~63, an SA(rs)bc \HII/LINER galaxy with a very large warped
\HI disk (Battaglia et al. 2006).  
GALEX shows that
the outer \HI disk has recently formed massive stars, principally
along ridges in the gas distribution.  One particular region of UV
emission to the SW of the galaxy is possibly a TDG (namely UGCA~342 -- Bremnes et al. 1999 suggest it is a ``condensation'' in
the outer gas disk).  NGC~5055 is a member of the M~51 group (7 members). NGC~5055 appears to be interacting with UGC~08313,
found to the NW and connected by a filament of \HI containing UV-bright
complexes. SDSS imaging recovers UGCA~342 and a few other XUV-clumps.

{{\em NGC~5236 [M~83]:}} The prototypical XUV-disk galaxy (SAB(s)c;\HII
Sbrst) exhibits clumpy SF complexes associated with local maxima
throughout an extensive, filamentary \HI disk (Thilker et al. 2005).
The distribution of UV complexes is sparse, with a low surface
covering factor, though there are many sites of recent star formation.
Some of the XUV complexes do contain faint \HII regions, which have low
abundance ($\sim$ Z$_\odot/10$) and are often ionized by single
massive stars (Gil de Paz et al. 2007b).  Past interaction with other
members of the CentaurusA group, including NGC~5253 and KK~208
(Karachentseva \& Karachentsev 1998), is considered likely. More details on NGC~5236 are presented by Thilker et al. (2005) and Gil de
Paz et al. (2007b).

{{\em NGC~5457 [M~101]:}} This giant SAB(rs)cd galaxy dominates a
group having at least 8 members, including NGC~5474 and NGC~5477 with
both of which it is probably interacting.  Our GALEX observations show
a radial gradient in $FUV-NUV$ (Gil de Paz et al. 2007), getting
clearly bluer at large galactocentric distances.  Although discrete SF
complexes do appear redder in the inner disk, some of the central NUV
flux originates in a diffuse slightly older stellar population.  This
component gradually fades leaving only the very blue, prominent
branching spiral arms at large galactocentric distance.  Comparison
with SDSS imagery confirms that a wealth of filamentary UV-bright
cluster complexes in the outermost disk can be traced easily in the
GALEX observation (Bianchi et al. 2005), but only intermittently at
optical wavelengths.  The previously unrecognized extent of recent
star formation is greatest on the E side of the galaxy, reaching most
of the way to NGC~5477.  Prior UV observations of Waller et al. (1997)
suggested tidal interaction with companions and an internal perturber
(NGC~5471) have produced the kinked, linear morphology spiral arms in
M~101.  Low metallicity of the outer disk has been demonstrated by
Garnett \& Kennicutt for \HII region H681. The \HI distribution (Beale
\& Davies 1969, Rogstad \& Shostak 1971, van der Hulst \& Sancisi
1988, Braun 1995) is lopsided and conceivably distorted by tidal
forces.

{{\em NGC~6239:}} GALEX observations of this SB(s)b pec? \HII galaxy
show a pair of filaments stretching NW of the galaxy and a loop-like
structure to the SE, both highlighted in the UV.  SDSS imaging reveals
only the innermost portions of the NW filaments.  H$\alpha$
observations of Garcia-Barreto et al. (1996) show only very limited
line emission from beyond the inner disk. The \HI study of Hogg \&
Roberts (2001) show that NGC~6239 is interacting with a small companion
to the E (SDSS J165045.93+424320.4) even though it is isolated
according to both the NOG catalog and our tidal perturbation parameter
analysis.

{{\em UGC~10791:}} This Sm: galaxy has extended morphology in the form
of scattered UV-bright clumps, arranged with no discernable
resemblance of spiral arms.  \HI observations are available through the
WHISP project (van der Hulst et al. 2001, van der Hulst 2002, Swaters
et al. 2002, Noordermeer et al. 2005). UGC~10791 does not belong to a
NOG group (it was too faint to be considered by Giuricin et al. 2000),
but is possibly interacting with NGC~6340.

{{\em NGC~6902:}} Optical imaging of this SA(r)b LINER shows a LSB
disk-component with tightly wrapped spiral structure outside of a
bright inner region, which hosts a ring and bar.  GALEX reveals that
this faint outer disk is actively forming stars, even at locations
where the LSB optical disk is not apparent.  The UV disk fades away in
a ragged, but continuous manner.  This relatively understudied galaxy
(considering its angular size) was considered in detail by Gallagher
(1979).  NGC~6902 is the brightest member of a triplet, also containing
NGC~6902B and IC~4946, but is only weakly interacting (if at all).

\subsection{Mixed-type XUV-disks}

On occasion a galaxy may satisfy both XUV-disk classification
criteria, exhibiting a large, blue LSB zone {\em and} structured
UV-bright complexes at extreme galactocentric distances beyond the
traditional SF threshold.  We call such objects
mixed-type XUV-disks.  They represent a population of galaxies
undergoing significant structural change like Type~2 XUV-disks but
they most likely have a comparatively extended gas distribution in
addition.

{{\em NGC~1051 [NGC~0961]:}} This SB(rs)m pec galaxy exhibits
flocculent, clumpy XUV disk structure, extending to about 3.5\arcmin.
In contrast, the $K_s$-band emission is compact, thereby qualifying
the galaxy also as a Type~2 XUV disk.  NGC~1051 is a member of the
NGC~988 group (see entry for NGC~1042).  SDSS imagery shows LSB
emission over some of NGC~1051's XUV disk, particularly to the NE.

{{\em NGC~2403:}} Often compared to M~33, this SAB(s)cd \HII galaxy
satisfies both our criteria for being considered an XUV-disk.  The
outermost XUV disk complexes lie near local maxima within the HI
filament which begins near the NW end of the major axis and wraps
counterclockwise around the N side of the galaxy (Fraternali et
al. 2002).  In at least one case, there appears to be an \HI shell
surrounding the XUV complex (at 07:34:47.5, +65:41:29) -- highlighting
the feedback of outer disk star formation on the gas morphology.  The
combination of NGC~2403's large angular extent (projecting background
galaxies onto the XUV disk) and proximity (breaking up complexes),
makes it difficult to tell the difference between isolated XUV clumps
and background sources seen through the most rarified portions of the
\HI disk.  Gas accretion has been suggested as the origin of
extraplanar \HI (Fraternali \& Binney 2006) in this object.
Known or likely companions include NGC~2366 (NOG catalog), DDO44
(kk61), Holmberg II, UGC~04483, and K52 (Karachentsev 1994, Karachentsev et
al. 2000). NGC~2403 may be an outlier galaxy within the M~81 group.

{{\em UGC~04393:}} Our GALEX observations of this SBc? \HII galaxy show
it to have an asymmetric distribution of star formation activity,
suggesting interaction with an unidentified companion.  In particular,
the inner disk has a spiral-like arc of UV-bright emission surrounding
it to the SE.  This feature, however, is contained within the body of
$\mu_{NUV}$=27.35 and does not place it into our XUV class -- despite
the peculiar morphology.  SDSS imagery of UGC~04393 is peculiar in the
sense that the distribution of blue star-forming clumps appears
projected against an underlying redder stellar population without
morphological correspondence.  Haynes et al. (1998) show an \HI profile
with no notable velocity asymmetry in this seemingly isolated galaxy.
Thus our suggestion of interaction must remain unconfirmed at present.
The most extreme galactocentric distance GALEX detection is
a collection of very faint clumps to the NW of the galaxy 
 which may be a continuation of the inner spiral feature.
Additionally, a possible curved filament of faint UV emission is
found just beyond the NE end of the major axis.  

{{\em NGC~5474:}} This companion of M~101 (NGC~5457) is well known
optically for hosting an offcenter nucleus, most likely indicative of
prior interaction (with M~101?).  Imaging from GALEX shows that
scattered SF complexes exist in the outer disk of this SA(s)cd pec \HII
galaxy, most especially to the W and SE.  NGC~5474 is a galaxy in which
the distribution of UV-bright sources falls off smoothly with
galactocentric distance, even if extended. The \HI structure of NGC~5474
was studied in Rownd, Dickey, \& Helou (1994), noting the clumpy nature
of the gas distribution without any clear evidence for spiral
features.  At a limit of $\sim 2\times10^{20}$~cm$^{-2}$ the \HI disk
covers an area of $9\arcmin\times7\arcmin$.  Curiously, the GALEX data show
spiral arm structure, as do H$\alpha$ observations (Hodge \& Kennicutt
1983).  Moreover, these features begin at the position of the offset
nucleus -- rather than the center of the overall optical distribution.
SDSS data again suggest a dramatic disconnection between the stars
contributing to the visible light morphology and the (spiral) pattern
of recent starbirth, similar to what we found for UGC~04393.

{{\em NGC~5832:}} NUV-only GALEX observations of this SB(rs)b?  galaxy
reveal faint XUV emission primarily in the SW quadrant of the galaxy.
There does not appear to be coherent spiral morphology in the outer
disk.  Kazarian (1996) noted that NGC~5832 belongs to a class of
galaxies having UV excess.
This galaxy is isolated according to both the NOG catalog of groups
and the tidal perturbation parameter (no possible
perturbers). 

{{\em UGC~10445:}} GALEX observations of UGC~10445 present an outer disk
with numerous clumps and diffuse emission, jointly resulting in a
fairly high surface covering factor.  The distribution of XUV emission
is lopsided in the sense that there is very little extended UV on the
SW side of the galaxy. H$\alpha$ imaging of this isolated SBc Sbrst
galaxy (van Zee 2001) shows relatively few \HII regions in outer UV
area.  This starbursting dwarf was also studied with Spitzer by Hinz
et al. (2006), who present evidence for an extended dust distribution.
We differ slightly regarding the Hinz et al. interpretation of the
GALEX imagery.  They contend that the dust distribution is more
extended than UV, but this appears to be due to the smoothing scale
used for the GALEX data.  \HI morphology (available through WHISP, van der Hulst et al. 2001, van der Hulst 2002, Swaters et al. 2002, Noordermeer et al. 2005) is
unremarkable, although the XUV-disk and gas appear roughly
coextensive.  Some of the XUV stellar complexes are visible in the
blue SDSS bands, especially at modest galactocentric distance in the LSB zone.

{{\em NGC~7418A:}} This isolated SA(rs)d: galaxy lies about 15.5\arcmin~
from NGC~7418, but at a velocity separation over $>600$ km/s.  Our
GALEX data show a two armed feathery-looking, spiral pattern of
UV-bright clumps stretching from the NE to SW.  Within a radius of
about 1$\arcmin$ the UV disk takes on a more filled appearance, and a
LSB optical component of the disk starts to be visible.  The nucleus
is bright and compact at UV and optical wavelengths.  The brightest of
NGC~7418A's XUV complexes can be discerned in DSS2-red plates.

\subsection{Type~2 XUV-disks}

{{\em NGC~991:}} Our GALEX observations of this SAB(rs)c galaxy show
fairly sharp, outer UV edge to the well-defined pattern of spiral
arms, in contrast to the Type~1 XUV-disks.  In fact,
the general UV morphology is quite similar to the appearance of the
galaxy in SDSS imaging at u,g.  The distribution of \HI has been
studied by Kornreich et al. (2000).  Their data, in combination with
our UV imagery, suggest that recent SF in NGC~991 is confined to areas
with N(HI) $>$ few $\times 10^{20}$~cm$^{-2}$. NGC~991 belongs to a
group of 11 objects dominated by NGC~988, but is dynamically
unperturbed at present.

{{\em NGC~1140:}} GALEX imaging shows additional structure in this IBm
pec:;\HII Sy2 galaxy, unrecognized at optical wavelengths, despite its
classification as a Type~2 XUV-disk.  The UV-optical morphology is
complex, with a curved (out-of-plane?), clumpy feature oriented to the
SW visible in both the GALEX and DSS images, plus a set of UV-only
(not seen in the DSS) sources extending N of the galaxy center for
nearly 2\arcmin.  The inner disk has a diffuse-looking background of
NUV bright emission, with very bright central surface brightness at
the nucleus.  The peculiar UV-optical morphology of NGC~1140 is
suggestive of interaction, altough the galaxy is considered isolated
in the NOG catalog.  As noted earlier, NGC~4383 and NGC~1140 are similar objects even though our classification criteria put them into different XUV classes.

{{\em NGC~2090:}} Our GALEX imaging of this isolated SA:(rs)b galaxy
 reveals a LSB outer disk barely detected in DSS2-red plates and not
at all in the NIR (2MASS-$K_s$), but prominent at UV wavelengths.  This
object is a prototypical example of a disk in formation at the present
epoch.  The GALEX data show a redder inner disk, surrounded by a much
bluer, flocculent outer region characterized by discontinuous spiral segments.    \HII regions are found in the Type~2 XUV-disk of NGC~2090 (Koopmann \& Kenney 2006).  In the bright inner part of the galaxy, some of these \HII regions (and associated UV emission) take the form of a ring structure.

{{\em ESO~556-012:}} This SB(s)m galaxy appears unremarkable in our GALEX NUV imagery or optical plates, but comparison with 2MASS imagery shows that almost all of the UV disk is extremely faint at $K_s$-band.  Little more than the nucleus is detectable in the NIR.  
The GALEX data show a spiral arm wrapping around the disk to N
side.

{{\em NGC~2541:}} The conspicuous outer spiral structure of this
SA(s)cd LINER galaxy detected in our GALEX observations is not seen at
all in the 2MASS-$K_s$ imagery and only partially in visible light.  SDSS
imaging shows a few faint sources in the XUV-disk LSB zone. \HI observations from
WHISP and Broeils \& van Woerden (1994) show that the galaxy has an extended gas distribution.  NGC~2541 is the brightest galaxy in a small
group (of 3+?), but isolated according to tidal perturbation parameter
analysis.

{{\em UGC~04390:}} GALEX imaging of this SBd galaxy shows a clumpy UV disk which begins crowded at small galactocentric distances but becomes progressively more sparse in the outer region.  The UV disk is notably lopsided, with greater extension to the E.  Such asymmetry is not as evident in the DSS2-red image. In the 2MASS-$K_s$ data, essentially only the nucleus and bar are detected.  

{{\em UGC~04800:}} Our GALEX data for this inclined, isolated SB(s)cd? galaxy 
show a bright UV disk with a sharp outer edge.  The disk is also bright in SDSS imaging, but it has a small $K_s$-band size in comparison to GALEX-SDSS extent.

{{\em IC~2574:}} This SAB(s)m member of the M~81 group presents a
dramatic integral shaped spiral morphology in our GALEX imaging, which
cannot be seen in the DSS2-red plates.  At $K_s$-band the galaxy is only
detected as a fuzzy patch of emission.  IC~2574 has received attention,
even in the UV, for the presence of a supergiant [HI] shell in the
northern part of the galaxy (Cannon et al. 2005, Walter 2001, Stewart
\& Walter 2000, Walter \& Brinks 1999).

{{\em NGC~3319:}} GALEX observations of this SB(rs)cd \HII: galaxy,
belonging to the NGC~3198 group of at least 5 optically bright
galaxies, show a highly filled UV disk with prominent spiral arms
beginning at the end of the bar structure.  SDSS imaging shows the
brightest parts of the spiral arms in the form of a string of blue
clumps with some more diffuse underlying emission at redder
wavelengths.  The \HI distribution of NGC~3319 was studied by Moore \&
Gottesman (1998) using the VLA.  They detected a gas-rich companion
lying 11\arcmin~ south of NGC~3319 and an \HI tail (also to the S), which
they postulate was created by an interaction of the two galaxies.

{{\em NGC~4116:}} Our GALEX imaging of this SB(rs)dm galaxy paired with
NGC~4123 demonstrate that the UV disk is highly structures into spiral
segments but lacks symmetry.  This disturbed structure is also visible
in SDSS images, but the GALEX sensitivity reveals more morphological
features in the outer disk.  For example, our NUV image shows a
tightly wrapped spiral arm encircling the S end of the major axis.  As
is the case for many of our Type~2 XUV-disks, the 2MASS-$K_s$ data
reveal essentially just the innermost portion of the galaxy (nucleus
and bar).  Both NGC~4116 and NGC~4123 are part of a quartet dominated by
NGC~4179, but it is not certain that NGC~4116 is interacting with any
galaxy (even NGC~4123).

{{\em NGC~4236:}} This SB(s)dm galaxy has very large angular extent
($> 20$\arcmin) in our GALEX survey images.  Spiral structure
accentuated by large clumpy complexes is apparent in the UV.  Much of
the outer UV spiral is barely discernable on the DSS2-red imagery.
The VLA \HI observations of Braun (1995) show the gas distribution has
a highly structured morphology even in the outer disk, matching the UV
emission quite closely. NGC~4236 lies in a grouping with UGC~08201,
and often associated with M~81 group.  Holm~357b also lies to the NE of
NGC~4236, at unknown redshift.

{{\em UGC~08365:}} GALEX imagery of this SB(s)d galaxy shows a disk
characterized by scattered UV-bright clumps.  The highest surface brightness portion of
the UV morphology is elongated in nearly the same direction as the bar
visible at optical (SDSS) wavelengths.  The $K_s$-band data from 2MASS only
faintly reveal the bar, and nothing of the remaining disk. UGC~08365
belongs to a group of three objects dominated by NGC~5013.  

{{\em NGC~5705:}} Our UV imagery of this SB(rs)d galaxy in a large
group of 25+ galaxies (brightest member NGC~5746) shows loose spiral
structure composed of numerous bright clumps with underlying diffuse emission.  SDSS imagery shows that
at least some of these clumps are massive enough to be detected
optically.  NGC~5705 may have close companions (SHOC~476, 2QZ J143953.9-004318, and SDSS~J143956.60-004439.3).

{{\em NGC~5727:}} This UV-bright, optically-faint SABdm galaxy has blue UV-optical color over almost the entire disk, with only slightly redder colors toward the center, as illustrated by SDSS imaging.  
Pisano, Wilcots, \& Liu (2001) indicate that it has a companion to the NW,
which they name NGC~5727A.  

{{\em NGC~6255:}} GALEX and SDSS observations of this SBcd: galaxy show
that the disk has well-defined, blue spiral arms on the N and S sides
of the target.  NGC~6255 belongs to a triplet of bright galaxies,
dominated by NGC~6207.  The galaxy has been noted to have a possible companion (HS~1653+3634) which lies 75'' to the E. It seems more likely from GALEX data that this object is just a particularly bright, blue
cluster complex in the XUV-disk of NGC~6255.   

{{\em ESO~406-042:}} GALEX observations of this comparatively isolated SAB(s)m galaxy show a clumpy structure with little organization into spiral features.  The disk (even the center) is
  almost undetected by 2MASS.

{{\em ESO~407-014:}} Our GALEX imaging of this isolated SB(s)c? \HII galaxy show obvious spiral structure in an outer disk which has surface brightness at visible wavelengths distinctly fainter than the inner galaxy.  In this manner it resembles NGC~2090.  Indeed, the disk has a bright central 2MASS-$K_s$ nuclear region with seemingly sharp edge.

\section{Inside-out disk formation models}

As mentioned in Section 3.1.2, we compared measured characteristics of
the LSB zone against predictions derived from the linear inside-out disk
formation models of Mu\'noz-Mateos et al. (2007).  In particular, the
models were used to estimate the anticipated locus of 'normal'
galaxies growing from the inside-out in Fig.~2, which plots the
relative area of the LSB zone versus GALEX--2MASS color measured in
the LSB zone ($S(LSB)/S(K_{80})$ versus $UV(AB)-K_s(AB)$).  Details on
the modeling are given in Mu\'noz-Mateos et al. (2007).  Here we
describe the most important points and any modifications for our
current purpose.

We assume that the SFR surface density of a disk can be parameterized
as:

$\Sigma_{SFR}(r,t) \sim exp(-t/\tau) exp(-r/(a_0 + b t))$

The first term imposes an exponential SFH with a time-scale $\tau$,
and whereas the second term introduces the radial exponential profile,
whose scale-length is supposed to vary linearly with time. $a_0$ is
the initial size [in kpc] at the redshift where the thin-disk is
formed ($z\sim$1-2) and $b$ is the growth rate of the scalelength
[kpc/Gyr]. By integrating $\Sigma_{SFR}(r,t)$ since $z\sim$1-2 until
$z$=0, the model computes the current stellar mass surface density
profile for a grid of values of $a_0$ and $b$. We assume that the
fraction of recycled gas is $R=0.3$, for a typical IMF. We also feed the
model with the central SFR surface densities of the disks observed today, in
order to properly normalize $\Sigma_{SFR}(r,t)$. Finally, an average
internal extinction of 1 mag in the FUV is used both when locating the
the UV and K$_{80}$ contours and when measuring the colors.

The original models of Mu\'noz-Mateos et al. (2007) did not explicitly
take into account the formation and development of the bulge. Since
the bulge is necessary to delineate the LSB zone in the same way we
did observationally, we implemented a bulge component assuming a de
Vaucouleurs profile, with a given effective surface brightness
$\mu_e$. [The effective radius is not needed, since it can be
internally computed from assumed bulge-to-disk ratio, $\mu_e$, and the
predicted disk-luminosity]. It seems that the bulge is necessary to
determine where to measure, but the measurements themselves depend
more on the properties of the disk generated by the model.

\clearpage
\clearpage
\begin{deluxetable}{lcrrrrrrrccr}
\tabletypesize{\scriptsize}
\tablecaption{40 Mpc, Disk Galaxy Sample\label{table1}}
\tablewidth{0pt}
\tablehead{
\colhead{Object} &
\colhead{$T$} &
\colhead{$D$} &
\colhead{$D_{25}$} &
\colhead{log $SFR$} &
\colhead{log $sSFR$} &
\colhead{log $\Sigma_{SFR}$} &
\colhead{log M$_{HI}$} &
\colhead{log M$_{HI}$/L$_B$} &
\colhead{N$_{group}$} &
\colhead{NOG} &
\colhead{$f$} \\
\colhead{Name} &
\colhead{} &
\colhead{(Mpc)} &
\colhead{(arcmin)} &
\colhead{(M$_\odot$ yr$^{-1}$)} &
\colhead{(yr$^{-1}$)} &
\colhead{(M$_\odot$ yr$^{-1}$ kpc$^{-2}$)} &
\colhead{(M$_\odot$)} &
\colhead{(M$_\odot$/L$_\odot$)} &
\colhead{} &
\colhead{Group} &
\colhead{} \\
\colhead{(1)} &
\colhead{(2)} &
\colhead{(3)} &
\colhead{(4)} &
\colhead{(5)} &
\colhead{(6)} &
\colhead{(7)} &
\colhead{(8)} &
\colhead{(9)} &
\colhead{(10)} &
\colhead{(11)} &
\colhead{(12)} }
\startdata
       UGC 00017 &9.0 $\pm$ 0.5 &      13.3 &       2.4 &     -1.52 &     -8.74 &     -3.35 &      8.33 &     -0.25 &   \nodata &     \nodata &      \nodata
             \\
        NGC 0024 &5.0 $\pm$ 0.5 &       8.2 &       5.5 &     -0.75 &     -9.36 &     -2.88 &      8.82 &     -0.77 &        13 &    Sculptor &     -4.63
             \\
        NGC 0115 &3.8 $\pm$ 0.8 &      23.4 &       2.0 &     -0.26 &     -8.88 &     -2.42 &      9.32 &     -0.37 &   \nodata &     \nodata &     -4.13
             \\
        NGC 0131 &3.0 $\pm$ 0.6 &      17.5 &       1.6 &     -0.41 &     -9.07 &     -2.15 &      9.34 &     -0.13 &   \nodata &     \nodata &     -1.99
             \\
        NGC 0247 &6.9 $\pm$ 0.5 &       3.1 &      19.9 &     -0.64 &     -9.02 &     -3.05 &      9.08 &     -0.68 &        13 &    Sculptor &     -3.60
             \\
        NGC 0253 &5.1 $\pm$ 0.5 &       3.9 &      28.3 &      1.03 &     -9.02 &     -1.89 &      9.19 &     -1.50 &        13 &    Sculptor &     -4.55
             \\
        NGC 0300 &6.9 $\pm$ 0.5 &       2.0 &      20.3 &     -0.55 &     -8.97 &     -2.59 &      9.25 &     -0.20 &        13 &    Sculptor &     \nodata
             \\
        NGC 0337 &6.7 $\pm$ 0.8 &      24.7 &       2.9 &      0.65 &     -8.85 &     -1.88 &      9.73 &     -0.79 &         4 &      NGC 337 &     -6.21
             \\
       NGC 0337A &7.9 $\pm$ 0.6 &      14.2 &       3.0 &     -0.28 &     -7.55 &     -2.37 &      9.47 &     -0.19 &         2 &  MCG-1-3-85 &     -5.44
             \\
        NGC 0470 &3.1 $\pm$ 0.4 &      33.5 &       2.4 &      0.80 &     -9.07 &     -1.84 &      9.56 &     -0.85 &         5 &      NGC 488 &     -1.74
             \\
        NGC 0514 &5.3 $\pm$ 0.5 &      35.5 &       3.3 &      0.47 &     -9.34 &     -2.48 &      9.79 &     -0.75 &   \nodata &     \nodata &     -5.97
             \\
        NGC 0520 &0.8 $\pm$ 2.6 &      31.3 &       4.0 &      1.17 &     -8.83 &     -1.84 &      9.68 &     -0.80 &         5 &      NGC 488 &    -0.994
             \\
        NGC 0586 &1.0 $\pm$ 1.1 &      27.2 &       1.5 &     -0.34 &     -9.54 &     -2.37 &      \nodata &      \nodata &         9 &      NGC 584 &     -1.03
             \\
        NGC 0628 &5.2 $\pm$ 0.5 &      11.1 &       9.2 &      0.60 &     -9.12 &     -2.25 &      9.82 &     -0.73 &         7 &      NGC 628 &     -4.95
             \\
        NGC 0660 &1.3 $\pm$ 1.2 &      12.3 &       5.9 &      0.73 &     -8.89 &     -1.82 &      9.62 &     -0.26 &         7 &      NGC 628 &     -3.39
             \\
        NGC 0693 &0.1 $\pm$ 0.8 &      21.9 &       2.3 &      0.22 &     -9.13 &     -2.01 &      9.11 &     -0.62 &         3 &      NGC 676 &     -3.10
             \\
        NGC 0755 &3.4 $\pm$ 1.0 &      21.9 &       2.8 &     -0.04 &     -9.01 &     -2.45 &      9.44 &     -0.52 &         2 &      NGC 755 &     -5.25
             \\
        NGC 0772 &3.0 $\pm$ 0.5 &      35.6 &       5.8 &      1.05 &     -9.54 &     -2.41 &      10.2 &     -1.17 &         3 &      NGC 772 &     -1.67
             \\
        NGC 0895 &5.9 $\pm$ 0.4 &      31.3 &       3.2 &      0.48 &     -9.11 &     -2.34 &      9.90 &     -0.52 &         2 &      NGC 895 &     -6.25
             \\
        NGC 0925 &7.0 $\pm$ 0.4 &       9.3 &      10.3 &      0.15 &     -9.00 &     -2.64 &      9.66 &     -0.55 &         7 &      NGC 925 &     -4.83
             \\
        NGC 0959 &7.8 $\pm$ 1.1 &      11.4 &       2.0 &     -0.64 &     -9.28 &     -2.20 &      8.44 &     -0.97 &         7 &      NGC 925 &     -5.41
             \\
        NGC 0986 &2.3 $\pm$ 0.9 &      24.9 &       4.0 &      1.00 &     -9.04 &     -1.82 &      9.13 &     -1.31 &   \nodata &     \nodata &     -4.94
             \\
        NGC 0991 &5.0 $\pm$ 0.5 &      20.4 &       1.6 &     -0.05 &     -8.57 &     -1.92 &      9.15 &     -0.60 &        11 &      NGC 988 &     -5.04
             \\
        NGC 1022 &1.1 $\pm$ 0.5 &      19.3 &       2.5 &      0.62 &     -8.92 &     -1.56 &      8.53 &     -1.50 &        11 &      NGC 988 &     -4.99
             \\
        NGC 1035 &5.0 $\pm$ 0.5 &      16.1 &       1.9 &     -0.03 &     -9.16 &     -1.82 &      8.88 &     -0.97 &        11 &      NGC 988 &     -3.64
             \\
        NGC 1042 &6.0 $\pm$ 0.4 &      18.0 &       3.8 &      0.15 &     -9.19 &     -2.34 &      9.49 &     -0.76 &        11 &      NGC 988 &     -2.57
             \\
        NGC 1051$^{*}$ &8.7 $\pm$ 1.1 &      17.1 &       1.8 &     -0.51 &     -7.94 &     -2.32 &      9.05 &     -0.43 &        11 &      NGC 988 &     -4.40
             \\
        NGC 1055 &3.2 $\pm$ 0.6 &      13.2 &       6.2 &      0.54 &     -9.21 &     -2.11 &      9.56 &     -0.55 &         5 &     NGC 1068 &     -2.31
             \\
        NGC 1068 &3.0 $\pm$ 0.6 &      14.4 &       6.7 &      1.59 &     -5.69 &     -1.20 &      9.15 &     -1.64 &         5 &     NGC 1068 &     -3.62
             \\
        NGC 1084 &4.9 $\pm$ 0.5 &      18.5 &       3.1 &      0.83 &     -8.90 &     -1.51 &      9.57 &     -0.86 &        11 &      NGC 988 &     -5.29
             \\
        NGC 1097 &3.3 $\pm$ 0.5 &      15.2 &       9.4 &      1.03 &     -9.20 &     -2.11 &      9.75 &     -0.93 &         4 &     NGC 1097 &     -1.14
             \\
        NGC 1140 &9.0 $\pm$ 2.4 &      18.2 &       1.8 &      0.10 &     -8.58 &     -1.76 &      9.34 &     -0.56 &   \nodata &     \nodata &     -6.18
             \\
        NGC 1291 &0.1 $\pm$ 0.5 &       9.7 &       9.7 &     -0.42 &    -10.49 &     -3.19 &      9.11 &     -1.33 &         2 &     NGC 1291 &     -5.28
             \\
        NGC 1310 &5.0 $\pm$ 0.5 &      22.4 &       2.0 &     -0.11 &     -9.20 &     -2.25 &      \nodata &     \nodata &        54 &      Fornax &     -2.65
             \\
        NGC 1317 &0.8 $\pm$ 0.6 &      18.6 &       3.1 &      0.04 &     -9.76 &     -2.31 &      \nodata &      \nodata &        54 &      Fornax &    -0.864
             \\
        NGC 1365 &3.1 $\pm$ 0.6 &      20.0 &      10.3 &      1.46 &     -8.96 &     -1.99 &      10.0 &     -0.80 &        54 &      Fornax &     -2.90
             \\
        NGC 1385 &5.9 $\pm$ 0.5 &      18.7 &       3.6 &      0.70 &     -8.78 &     -1.78 &      9.29 &     -1.04 &        51 &    Eridanus &     -4.15
             \\
        NGC 1512 &1.1 $\pm$ 0.6 &      10.4 &       7.0 &     -0.22 &     -9.62 &     -2.76 &      9.46 &     -0.43 &         3 &     NGC 1512 &     -1.37
             \\
        NGC 1546 &-0.4 $\pm$ 1.7 &      17.2 &       3.2 &      0.29 &     -9.32 &     -2.00 &      \nodata &    \nodata &        16 &      Dorado &     -2.64
             \\
        NGC 1566 &4.0 $\pm$ 0.1 &      17.4 &       8.2 &      0.84 &     -9.26 &     -2.30 &      9.89 &     -0.81 &        16 &      Dorado &     -3.51
             \\
        NGC 1672 &3.3 $\pm$ 0.6 &      15.1 &       6.4 &      0.83 &     -9.09 &     -1.96 &      10.0 &     -0.46 &         8 &     NGC 1672 &     -3.92
             \\
        NGC 1800 &8.0 $\pm$ 4.0 &       8.3 &       1.9 &     -1.06 &     -9.18 &     -2.27 &      8.17 &     -0.79 &   \nodata &     \nodata &     -5.45
             \\
        NGC 1808 &1.2 $\pm$ 0.6 &      10.8 &       6.6 &      0.83 &     -8.95 &     -1.70 &      9.18 &     -0.95 &         2 &     NGC 1808 &     -3.23
             \\
        NGC 1964 &3.2 $\pm$ 1.0 &      21.1 &       5.2 &      0.56 &     -9.38 &     -2.33 &      9.65 &     -0.86 &         5 &     NGC 1964 &     -5.09
             \\
        NGC 2090 &4.5 $\pm$ 2.0 &      11.3 &       4.3 &     -0.14 &     -9.39 &     -2.33 &      9.35 &     -0.67 &   \nodata &     \nodata &     -6.18
             \\
        NGC 2207 &4.5 $\pm$ 0.9 &      36.6 &       4.6 &      1.31 &     -8.91 &     -1.96 &      10.1 &     -0.65 &   \nodata &     \nodata &     0.877
             \\
         IC 2163 &5.0 $\pm$ 0.6 &      36.0 &       3.0 &      0.81 &     -9.24 &     -2.08 &      9.93 &     -1.13 &   \nodata &     \nodata &    0.0490
             \\
     ESO 556-012$^{*}$ &8.2 $\pm$ 1.5 &      34.2 &       1.5 &      0.01 &     -7.89 &     -2.25 &      9.25 &     -0.31 &   \nodata &     \nodata &     -3.31
             \\
        NGC 2146 &2.3 $\pm$ 0.7 &      16.4 &       5.3 &      1.27 &     -8.71 &     -1.44 &      9.51 &     -1.37 &   \nodata &     \nodata &     -7.08
             \\
       NGC 2146A &5.1 $\pm$ 0.6 &      25.0 &       2.8 &     -0.18 &     -9.17 &     -2.69 &      9.54 &     -0.45 &   \nodata &     \nodata &     -6.52
             \\
        NGC 2403 &6.0 $\pm$ 0.4 &       3.2 &      19.8 &     -0.04 &     -8.94 &     -2.47 &      9.34 &     -0.61 &         2 &     NGC 2403 &     -5.12
             \\
        NGC 2500 &6.9 $\pm$ 0.4 &       9.9 &       2.5 &     -0.40 &     -9.05 &     -2.02 &      8.76 &     -0.63 &         3 &     NGC 2541 &     -6.29
             \\
        NGC 2543 &3.1 $\pm$ 0.4 &      36.9 &       1.9 &      0.52 &     -9.21 &     -1.97 &      9.84 &     -0.72 &   \nodata &     \nodata &     -2.85
             \\
        NGC 2537 &8.4 $\pm$ 1.6 &       6.9 &       2.0 &     -0.75 &     -9.14 &     -1.83 &      8.24 &     -0.91 &   \nodata &     \nodata &     -3.73
             \\
        NGC 2541 &6.0 $\pm$ 0.4 &      10.3 &       4.1 &     -0.39 &     -8.75 &     -2.48 &      9.39 &     -0.27 &         3 &     NGC 2541 &     -4.82
             \\
        NGC 2552 &9.0 $\pm$ 0.5 &      10.0 &       3.2 &     -0.77 &     -7.70 &     -2.62 &      8.72 &     -0.67 &         3 &     NGC 2541 &     -4.75
             \\
       UGC 04393 &4.6 $\pm$ 1.3 &      32.6 &       1.6 &      0.09 &     -8.82 &     -2.18 &      9.61 &     -0.51 &   \nodata &     \nodata &     -4.59
             \\
       UGC 04390 &6.6 $\pm$ 0.7 &      34.6 &       1.5 &      0.01 &     -8.04 &     -2.28 &      9.51 &      0.19 &   \nodata &     \nodata &     -2.97
             \\
       UGC 04499 &7.9 $\pm$ 0.8 &      12.5 &       2.0 &     -0.74 &     -7.91 &     -2.36 &      8.89 &     -0.12 &         3 &     NGC 2681 &     -5.28
             \\
       UGC 04514 &5.9 $\pm$ 0.5 &      12.7 &       1.8 &     -0.91 &     -8.08 &     -2.46 &      8.75 &     -0.34 &         3 &     NGC 2681 &     -5.73
             \\
        NGC 2681 &0.4 $\pm$ 0.6 &      12.6 &       3.5 &     -0.23 &     -9.82 &     -2.34 &      7.22 &     -2.76 &         3 &     NGC 2681 &     -4.85
             \\
        NGC 2710 &3.1 $\pm$ 0.7 &      39.0 &       1.7 &      0.15 &     -9.22 &     -2.31 &      9.56 &     -0.65 &   \nodata &     \nodata &     -6.17
             \\
       UGC 04800 &5.9 $\pm$ 0.4 &      37.7 &       1.6 &     -0.20 &     -8.32 &     -2.57 &      9.38 &     -0.51 &   \nodata &     \nodata &     -5.76
             \\
        NGC 2775 &1.7 $\pm$ 0.8 &      19.2 &       4.0 &      0.06 &    -10.05 &     -2.53 &      8.68 &     -1.75 &         2 &     NGC 2775 &     -3.36
             \\
        NGC 2782 &1.1 $\pm$ 0.5 &      38.7 &       3.4 &      0.99 &     -9.00 &     -2.06 &      9.61 &     -0.93 &         4 &     NGC 2782 &     -4.81
             \\
        NGC 2841 &3.0 $\pm$ 0.7 &      14.1 &       8.1 &      0.28 &     -9.96 &     -2.66 &      9.75 &     -0.93 &         2 &     NGC 2841 &     -4.62
             \\
        NGC 2976 &5.3 $\pm$ 0.8 &       3.6 &       6.0 &     -0.74 &     -9.20 &     -2.23 &      8.09 &     -1.18 &         9 &         M81 &     -3.11
             \\
        NGC 3023 &5.4 $\pm$ 0.8 &      26.4 &       1.9 &      0.27 &     -8.75 &     -1.95 &      9.53 &     -0.49 &         7 &     NGC 2967 &     0.211
             \\
        NGC 3049 &2.5 $\pm$ 0.7 &      21.6 &       2.1 &     0.00 &     -9.05 &     -2.14 &      9.03 &     -0.63 &   \nodata &     \nodata &     -5.90
             \\
        NGC 3031$^{*}$ &2.4 $\pm$ 0.6 &       3.6 &      21.7 &      0.03 &     -9.92 &     -2.59 &      9.46 &     -1.05 &         9 &         M81 &     -2.05
             \\
     ESO 435-016$^{*}$ &3.4 $\pm$ 2.3 &      11.7 &       1.8 &     -0.38 &     -8.81 &     -1.85 &      8.80 &     -0.38 &         9 &     NGC 3175 &     -4.93
             \\
        NGC 3089 &3.2 $\pm$ 0.9 &      36.4 &       1.8 &      0.66 &     -9.10 &     -1.79 &      8.82 &     -1.49 &         3 &      IC 2537 &     -4.74
             \\
     ESO 499-037$^{*}$ &6.2 $\pm$ 1.5 &      11.5 &       3.3 &     -0.43 &     -7.84 &     -2.40 &      9.00 &     -0.37 &         9 &     NGC 3175 &     -4.54
             \\
         IC 2537 &5.0 $\pm$ 0.4 &      37.6 &       2.5 &      0.63 &     -9.13 &     -2.13 &      9.75 &     -0.74 &         3 &      IC 2537 &     -5.21
             \\
        NGC 3185 &1.1 $\pm$ 1.0 &      17.4 &       2.0 &     -0.40 &     -9.59 &     -2.29 &      8.42 &     -1.23 &         8 &     NGC 3190 &     -2.57
             \\
        NGC 3187 &5.0 $\pm$ 0.8 &      17.4 &       2.3 &     -0.04 &     -8.55 &     -2.08 &      8.85 &     -0.68 &   \nodata &     \nodata &     -1.12
             \\
        NGC 3198 &5.2 $\pm$ 0.6 &      16.6 &       7.6 &      0.41 &     -9.29 &     -2.62 &      9.99 &     -0.66 &         5 &     NGC 3198 &     -5.00
             \\
     ESO 317-023$^{*}$ &1.1 $\pm$ 0.6 &      37.9 &       1.9 &      0.99 &     -8.33 &     -1.56 &      \nodata &    \nodata &   \nodata &     \nodata &     -4.18
             \\
        NGC 3225 &5.9 $\pm$ 0.6 &      33.7 &       1.6 &      0.06 &     -8.90 &     -2.23 &      9.57 &     -0.52 &         2 &     NGC 3225 &     -4.91
             \\
        NGC 3244 &5.6 $\pm$ 0.8 &      36.5 &       2.0 &      0.53 &     -9.11 &     -2.04 &      9.53 &     -0.88 &   \nodata &     \nodata &     -3.10
             \\
         IC 2574 &8.9 $\pm$ 0.4 &       4.0 &      11.8 &     -0.76 &     -8.05 &     -2.94 &      9.12 &     -0.26 &         9 &         M81 &     -3.25
             \\
       UGC 05675 &8.6 $\pm$ 1.1 &      16.8 &       1.6 &     -1.48 &     -8.90 &     -3.16 &      8.42 &     -6.22 &   \nodata &     \nodata &     -3.30
             \\
        NGC 3277 &1.8 $\pm$ 0.8 &      21.8 &       1.9 &     -0.34 &     -9.82 &     -2.42 &      8.66 &     -1.32 &         5 &     NGC 3254 &     -4.89
             \\
        NGC 3319 &5.9 $\pm$ 0.4 &      14.6 &       4.9 &      0.22 &     -8.45 &     -2.32 &      9.49 &     -0.56 &         5 &     NGC 3198 &     -5.42
             \\
        NGC 3344 &4.0 $\pm$ 0.3 &       6.9 &       6.7 &     -0.14 &     -9.20 &     -2.29 &      9.14 &     -0.60 &         3 &     NGC 3344 &     -6.57
             \\
        NGC 3351$^{*}$ &3.0 $\pm$ 0.6 &      11.8 &       7.1 &      0.36 &     -9.48 &     -2.30 &      9.15 &     -1.12 &        24 &        LeoI &     -3.01
             \\
        NGC 3359 &5.2 $\pm$ 0.5 &      18.0 &       5.7 &      0.45 &     -8.98 &     -2.39 &      9.98 &     -0.46 &   \nodata &     \nodata &     -5.51
             \\
        NGC 3368$^{*}$ &1.8 $\pm$ 0.5 &      13.5 &       7.4 &      0.25 &     -9.85 &     -2.58 &      9.45 &     -1.08 &        24 &        LeoI &     -3.31
             \\
       NGC 3377A &8.8 $\pm$ 0.5 &       9.1 &       1.6 &     -1.62 &     -8.49 &     -2.74 &      8.01 &     -0.41 &   \nodata &     \nodata &     -1.57
             \\
        NGC 3486 &5.2 $\pm$ 0.8 &      11.6 &       6.1 &      0.15 &     -9.15 &     -2.37 &      9.46 &     -0.58 &         2 &     NGC 3486 &     -4.39
             \\
        NGC 3521 &4.0 $\pm$ 0.3 &       9.0 &       9.6 &      0.52 &     -9.44 &     -2.18 &      9.54 &     -0.92 &   \nodata &     \nodata &     -6.06
             \\
        NGC 3621 &6.9 $\pm$ 0.4 &       8.3 &      10.4 &      0.47 &     -9.09 &     -2.23 &      9.86 &     -0.53 &   \nodata &     \nodata &     \nodata
             \\
        NGC 3627 &3.0 $\pm$ 0.4 &       9.1 &       8.9 &      0.63 &     -9.30 &     -2.00 &      8.88 &     -1.55 &        24 &        LeoI &     -2.12
             \\
        NGC 3705 &2.4 $\pm$ 0.6 &      15.3 &       4.3 &      0.00 &     -9.56 &     -2.46 &      9.32 &     -0.83 &        24 &        LeoI &     -5.39
             \\
        NGC 3885 &0.2 $\pm$ 1.0 &      23.1 &       2.4 &      0.49 &     -9.25 &     -1.82 &      9.27 &     -0.70 &         3 &     NGC 3885 &     -4.46
             \\
     ESO 440-011$^{*}$ &6.9 $\pm$ 0.5 &      23.1 &       2.6 &      0.04 &     -7.65 &     -2.34 &      9.37 &     -0.58 &         3 &     NGC 3936 &     -3.64
             \\
        NGC 3938 &5.1 $\pm$ 0.5 &      12.2 &       4.4 &      0.28 &     -9.13 &     -1.99 &      9.34 &     -0.71 &        25 &     NGC 4151 &     -4.94
             \\
        NGC 4030 &4.0 $\pm$ 0.2 &      21.0 &       3.6 &      0.92 &     -9.16 &     -1.67 &      9.62 &     -0.88 &         3 &     NGC 4030 &     -4.09
             \\
        NGC 4038 &8.9 $\pm$ 0.5 &      22.2 &       4.4 &      0.54 &     -7.11 &     -2.26 &      9.55 &     -1.21 &        17 &     NGC 4038 &     0.779
             \\
        NGC 4039 &8.9 $\pm$ 0.2 &      22.2 &       4.0 &     -0.11 &     -7.77 &     -2.83 &      \nodata &    \nodata &        17 &     NGC 4038 &     0.907
             \\
        NGC 4116 &7.5 $\pm$ 1.1 &      17.0 &       2.9 &     -0.01 &     -8.73 &     -2.22 &      9.40 &     -0.48 &         4 &     NGC 4179 &     -2.72
             \\
        NGC 4136 &5.2 $\pm$ 0.7 &      10.9 &       2.6 &     -0.36 &     -9.08 &     -2.08 &      8.98 &     -0.56 &        27 &       ComaI &     -4.45
             \\
        NGC 4189 &6.0 $\pm$ 0.9 &      32.0 &       2.8 &      0.57 &     -9.24 &     -2.16 &      9.37 &     -0.99 &         9 &      VirgoM &     -3.38
             \\
        NGC 4192$^{*}$ &2.5 $\pm$ 0.8 &      17.0 &       9.5 &      0.36 &     -9.71 &     -2.88 &      9.68 &     -0.99 &       159 &      VirgoA &     \nodata
             \\
        NGC 4193 &4.1 $\pm$ 1.8 &      32.0 &       2.0 &      0.30 &     -9.34 &     -2.14 &      9.15 &     -1.04 &         9 &      VirgoM &     -3.14
             \\
        NGC 4236 &7.9 $\pm$ 0.5 &       4.5 &      20.4 &     -0.55 &     -8.61 &     -3.29 &      9.22 &     -0.66 &         2 &     NGC 4236 &     -5.51
             \\
        NGC 4242 &8.0 $\pm$ 0.4 &      10.3 &       4.2 &     -0.52 &     -7.51 &     -2.62 &      8.96 &     -0.78 &        21 &     NGC 4258 &     -4.09
             \\
        NGC 4248 &2.9 $\pm$ 2.7 &       7.2 &       2.3 &     -1.07 &     -8.91 &     -2.31 &      7.77 &     -1.11 &        21 &     NGC 4258 &     -1.58
             \\
        NGC 4254$^{*}$ &5.2 $\pm$ 0.6 &      17.0 &       5.1 &      1.02 &     -9.03 &     -1.67 &      9.68 &     -0.92 &       159 &      VirgoA &     -5.52
             \\
        NGC 4258$^{*}$ &4.0 $\pm$ 0.2 &       7.2 &      17.4 &      0.26 &     -9.63 &     -2.76 &      9.64 &     -0.90 &        21 &     NGC 4258 &     -2.20
             \\
        NGC 4274 &1.7 $\pm$ 0.6 &      16.0 &       5.0 &      0.04 &     -9.92 &     -2.59 &      8.67 &     -1.61 &        27 &       ComaI &     -2.61
             \\
        NGC 4298 &5.2 $\pm$ 0.5 &      17.0 &       2.7 &      0.41 &     -9.03 &     -1.75 &      8.95 &     -1.11 &       159 &      VirgoA &    -0.721
             \\
        NGC 4303 &4.0 $\pm$ 0.2 &      17.0 &       6.3 &      1.01 &     -9.08 &     -1.87 &      9.74 &     -0.93 &        55 &      VirgoW &     -2.78
             \\
        NGC 4314 &1.0 $\pm$ 0.4 &      16.0 &       3.8 &     -0.18 &     -9.97 &     -2.57 &      \nodata &   \nodata &        27 &       ComaI &     -3.60
             \\
        NGC 4321 &4.0 $\pm$ 0.3 &      17.5 &       7.5 &      0.93 &     -9.28 &     -2.13 &      9.52 &     -1.29 &       159 &      VirgoA &     -3.80
             \\
        NGC 4383 &1.0 $\pm$ 0.9 &      17.0 &       1.5 &      0.22 &     -8.81 &     -1.39 &      9.34 &     -0.38 &       159 &      VirgoA &     -3.99
             \\
        NGC 4396 &6.7 $\pm$ 0.9 &      17.0 &       3.1 &     -0.26 &     -8.95 &     -2.54 &      9.02 &     -0.82 &       159 &      VirgoA &     \nodata
             \\
        NGC 4414 &5.1 $\pm$ 0.8 &      18.6 &       2.7 &      0.91 &     -9.21 &     -1.32 &      9.58 &     -0.95 &        27 &       ComaI &     -5.68
             \\
        NGC 4413$^{*}$ &1.9 $\pm$ 1.0 &      17.0 &       1.8 &     -0.37 &     -9.27 &     -2.17 &      8.35 &     -1.29 &       159 &      VirgoA &     -2.73
             \\
        NGC 4421 &-0.2 $\pm$ 0.7 &      17.0 &       2.2 &     -1.55 &    -10.86 &     -3.50 &     \nodata &   \nodata &       159 &      VirgoA &     -4.02
             \\
        NGC 4438 &0.7 $\pm$ 1.6 &      17.0 &       8.9 &      0.03 &     -9.89 &     -3.14 &      8.38 &     -2.03 &       159 &      VirgoA &     -1.71
             \\
        NGC 4440 &1.1 $\pm$ 0.6 &      17.0 &       1.7 &     -1.01 &    -10.27 &     -2.74 &      \nodata &      \nodata &       159 &      VirgoA &     -2.13
             \\
        NGC 4450 &2.3 $\pm$ 0.7 &      17.0 &       5.6 &     -0.13 &    -10.13 &     -2.91 &      8.52 &     -1.94 &       159 &      VirgoA &     -4.44
             \\
        NGC 4454 &0.0 $\pm$ 0.6 &      34.6 &       2.0 &      0.03 &     -9.79 &     -2.47 &      8.94 &     -1.26 &   \nodata &     \nodata &     -3.44
             \\
        NGC 4490 &7.0 $\pm$ 0.2 &      11.0 &       6.3 &      0.66 &     -8.85 &     -1.84 &      9.82 &     -0.98 &        21 &     NGC 4258 &     -1.16
             \\
        NGC 4491 &1.1 $\pm$ 0.9 &      17.0 &       1.6 &     -0.38 &     -9.25 &     -2.09 &      \nodata &     \nodata &       159 &      VirgoA &     -4.72
             \\
        NGC 4506 &1.3 $\pm$ 1.2 &      17.0 &       1.4 &     -0.70 &     -9.42 &     -2.30 &      \nodata &     \nodata &       159 &      VirgoA &     -4.59
             \\
        NGC 4531 &-0.3 $\pm$ 2.0 &      17.0 &       3.0 &     -0.60 &     -9.93 &     -2.85 &      7.43 &     -2.33 &       159 &      VirgoA &     -3.22
             \\
        NGC 4536 &4.2 $\pm$ 0.8 &      15.3 &       7.3 &      0.65 &     -9.08 &     -2.27 &      9.56 &     -0.84 &        55 &      VirgoW &     -2.47
             \\
        NGC 4559 &6.0 $\pm$ 0.4 &      17.0 &      10.0 &      0.69 &     -9.10 &     -2.60 &      10.1 &     -0.68 &        27 &       ComaI &     -4.03
             \\
        NGC 4567 &4.0 $\pm$ 0.3 &      17.0 &       2.9 &      0.02 &     -9.48 &     -2.19 &      9.05 &     -0.90 &       159 &      VirgoA &     0.389
             \\
        NGC 4569 &2.4 $\pm$ 0.6 &      17.0 &       9.6 &      0.44 &     -9.75 &     -2.81 &      8.76 &     -2.13 &       159 &      VirgoA &     \nodata
             \\
        NGC 4579$^{*}$ &2.8 $\pm$ 0.7 &      17.0 &       5.6 &      0.32 &     -9.91 &     -2.46 &      8.80 &     -1.82 &       159 &      VirgoA &     -3.67
             \\
        NGC 4584 &1.1 $\pm$ 0.7 &      17.0 &       1.4 &     -0.57 &     -9.21 &     -2.12 &      7.45 &     -1.81 &       159 &      VirgoA &     -4.73
             \\
        NGC 4594 &1.1 $\pm$ 0.4 &       9.1 &       8.9 &     -0.21 &    -10.50 &     -2.84 &      8.45 &     -2.18 &         6 &     NGC 4594 &     -5.55
             \\
        NGC 4618 &8.6 $\pm$ 1.1 &       9.5 &       3.8 &     -0.11 &     -8.96 &     -2.04 &      9.11 &     -0.67 &        21 &     NGC 4258 &     -2.63
             \\
        NGC 4625 &8.8 $\pm$ 0.7 &       9.5 &       1.6 &     -0.70 &     -9.13 &     -1.88 &      8.68 &     -0.38 &        21 &     NGC 4258 &     -2.39
             \\
        NGC 4691 &0.4 $\pm$ 1.0 &      16.1 &       3.2 &      0.48 &     -8.88 &     -1.76 &      8.43 &     -1.59 &        21 &VirgoSE;high &     -4.31
             \\
        NGC 4736 &2.4 $\pm$ 0.7 &       5.2 &      10.0 &      0.16 &     -9.59 &     -2.10 &      8.50 &     -1.75 &        16 &     NGC 4736 &     -5.02
             \\
        NGC 4771 &6.4 $\pm$ 1.0 &      17.0 &       3.2 &     -0.37 &     -9.59 &     -2.66 &      8.96 &     -0.97 &        21 &VirgoSE;high &     -4.10
             \\
        NGC 4772 &1.1 $\pm$ 0.3 &      17.0 &       3.6 &     -0.07 &     -9.55 &     -2.46 &      8.99 &     -1.02 &        21 &VirgoSE;high &     -4.12
             \\
        NGC 4826$^{*}$ &2.4 $\pm$ 0.6 &      17.0 &       9.7 &      0.84 &     -9.86 &     -2.42 &      9.44 &     -1.72 &         2 &     NGC 4826 &     -6.21
             \\
       UGC 08313 &5.0 $\pm$ 0.4 &      12.1 &       1.4 &     -0.95 &     -8.71 &     -2.20 &      8.14 &     -0.72 &         7 &     NGC 5194 &     -2.41
             \\
        NGC 5055 &4.0 $\pm$ 0.2 &       8.2 &      12.5 &      0.58 &     -9.37 &     -2.27 &      9.61 &     -0.85 &         7 &     NGC 5194 &     -3.91
             \\
       UGC 08365 &6.5 $\pm$ 0.8 &      20.5 &       1.5 &     -0.63 &     -8.22 &     -2.44 &      8.82 &     -0.57 &         3 &     NGC 5103 &     -5.33
             \\
        NGC 5169 &4.0 $\pm$ 1.4 &      38.2 &       1.5 &      0.06 &     -9.04 &     -2.30 &      9.60 &     -0.09 &         4 &     NGC 5198 &     -2.09
             \\
        NGC 5194$^{*}$ &4.0 $\pm$ 0.3 &       8.4 &      10.1 &      0.92 &     -9.09 &     -1.76 &      9.42 &     -1.35 &         7 &     NGC 5194 &    -0.616
             \\
        NGC 5195$^{*}$ &1.0 $\pm$ 4.2 &       8.4 &       5.7 &     -1.30 &    -11.01 &     -3.48 &      \nodata &    \nodata &         7 &     NGC 5194 &     0.127
             \\
        NGC 5236$^{*}$ &5.0 $\pm$ 0.4 &       4.5 &      13.7 &      0.76 &     -9.07 &     -1.65 &      9.24 &     -1.28 &        10 &  CentaurusA &     -4.70
             \\
        NGC 5253 &7.5 $\pm$ 4.8 &       3.1 &       4.7 &     -0.50 &     -8.55 &     -1.66 &      7.93 &     -1.30 &        10 &  CentaurusA &     -3.82
             \\
        NGC 5398 &7.9 $\pm$ 0.8 &      15.9 &       2.8 &     -0.23 &     -8.86 &     -2.36 &      9.03 &     -0.22 &   \nodata &     \nodata &     -5.32
             \\
        NGC 5457$^{*}$ &5.9 $\pm$ 0.3 &       7.5 &      21.6 &      0.90 &     -9.01 &     -2.34 &      10.0 &     -0.61 &         8 &     NGC 5457 &     -3.04
             \\
        NGC 5474 &6.0 $\pm$ 0.3 &       6.8 &       2.8 &     -0.60 &     -8.83 &     -1.96 &      8.96 &     -0.46 &         8 &     NGC 5457 &     -3.29
             \\
        NGC 5566 &1.5 $\pm$ 0.7 &      22.7 &       5.8 &      0.06 &    -10.06 &     -3.00 &      9.26 &     -1.32 &         5 &     NGC 5566 &     -1.49
             \\
       UGC 09215 &6.4 $\pm$ 0.8 &      20.8 &       2.2 &     -0.06 &     -8.84 &     -2.21 &      9.29 &     -0.43 &         5 &     NGC 5566 &     -5.15
             \\
        NGC 5701 &-0.4 $\pm$ 0.9 &      22.8 &       2.2 &     -0.21 &    -10.03 &     -2.44 &      9.69 &     -0.49 &        26 &     NGC 5746 &     -5.81
             \\
        NGC 5705 &6.5 $\pm$ 1.1 &      26.9 &       1.4 &     0.00 &     -8.70 &     -2.01 &      9.52 &     -0.09 &        26 &     NGC 5746 &     -1.39
             \\
        NGC 5713 &4.0 $\pm$ 0.3 &      26.9 &       2.6 &      0.99 &     -8.90 &     -1.54 &      9.78 &     -0.80 &        26 &     NGC 5746 &     -3.44
             \\
        NGC 5727 &7.9 $\pm$ 0.9 &      24.3 &       1.1 &     -0.31 &     -8.04 &     -1.96 &      9.14 &     -0.16 &   \nodata &     \nodata &     -4.11
             \\
        NGC 5719 &2.4 $\pm$ 0.7 &      26.9 &       2.8 &      0.63 &     -9.30 &     -1.95 &      9.79 &     -0.21 &        26 &     NGC 5746 &     -1.95
             \\
        NGC 5832 &3.1 $\pm$ 0.7 &      10.6 &       2.4 &     -0.80 &     -7.81 &     -2.42 &      8.88 &     -0.21 &   \nodata &     \nodata &     \nodata
             \\
        NGC 5806 &3.1 $\pm$ 0.5 &      20.5 &       2.8 &      0.11 &     -9.50 &     -2.23 &      8.97 &     -1.17 &        13 &     NGC 5846 &     -4.32
             \\
        NGC 5962 &5.0 $\pm$ 0.8 &      30.2 &       2.5 &      0.87 &     -9.04 &     -1.71 &      9.53 &     -1.01 &         3 &     NGC 5962 &     -3.93
             \\
       UGC 10445 &6.0 $\pm$ 0.9 &      16.7 &       2.0 &     -0.50 &     -8.20 &     -2.37 &      9.11 &     -0.13 &   \nodata &     \nodata &     -6.97
             \\
        NGC 6239 &3.2 $\pm$ 0.7 &      16.9 &       1.9 &     -0.07 &     -8.87 &     -1.92 &      9.38 &     -0.38 &   \nodata &     \nodata &     \nodata
             \\
        NGC 6255 &6.0 $\pm$ 0.8 &      16.5 &       3.3 &     -0.36 &     -8.02 &     -2.66 &      9.11 &     -0.44 &         3 &     NGC 6207 &     -5.19
             \\
        NGC 6340 &0.4 $\pm$ 0.6 &      21.3 &       2.6 &     -0.75 &    -10.42 &     -3.05 &      9.10 &     -1.06 &         5 &     NGC 6340 &     -2.98
             \\
       UGC 10791 &8.8 $\pm$ 0.5 &      23.2 &       1.7 &     -1.08 &     -8.78 &     -3.08 &      8.68 &     -0.30 &   \nodata &     \nodata &     -2.87
             \\
        NGC 6902 &2.4 $\pm$ 1.1 &      37.6 &       3.3 &      0.59 &     -9.48 &     -2.43 &      10.2 &     -0.43 &         3 &    PGC 64632 &     -3.19
             \\
        NGC 7167 &5.1 $\pm$ 0.7 &      35.3 &       1.5 &      0.41 &     -9.07 &     -1.83 &      9.40 &     -0.96 &   \nodata &     \nodata &     -6.93
             \\
         IC 5156 &2.2 $\pm$ 0.8 &      37.4 &       2.3 &      0.36 &     -9.53 &     -2.35 &      9.44 &     -0.98 &        12 &    PGC 67883 &     -4.24
             \\
        NGC 7320 &6.6 $\pm$ 0.9 &      13.5 &       2.3 &     -0.56 &     -8.97 &     -2.35 &      8.47 &     -1.02 &         4 &    PGC 69327 &     -3.11
             \\
        NGC 7331 &3.9 $\pm$ 0.4 &      14.8 &       9.9 &      0.91 &     -9.38 &     -2.24 &      9.88 &     -0.97 &         4 &    PGC 69327 &     -3.69
             \\
        NGC 7418 &5.8 $\pm$ 0.5 &      18.2 &       3.8 &      0.32 &     -9.16 &     -2.19 &      9.22 &     -0.91 &   \nodata &     \nodata &     -3.38
             \\
       NGC 7418A &6.5 $\pm$ 1.9 &      27.6 &       1.9 &      0.04 &     -7.81 &     -2.23 &      9.58 &     -0.28 &   \nodata &     \nodata &     -2.87
             \\
        NGC 7421 &3.8 $\pm$ 1.1 &      23.7 &       2.0 &      0.38 &     -9.03 &     -1.81 &      8.84 &     -1.12 &         5 &    PGC 70090 &     -3.38
             \\
         IC 5271 &3.1 $\pm$ 0.6 &      22.6 &       2.5 &      0.20 &     -9.52 &     -2.13 &      9.13 &     -1.07 &   \nodata &     \nodata &     -5.60
             \\
     ESO 406-042$^{*}$ &8.8 $\pm$ 0.8 &      17.2 &       1.5 &     -0.73 &     -8.17 &     -2.40 &      8.43 &     -0.57 &   \nodata &     \nodata &     -4.34
             \\
        NGC 7479 &4.4 $\pm$ 0.9 &      34.9 &       3.5 &      1.21 &     -8.96 &     -1.80 &      9.93 &     -0.95 &   \nodata &     \nodata &     -6.39
             \\
     ESO 407-007$^{*}$ &3.1 $\pm$ 0.6 &      21.0 &       1.9 &     -0.22 &     -9.27 &     -2.27 &      8.82 &     -0.69 &         2 &    PGC 70582 &     -5.23
             \\
        NGC 7496 &3.1 $\pm$ 0.7 &      20.7 &       3.1 &      0.51 &     -9.02 &     -1.92 &      9.21 &     -0.93 &        13 &    PGC 71001 &     -4.63
             \\
       NGC 7496A &9.0 $\pm$ 0.5 &      20.7 &       1.5 &     -0.62 &     -8.22 &     -2.43 &      \nodata &    \nodata &   \nodata &     \nodata &     -3.55
             \\
     ESO 407-009$^{*}$ &6.8 $\pm$ 0.9 &      19.9 &       2.0 &     -0.77 &     -8.34 &     -2.82 &      \nodata &    \nodata &         2 &    PGC 70582 &     -5.20
             \\
        NGC 7552 &2.4 $\pm$ 0.6 &      22.3 &       3.9 &      1.32 &     -8.72 &     -1.38 &      9.55 &     -0.95 &        13 &    PGC 71001 &     -3.52
             \\
     ESO 407-014$^{*}$ &5.0 $\pm$ 0.6 &      37.0 &       1.5 &      0.30 &     -8.88 &     -2.03 &      9.53 &     -0.51 &   \nodata &     \nodata &     -6.63
             \\
        NGC 7582 &2.0 $\pm$ 0.5 &      22.3 &       6.0 &      1.12 &     -9.01 &     -1.95 &      9.58 &     -0.97 &        13 &    PGC 71001 &     -1.80
             \\
         IC 5325 &4.2 $\pm$ 0.5 &      18.7 &       2.8 &      0.31 &     -9.26 &     -1.96 &      8.82 &     -1.28 &        13 &    PGC 71001 &     -5.69
             \\
        NGC 7741 &6.0 $\pm$ 0.4 &      12.3 &       3.8 &     -0.15 &     -8.84 &     -2.32 &      9.16 &     -0.77 &   \nodata &     \nodata &     -5.03
             \\
        NGC 7793 &7.4 $\pm$ 0.6 &       2.0 &       9.9 &     -0.82 &     -9.04 &     -2.23 &      8.25 &     -0.85 &        13 &    Sculptor &     \nodata
             \\
\enddata
\tablecomments{
(1) Galaxies noted with an asterisk have a different name in Gil de Paz et al. (this volume).
(2) from the Virgo-infall corrected radial velocity adopting H$_0$=70 km s$^{-1}$ Mpc$^{-1}$, except as noted in Gil de Paz et al. (this volume).
}
\end{deluxetable}

\clearpage
\pagestyle{empty}
\begin{deluxetable}{lccrrrrcl}
\rotate
\tabletypesize{\scriptsize}
\tablecaption{XUV-disk Classification Results\label{table2}}
\tablewidth{0pt}
\tablehead{
\colhead{Object} & \multicolumn{2}{c}{-------- Type~1 --------} & \multicolumn{5}{c}{----------------------------------------- Type~2 -------------------------------------------} \\
\colhead{Name}     &\colhead{Assessment} &\colhead{Band} & \colhead{$S(K_{80})$} & \colhead{$S(UV)$} &\colhead{$S(LSB)/S(K_{80})$}& \colhead{$UV(AB)-K_{s}(AB)$} & \colhead{Assessment} & \colhead{Interaction comment}\\
\colhead{} & \colhead{(Y/N)} & \colhead{}&\colhead{(kpc$^2$)} & \colhead{(kpc$^2$)} &\colhead{} &\colhead{(mag)} &\colhead{(Y/N)} & \colhead{}\\
\colhead{(1)} & \colhead{(2)} &\colhead{(3)} &\colhead{(4)} &\colhead{(5)} &\colhead{(6)} &\colhead{(7)} &\colhead{(8)} &\colhead{(9)}\\}
\startdata
       NGC 0300  &    Y &  FUV &  19.4 & 102.2 & 4.20 &  1.4 &    N & \nodata                                          \\
       NGC 0628  &    Y &  FUV & 179.3 &1103.3 & 5.09 &  1.9 &    N & HVCs                                             \\
       NGC 0772  &    Y &  FUV &1163.3 &1262.8 & 0.20 &  1.0 &    N & asymmetric arm, NGC 770                           \\
       NGC 0991  &    N &  FUV &  11.8 & 207.0 &16.50 &  2.3 &    Y & asymmetric disk, asymmetric HI profile           \\
       NGC 1042  &    Y &  FUV & 148.1 & 546.2 & 2.25 &  \nodata &    N & asymmetric gas, NGC 1052                          \\
       NGC 1051  &    Y &  FUV &   5.9 & 103.0 &16.26 &  2.3 &    Y & \nodata                                          \\
       NGC 1140  &    N &  FUV &   6.6 & 127.2 &11.15 &  0.4 &    Y & pec UV-opt morphology                            \\
       NGC 1512  &    Y &  FUV &  50.8 & 138.1 & 1.67 &  2.8 &    N & NGC 1510                                          \\
       NGC 1672  &    Y &  NUV & 116.6 & 646.7 & 4.54 &  2.8 &    N & warped?                                          \\
       NGC 2090  &    N &  FUV &  19.9 & 249.6 &11.27 &  1.6 &    Y & inner ring                                       \\
    ESO 556-012  &    N &  NUV &   4.5 & 203.9 &23.90 &  0.5 &    Y & \nodata                                          \\
      NGC 2146A  &    Y &  FUV &  44.7 & 152.4 & 2.39 &  2.6 &    N & \nodata                                          \\
       NGC 2403  &    Y &  FUV &  21.3 & 209.0 & 8.57 &  1.2 &    Y & extraplanar gas accretion                        \\
       NGC 2541  &    N &  FUV &  14.6 & 180.0 &11.34 &  1.3 &    Y & \nodata                                          \\
      UGC 04393  &    Y &  NUV &  30.6 & 262.4 & 7.35 &  0.4 &    Y & asymmetric arm, UV-opt morphology                \\
      UGC 04390  &    N &  FUV &  10.3 & 381.8 &36.16 &  2.8 &    Y & lopsided UV disk                                 \\
       NGC 2710  &    Y &  FUV &  79.2 & 262.5 & 2.32 &  2.5 &    N & \nodata                                          \\
      UGC 04800  &    N &  FUV &   6.8 & 114.6 &15.87 &  3.0 &    Y & \nodata                                          \\
       NGC 2782  &    Y &  FUV & 121.3 &  \nodata & \nodata & \nodata &    N & tidal arms, merger remnant                       \\
       NGC 2841  &    Y &  FUV & 123.1 & 281.7 & 1.29 &  5.0 &    N & outer HI ring                                    \\
       NGC 3031  &    Y &  FUV &  58.8 & 354.0 & 4.95 &  3.6 &    N & tidal system                                     \\
       NGC 3185  &    Y &  FUV &  26.9 &  47.2 & 0.63 &  4.8 &    N & inner UV/opt ring                                \\
       NGC 3198  &    Y &  FUV & 133.7 & 614.8 & 3.59 &  2.4 &    N & NE HI tail                                       \\
        IC 2574  &    N &  FUV &   7.2 & 106.0 &13.76 &  -0.3 &    Y & \nodata                                          \\
       NGC 3319  &    N &  FUV &  36.1 & 326.9 & 8.05 &  2.0 &    Y & HI tail, asymmetry from companion                \\
       NGC 3344  &    Y &  FUV &  28.9 & 190.9 & 5.49 &  1.4 &    N & asymmetric HI warp                               \\
       NGC 3359  &    Y &  NUV & 102.2 & 604.1 & 4.91 &  1.7 &    N & accretion via HI companion                       \\
       NGC 3621  &    Y &  FUV &  62.6 & 459.6 & 5.86 &  2.4 &    N & linear HI filament across disk                   \\
       NGC 3705  &    Y &  NUV & 158.5 & 187.2 & 0.37 &  \nodata &    N & \nodata                                          \\
       NGC 4116  &    N &  NUV &  20.9 & 209.5 & 8.73 &  0.5 &    Y & \nodata                                          \\
       NGC 4236  &    N &  FUV &  14.6 & 157.8 & 9.76 &  1.8 &    Y & \nodata                                          \\
       NGC 4254  &    Y &  NUV & 154.5 & 481.8 & 2.11 &  2.8 &    N & asymmetric disk, VIRGOHI21                       \\
       NGC 4258  &    Y &  NUV &  80.5 & 422.7 & 4.30 &  3.6 &    N & anomalous arms?                                  \\
       NGC 4383  &    Y &  FUV &  \nodata &  \nodata & \nodata &  \nodata &    N & UV-opt morphology                                \\
       NGC 4414  &    Y &  FUV & 108.9 & 260.2 & 1.34 &  3.7 &    N & radio cont. asymmetries                          \\
       NGC 4438  &    Y &  FUV &  41.5 & 158.2 & 2.83 &  4.4 &    N & interaction/merger, NGC 4435                      \\
       NGC 4559  &    Y &  FUV & 212.7 &1029.9 & 3.75 &  1.9 &    N & \nodata                                          \\
       NGC 4625  &    Y &  FUV &   8.7 &  53.8 & 5.17 &  2.2 &    N & NGC 4618/NGC 4625A                                 \\
       NGC 5055  &    Y &  FUV &  85.9 & 260.7 & 2.02 &  4.2 &    N & TDG?(UGCA 342) and UGC 8313 (HI filament?)         \\
      UGC 08365  &    N &  FUV &   3.9 &  67.0 &15.65 &  0.6 &    Y & \nodata                                          \\
       NGC 5236  &    Y &  FUV & 140.3 & \nodata &\nodata &  \nodata &    N & NGC 5253                                          \\
       NGC 5457  &    Y &  FUV & 338.2 &1839.2 & 4.39 &  1.4 &    N & HI lopsided, companions N5477/N5474              \\
       NGC 5474  &    Y &  FUV &   9.0 & 110.0 &11.27 &  1.9 &    Y & offset nucleus, NGC 5457                          \\
       NGC 5705  &    N &  FUV &  23.7 & 322.7 &12.01 &  0.2 &    Y & companions to SE                                 \\
       NGC 5727  &    N &  FUV &  10.3 & 122.3 &10.85 &  1.9 &    Y & companion                                        \\
       NGC 5832  &    Y &  NUV &   6.1 &  76.7 &11.62 &  2.1 &    Y & \nodata                                          \\
      UGC 10445  &    Y &  FUV &   4.8 &  91.3 &10.04 &  0.4 &    Y & lopsided UV                                      \\
       NGC 6239  &    Y &  FUV &  12.6 &  99.4 & 6.26 &  2.5 &    N & tidal filaments, SDSS165045.93+424320.4          \\
       NGC 6255  &    N &  FUV &   5.2 & 115.3 &16.40 &  0.9 &    Y & \nodata                                          \\
      UGC 10791  &    Y &  FUV &   7.5 &  49.6 & 5.59 &  3.6 &    N & NGC 6340                                          \\
       NGC 6902  &    Y &  FUV & 184.1 &1092.5 & 4.78 &  2.7 &    N & inner ring                                       \\
      NGC 7418A  &    Y &  FUV &   1.7 & 527.8 &22.62 &  0.9 &    Y & NE complex in UV                                 \\
    ESO 406-042  &    N &  FUV &   7.8 &  63.2 & 7.11 &  1.5 &    Y & \nodata                                          \\
    ESO 407-014  &    N &  FUV &  24.5 & 294.2 &10.83 &  1.6 &    Y & \nodata                                          \\
\enddata
\tablecomments{
(1) Mixed-category XUV-disks have a 'Y' in the Assessment column for both criteria. (2) Colors measured within the LSB zone, $UV(AB)-K_s(AB)$, have been corrected for Galactic extinction.
}
\end{deluxetable}

\clearpage
\pagestyle{plaintop}

\begin{figure}[htbp]
\epsscale{0.7}
\plotone{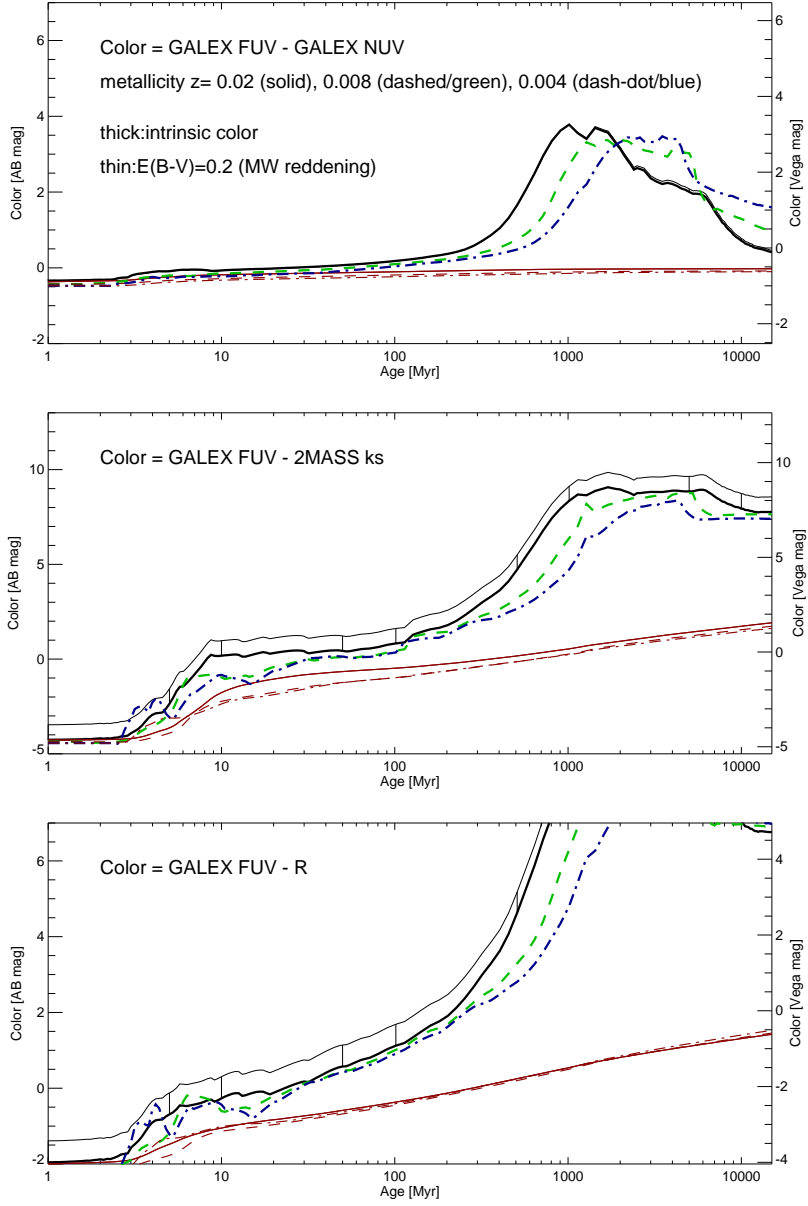}
\caption{Variation of $FUV-NUV$ and $FUV-$optical colors as a function
of age for Bruzual \& Charlot (2003) models of an instantaneous
starburst and continuous star formation (red lines).  In all panels, we show three
different metallicities (Z=0.02, solid; Z=0.008, dashed/green; Z=0.004
dash-dot/blue).  For the instantaneous starburst, solar metallicity case we also plot
(with a thin line) reddened colors for \ebv=0.2 and the MW
attentuation law (R$_V$ = 3.1).  Colors are specified in both ABmag and Vegamag
systems.  The panels
correspond to $FUV-NUV$ (top), $FUV-K_s$ (middle) and $FUV-R$ (bottom).  The
$FUV-K_s$ model colors imply that our requirement $FUV(AB)-K_s(AB) \le 4$ for Type~2
XUV-disks corresponds to a burst population younger than $\sim$500 Myr, provided that $FUV-NUV$ is significantly redder than about -0.5 (excluding a much older {\em continuous} star formation event).}
\end{figure}   

\clearpage
\begin{figure}[htbp]
\epsscale{1}
\vspace*{-8mm}
\plotone{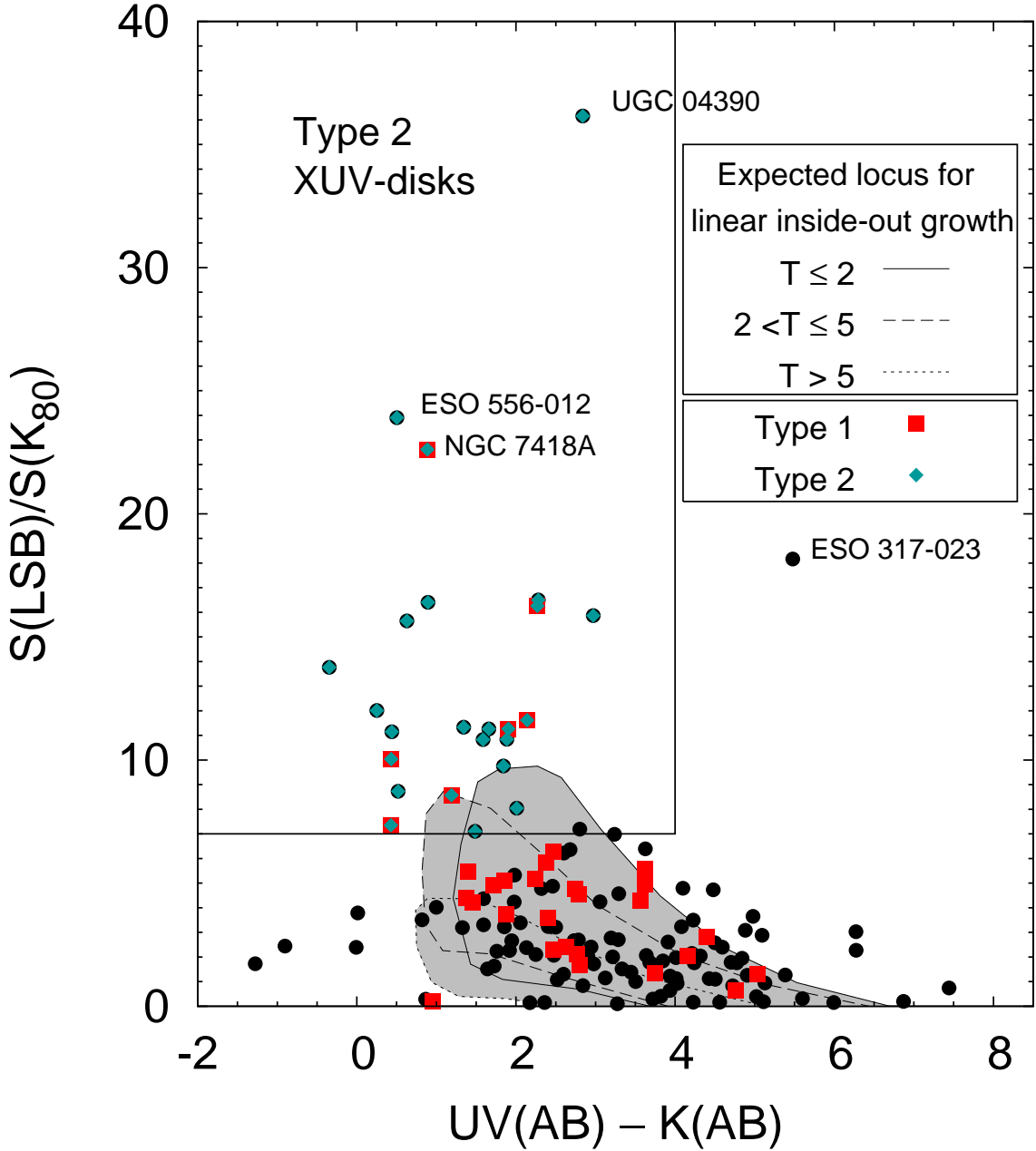}
\caption{The LSB zone to $K_s$-band 80\%-effective area ratio,
$S(LSB)/S(K_{80})$, versus $FUV-K_s$ [$NUV-K_s$] measured in the LSB
zone.  Note the excess of objects having blue color and a
comparatively large LSB zone.  We empirically define the objects
having $S(LSB)/S(K_{80})$ $\ge$7 and $FUV(AB)-K_s(AB)$ $\le 4$ to be
Type~2 (or possibly mixed-type) XUV-disk galaxies.  Symbols are
red squares for Type~1 galaxies, blue diamonds for Type~2 objects, and
black dots for other galaxies in the sample.  We overplot the
predicted locus of ``normal'' galaxies growing from inside-out for
three categories in T, according to the models of Mu\'noz-Mateos et
al. (2007).  The relative spatial extent of the star-forming LSB zone in Type~2 XUV-disk
galaxies is clearly extreme even for the inside-out disk building
process.  The single black data point inside the area designated at Type~2 XUV-disks is NGC~7479, which was classified as non-XUV because it fell much closer to the normal galaxies in the plot than the other Type~2 XUV-disks.}
\end{figure}   
\clearpage

\begin{figure}[htbp]
\plotone{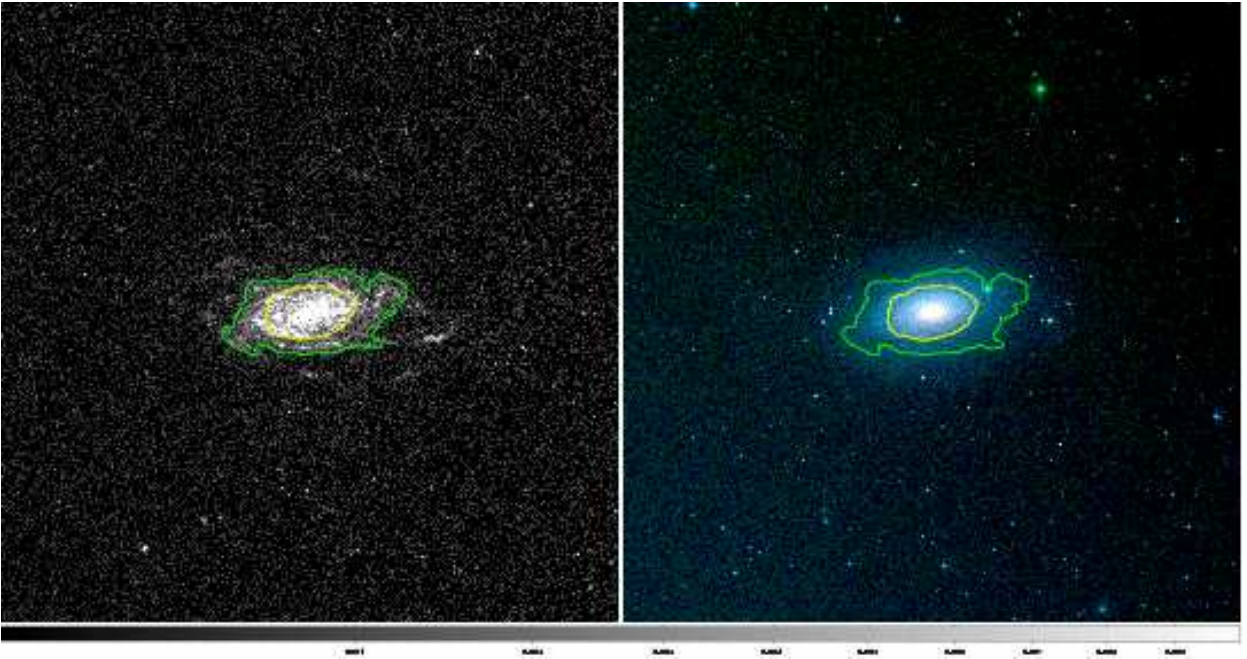}
\caption{FUV-NIR imagery and classification contours for XUV-disk
galaxy NGC~5055 (M~63), a prototype for our Type~1 class.  On the left
we show the GALEX FUV image of the galaxy.  On the right, we present
the 2MASS $K_s$-band, DSS2-red, and DSS2-blue imaging (as RGB
channels) for an identical field-of-view (3 $D_{25}$ = 37.5\arcmin =
89.4 kpc at 8.2 Mpc).  Contours are the same on both images.  In the
case of the green line, the FUV surface brightness (corrected for
Galactic foreground extinction and measured at 1~kpc resolution) is
$\mu_{FUV}$ = 27.25 AB mag / sq.arcsec.  This is the position at which
(apparent?) star formation threshold mechanisms are thought to become
important.  The yellow contour encloses 80\% of the $K_s$-band
luminosity of the galaxy, defining an effective extent for the old
stellar population.  Note the structured UV-bright emission features
beyond the green (UV) contour which give this galaxy the Type~1
XUV-disk designation.}
\end{figure}

\begin{figure}[htbp]
\plotone{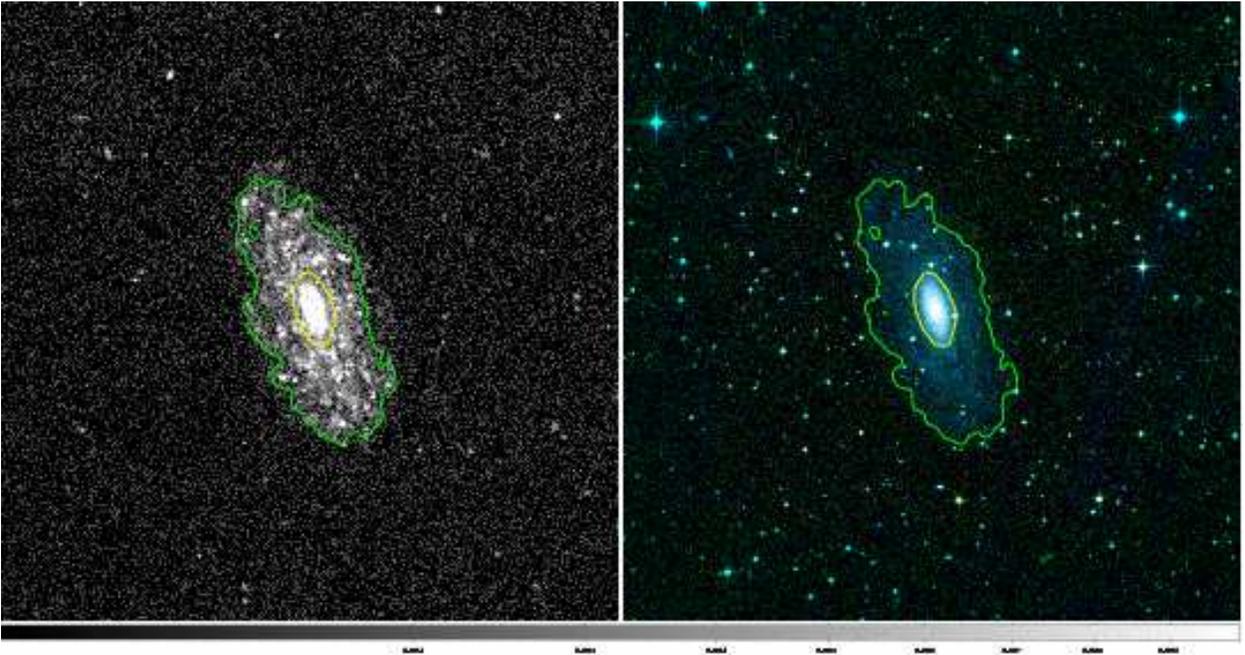}
\caption{FUV-NIR imagery and classification contours for Type~2
XUV-disk galaxy NGC~2090.  We observe a rather large blue LSB zone
which dominates the spatial extent of the galaxy despite being of low
[optical] surface brightness. The image passbands and contour types
are identical to those of Fig. 3.  The field of view spans 3 $D_{25}$
= 12.9\arcmin = 42.4 kpc at 11.3 Mpc. }
\end{figure}

\begin{figure}[htbp]
\plotone{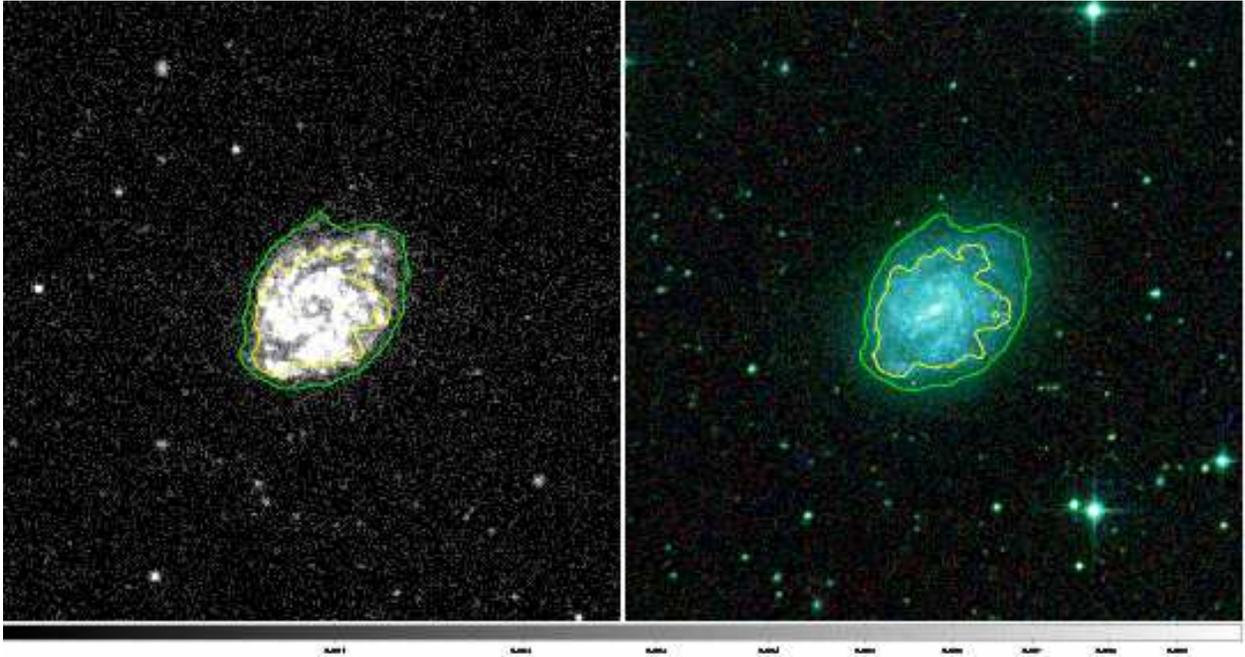}
\caption{FUV-NIR imagery and classification contours for {\em
non}-XUV-disk galaxy NGC~7418, presented as an example of a target
having a blue edge which is not extended enough to be considered
significant for classification purposes.  Furthermore, no structured
UV emission can be confidently associated with the target beyond the
FUV ``threshold'' contour.  NGC~7418, an SAB(rs)cd galaxy, has
$FUV-K_s$ = 2.8 in the LSB zone but $S(LSB)/S(K_{80})$ = 0.8.  The
image passbands and contour types are identical to those of
Fig. 3. The field of view spans 3 $D_{25}$ = 11.4\arcmin = 60.3 kpc at
18.2 Mpc. }
\end{figure}

\begin{figure}[htbp]
\plotone{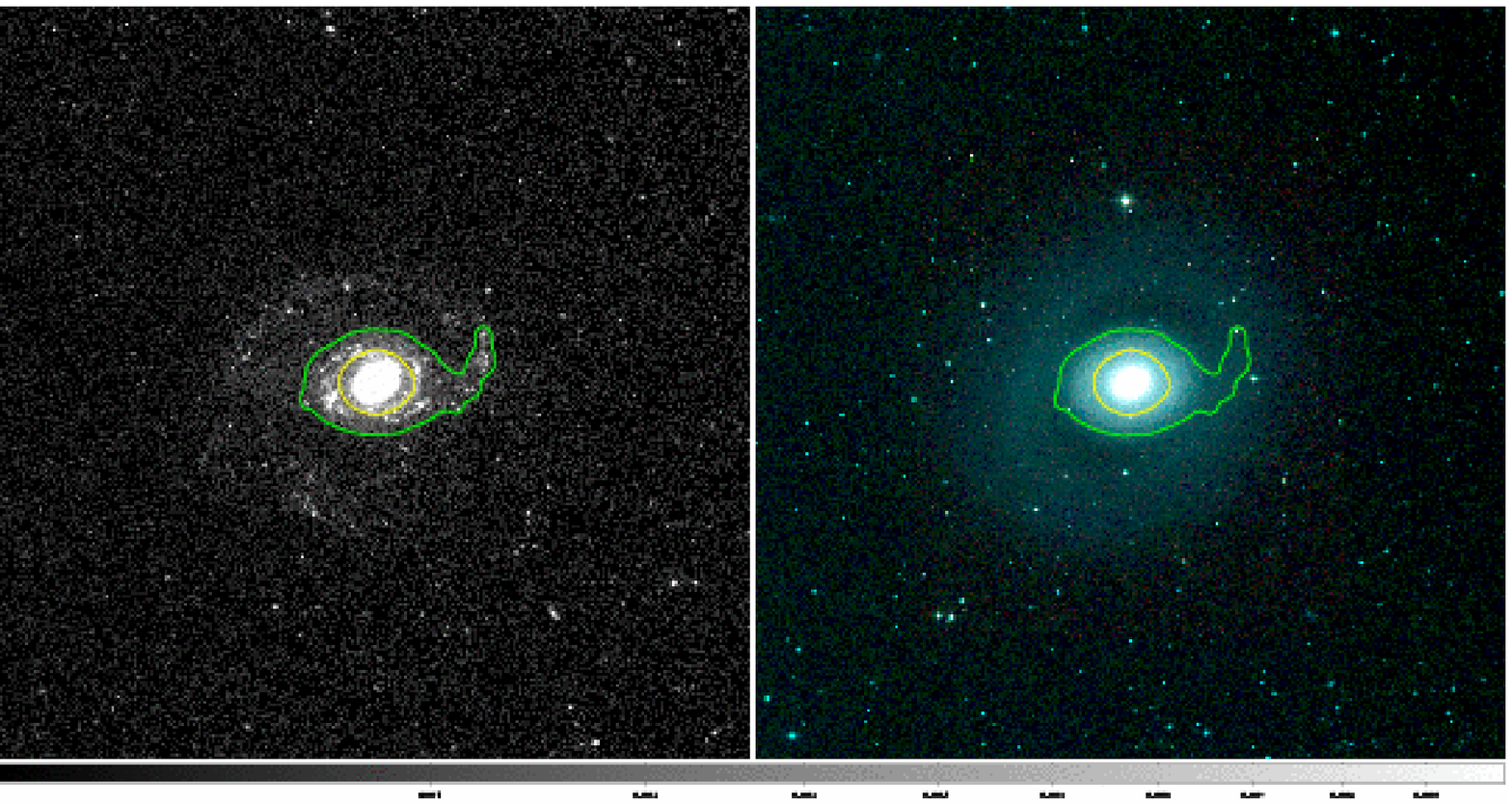}
\caption{FUV-NIR imagery and classification contours for {\em
non}-XUV-disk galaxy NGC~4736 (M~94, type (R)SA(r)ab;Sy2 LINER),
presented as an example of a target having significant, structured,
UV-only clumps beyond the threshold radius but with relatively-HSB
optical structure in nearly the same form.  The LSB zone in NGC~4736
is typical, with $S(LSB)/S(K_{80})$ = 2.9 and ($FUV-K_s$ = 5.1).  The
image passbands and contour types are identical to those of
Fig. 3. The field of view spans 3 $D_{25}$ = 30.0\arcmin = 136 kpc at
5.2 Mpc.}
\end{figure}

\eject

\begin{figure}[htbp]
\plotone{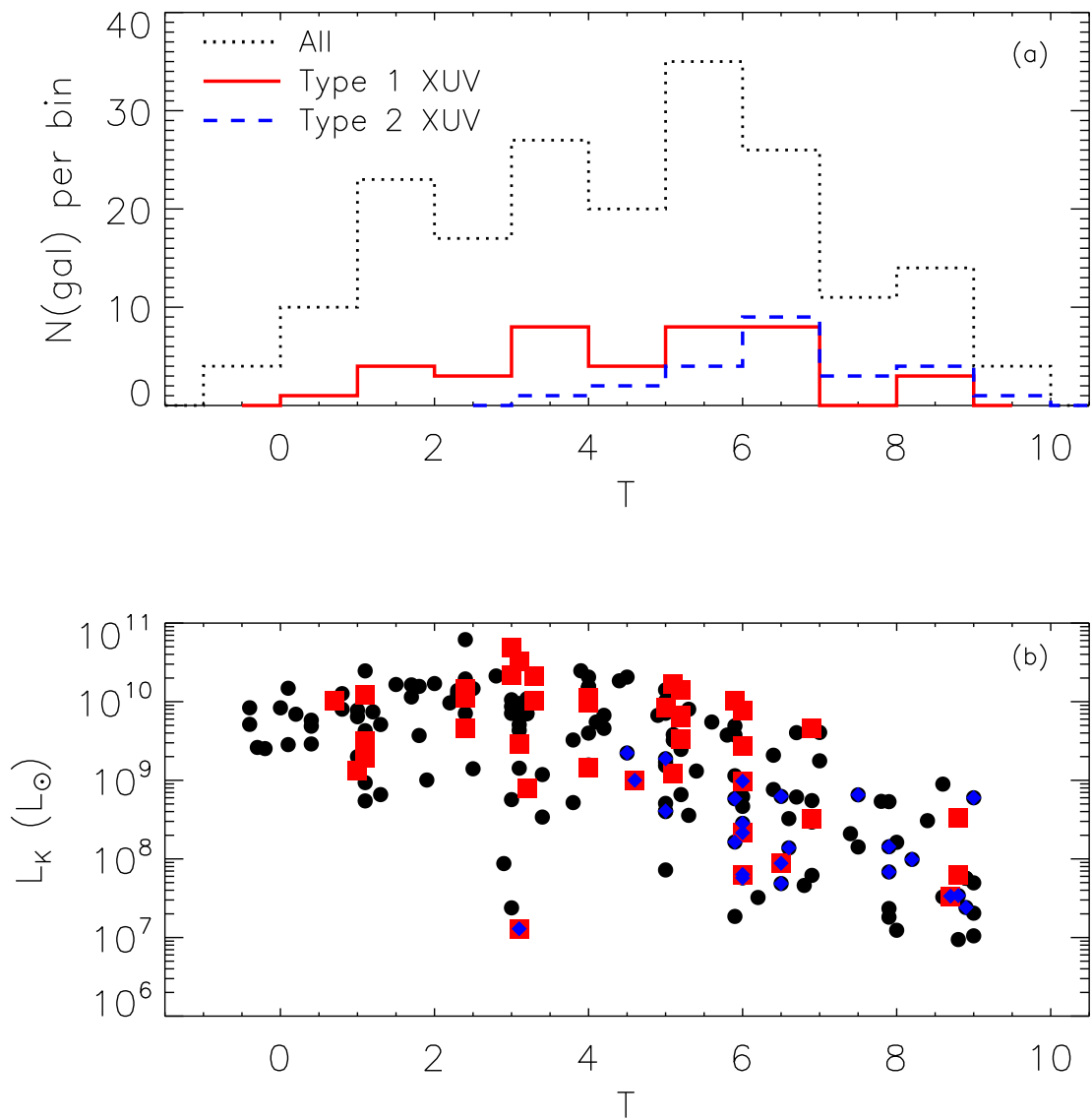}
\caption{(a) Distribution of Hubble type ($T$) for all galaxies in our sample (black dotted line), in comparison to Type~1 XUV-disks (red solid line) and Type~2 XUV-disks (blue dashed line).   (b) $L_K$ versus Hubble type, using the same color coding as in panel (a).  In addition, Type~1 XUV-disks have been drawn with a square and Type~2 XUV disks with a diamond allowing for identification of galaxies meeting one [or both] of our XUV criteria.}
\end{figure}

\begin{figure}[htbp]
\plotone{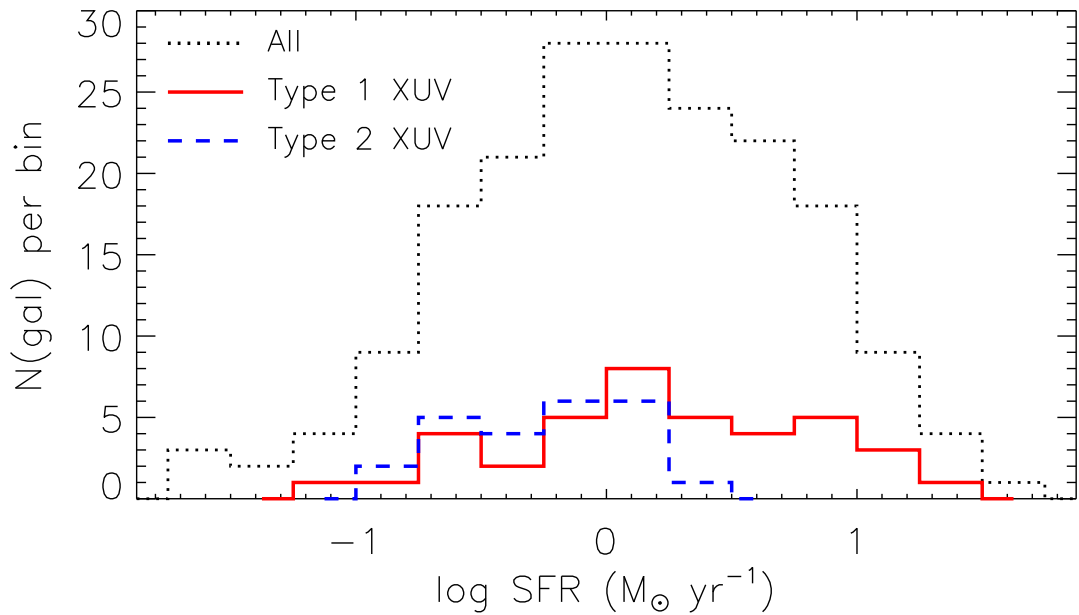}
\caption{Distribution of globally averaged SFR for the survey sample (black dotted line) and for galaxies classified as XUV-disks (red solid line - Type~1, blue dashed line - Type~2).  SFR has been estimated on the basis of UV and IR luminosity jointly.  }
\end{figure}

\begin{figure}[htbp]
\plotone{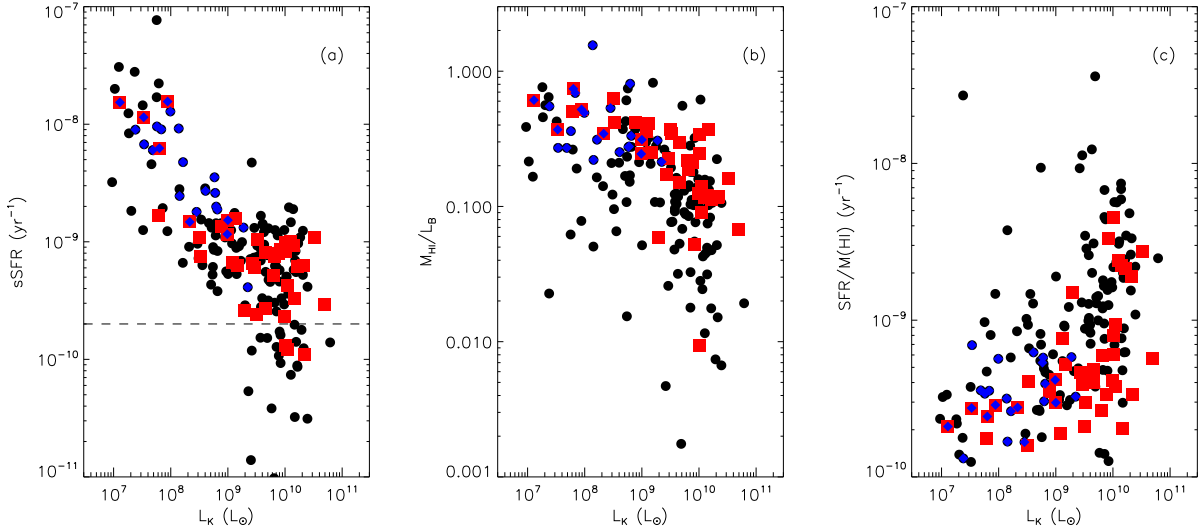}
\caption{(a) Distribution of globally averaged specific SFR ($sSFR$) versus $L_K$ for the survey sample (black dots) and
for galaxies classified as XUV-disks (red squares - Type~1, blue diamonds - Type~2) .  XUV-disk galaxies may become
increasingly rare below the $sSFR$ indicated with a dashed line,
though number statistics are small.   (b) Distribution of globally averaged $M(HI)/L_B$
versus $L_K$.  (c) Distribution of globally averaged $SFR/M(HI)$, a
measure of star formation efficiency, versus $L_K$. }
\end{figure}

\begin{figure}[htbp]
\plottwo{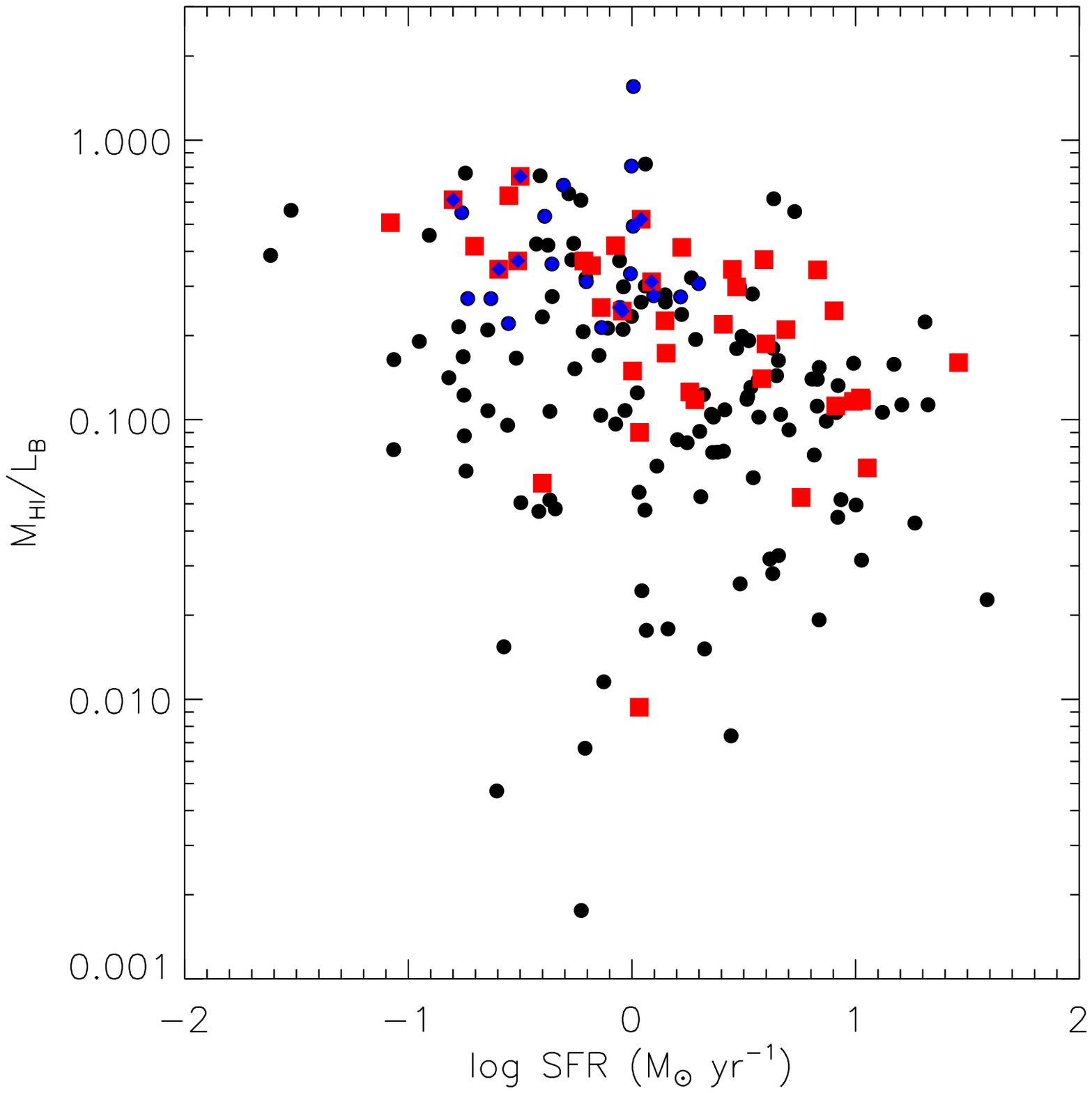}{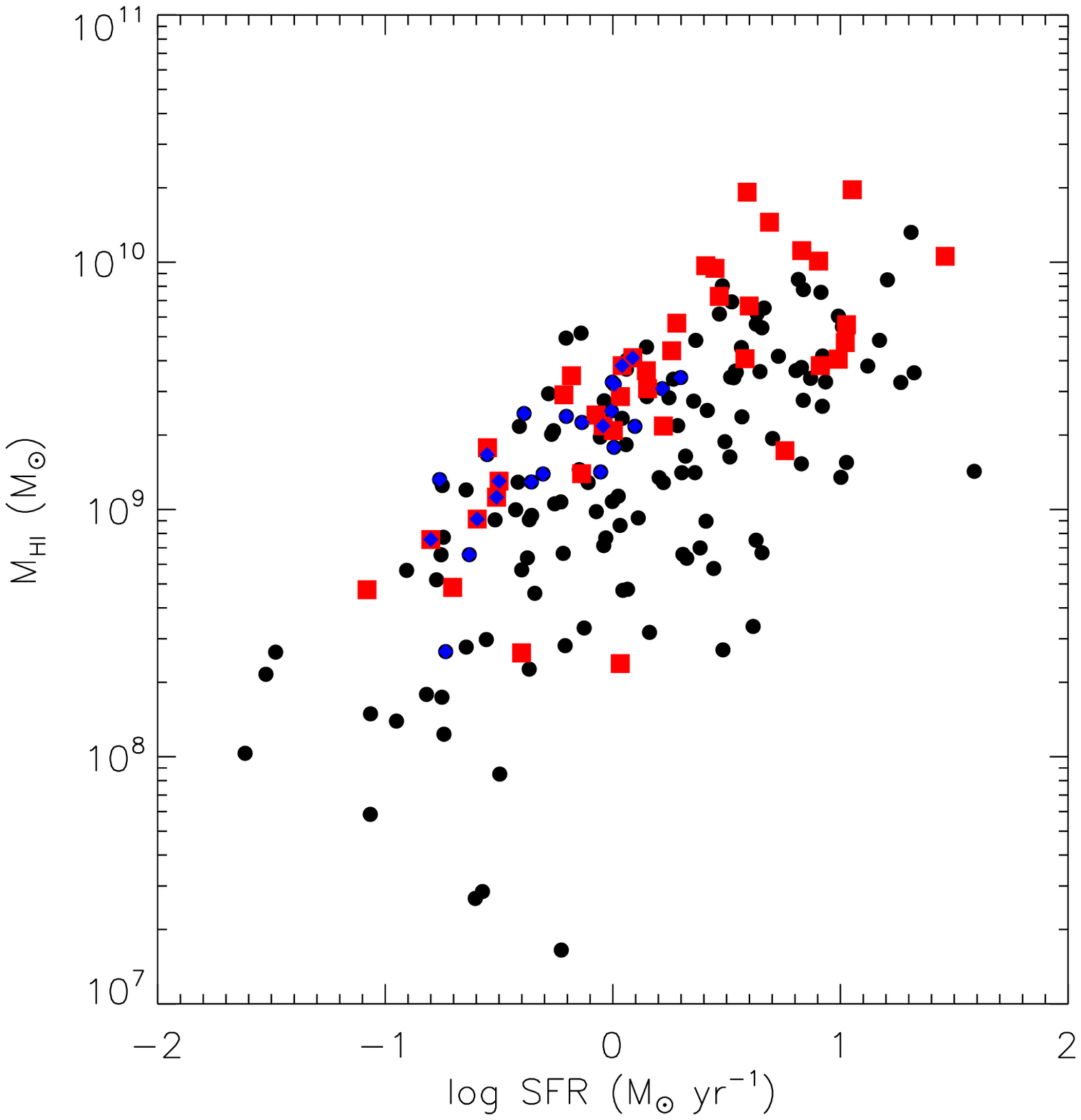}
\caption{(a) Distribution of globally averaged $M(HI)/L_B$
versus $SFR$ for the survey sample (black dots) and
for galaxies classified as XUV-disks (red squares - Type~1, blue diamonds - Type~2).  (b) Distribution of $M(HI)$ versus $SFR$. }
\end{figure}

\begin{figure}[htbp]
\plotone{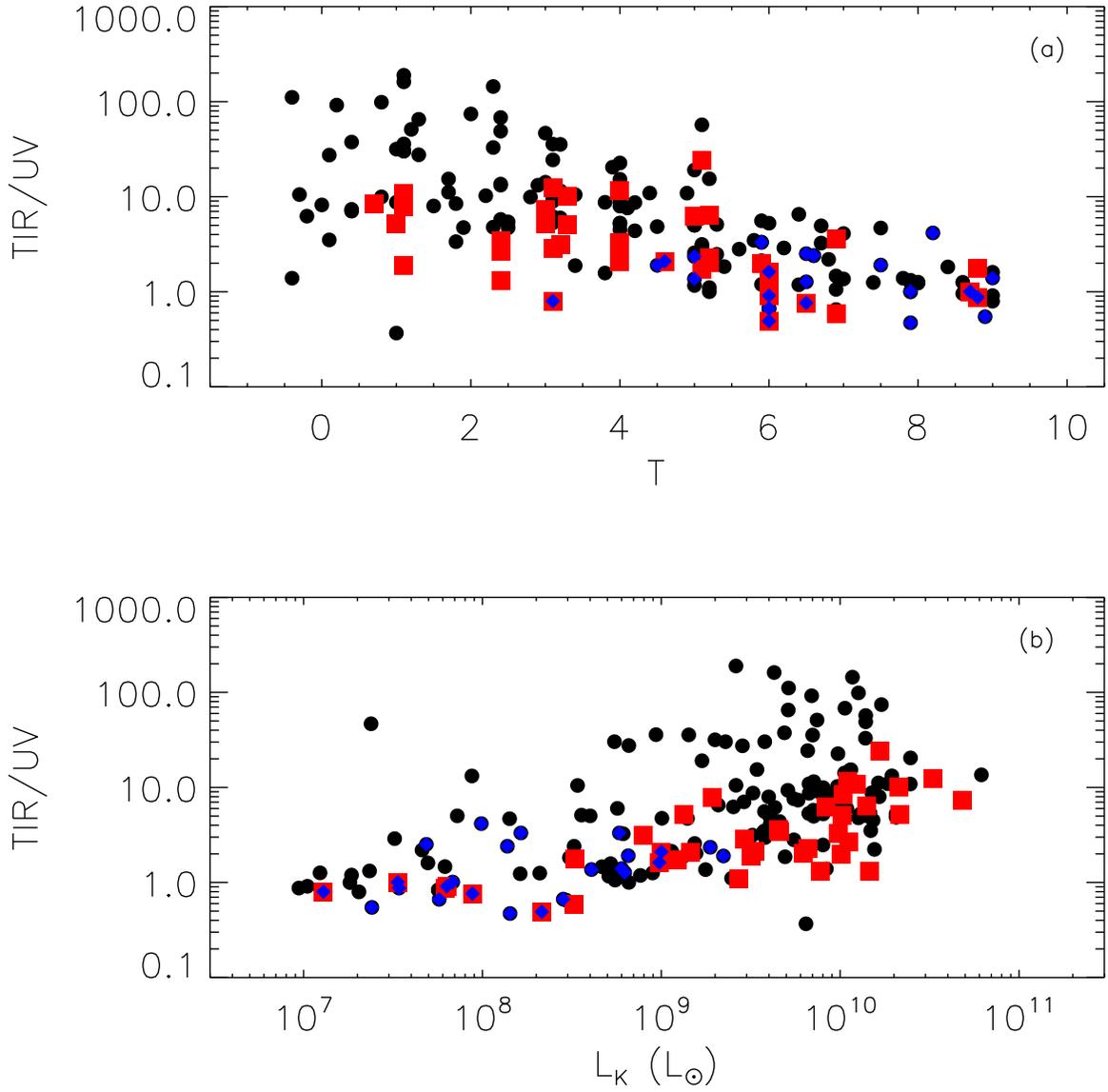}
\caption{(a) Distribution of globally averaged TIR/UV versus $T$ for the survey sample (black dots) and for galaxies classified
as XUV-disks (red squares - Type~1, blue diamonds - Type~2). (b) Distribution of globally averaged TIR/UV versus $L_K$.  }
\end{figure}

\begin{figure}[htbp]
\plotone{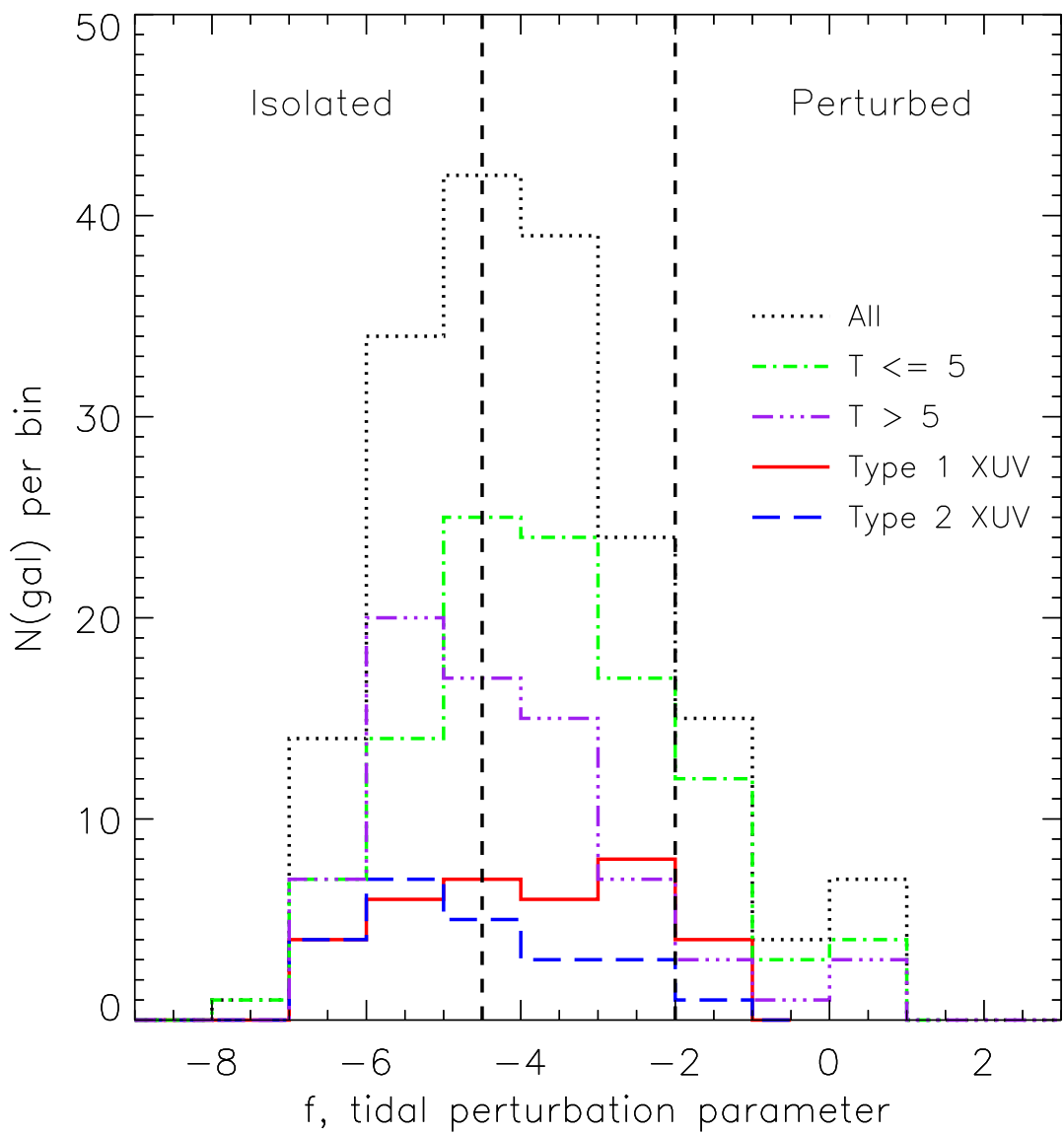}
\caption{The tidal perturbation parameter, $f$, for all galaxies in our current survey (black dotted line) compared to those classified as XUV-disks (red solid line - Type~1, blue dashed line - Type~2). We also plot the sample-wide distribution of $f$ for early- and late-type spirals with green dash-dot and purple dash-triple-dot lines, respectively.  The designation of isolated and perturbed labels follows the limits proposed by Varela et al. (2000).   }
\end{figure}

\begin{figure}[htbp]
\plotone{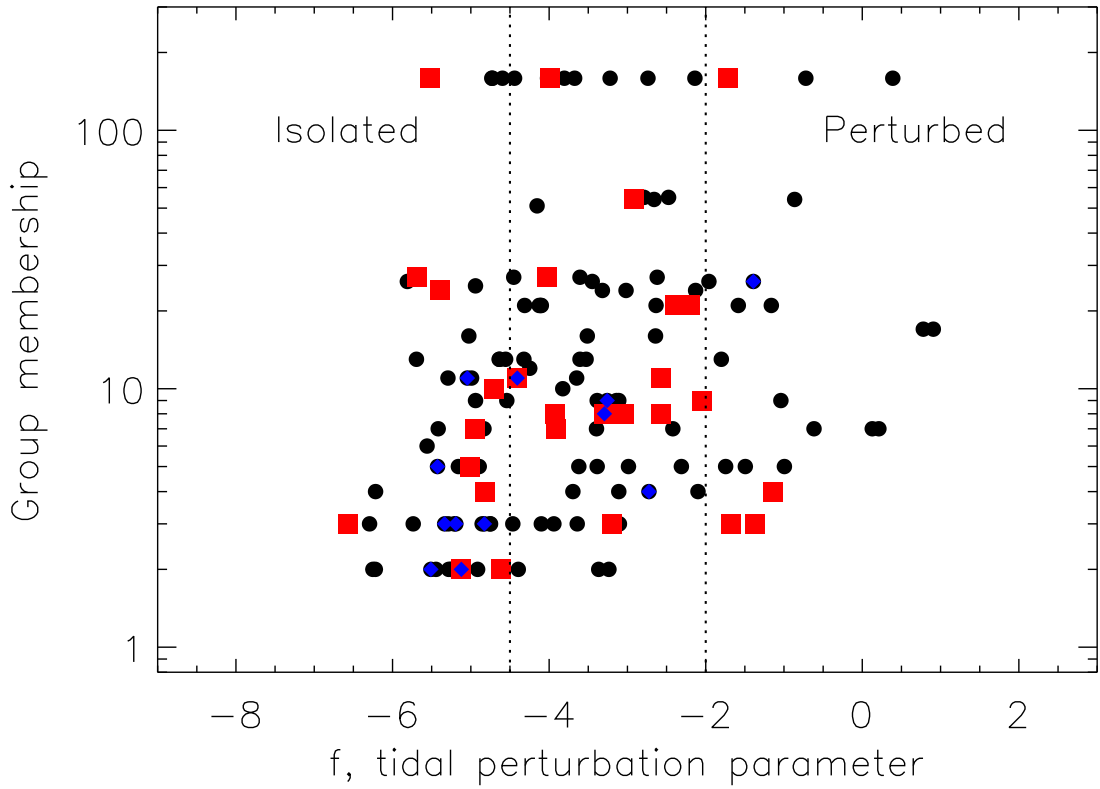}
\caption{Comparison of NOG group membership versus $f$, the tidal perturbation parameter.  Note that a fraction of our sample galaxies (black dots), and of the XUV disk subset (red squares - Type~1, blue diamonds - Type~2), are not represented in this panel either because they were not in a group/pair or they are too faint to have even been considered in the NOG grouping analysis of Giuricin et al. (2000).}
\end{figure}

\begin{figure}[htbp]
\plotone{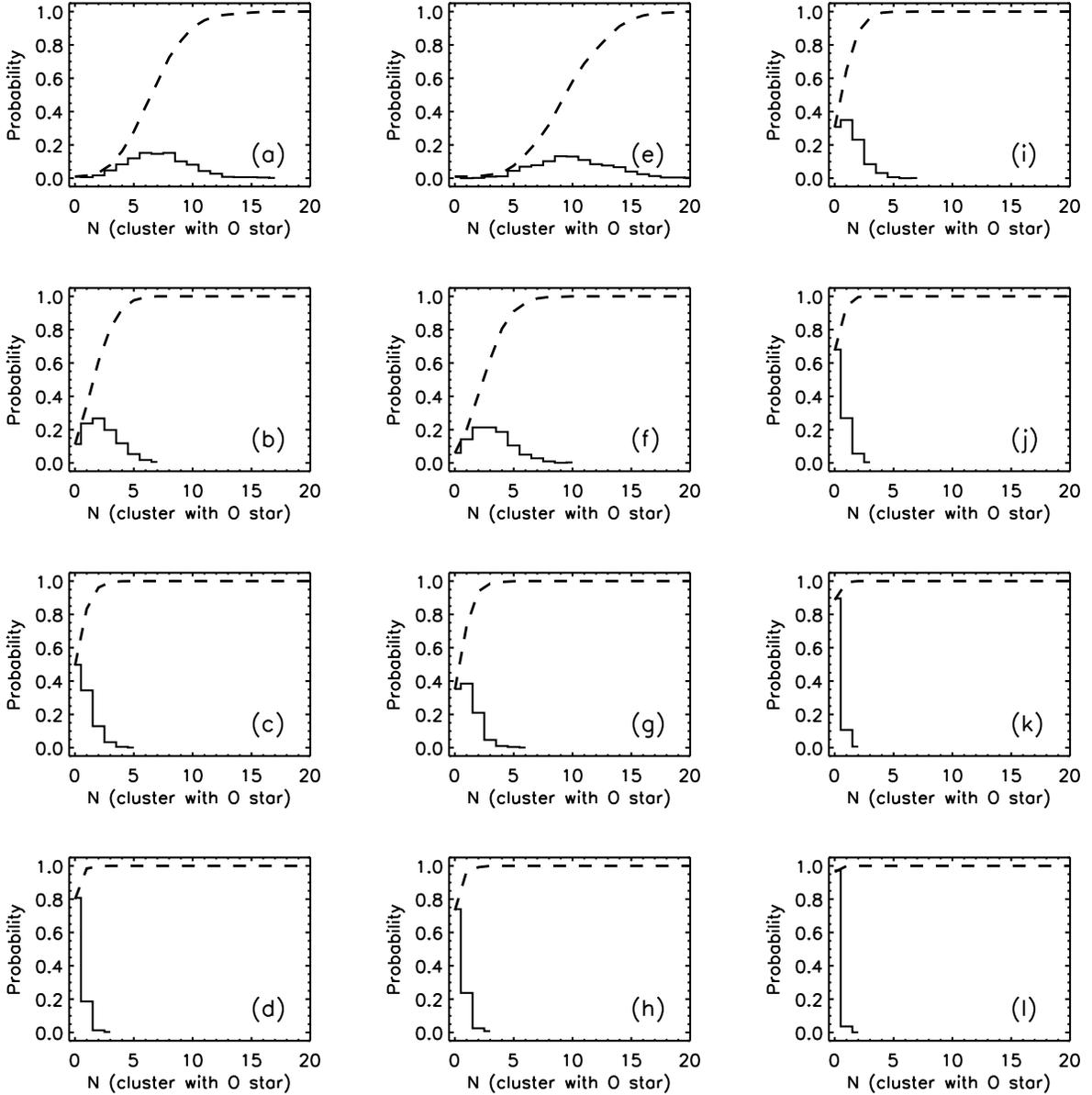}
\caption{Probability of N clusters each containing at least one surviving O star  for various Monte Carlo models. Probability distribution functions (solid histogram) and cumulative distributions (dashed line) are plotted.  The simulated log($SFR$) value is -3.0, -3.5, -4.0, and -4.5 M$_\odot$~yr$^{-1}$ from the top row to bottom row, with SF distributed throughout a region of arbitrary size.  We assume a Salpeter IMF for the left and middle columns (panels a-h) and a conditionally truncated Salpeter IMF for the right column (panels i-l).  In the later case, we assumed massive stars (M$\ge$10 M$_\odot$) may only form in clusters having initial membership of at least 100 stars with M$>$1 M$_\odot$.  The left column has a default CMF of slope -2, allowing clusters up to 10$^6$ M$_\odot$, whereas the middle and right columns have an upper limit to the CMF at 10$^3$ M$_\odot$.}
\end{figure}
\clearpage

\begin{figure}[htbp]
\plotone{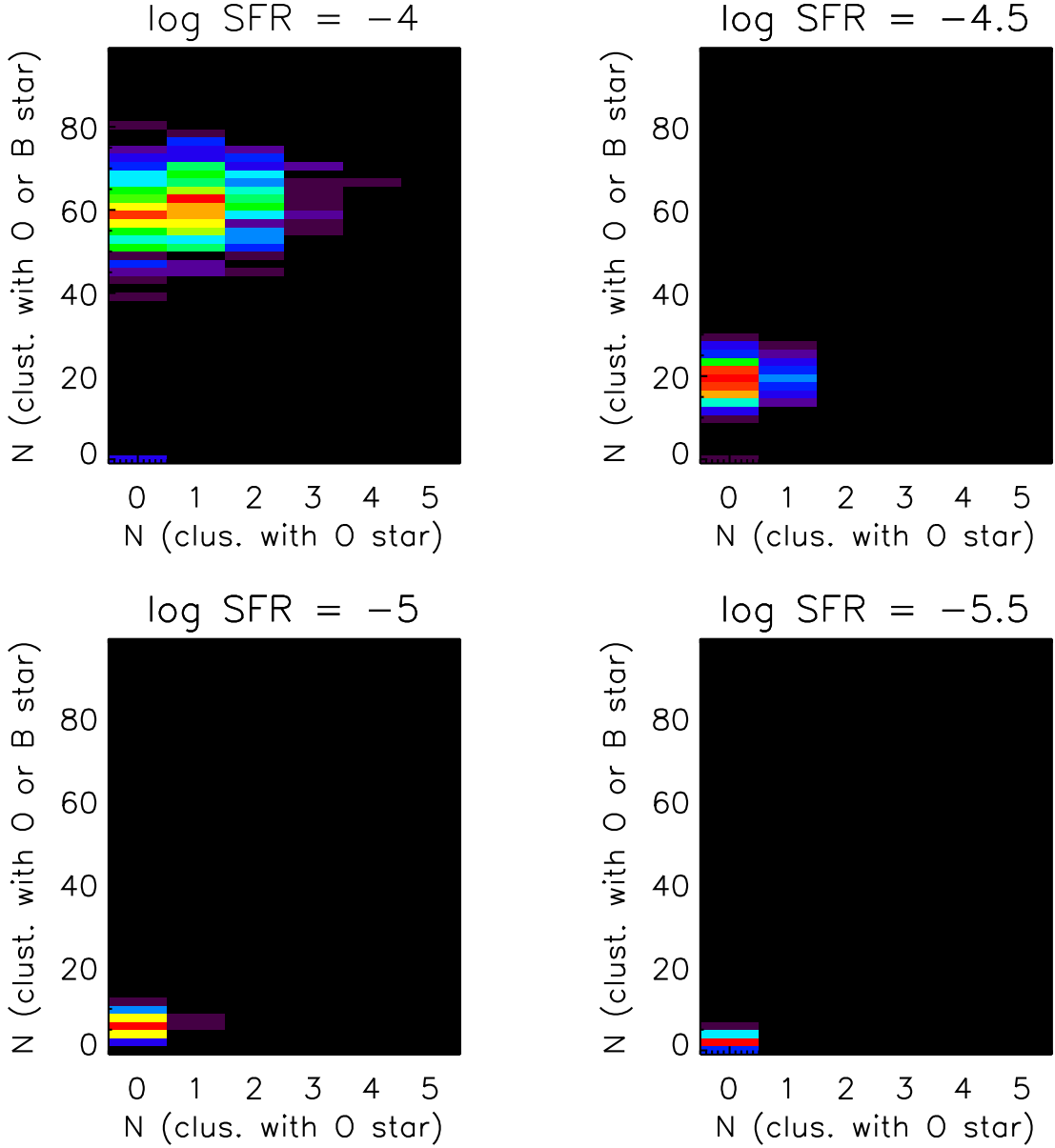}
\caption{The predicted number of clusters containing at least one O
star and the number of clusters containing one or more OB stars, for
log(SFR) = -4.0, -4.5, -5.0, and -5.5 M$_\odot$ yr$^{-1}$ integrated
over an arbitrary region of interest. The probability distribution function shows that the UV SFR tracer (sensitive to OB stars) is stochastically limited for log(SFR) $\le$ -5.5.}
\end{figure}

\clearpage

\figsetstart
\figsetnum{16}
\figsettitle{Multiwavelength imaging of XUV-disk galaxies}

\figsetend

\renewcommand{\thefigure}{ 16}
\begin{figure}[htbp]
\epsscale{0.9}
\figurenum{16}
\plotone{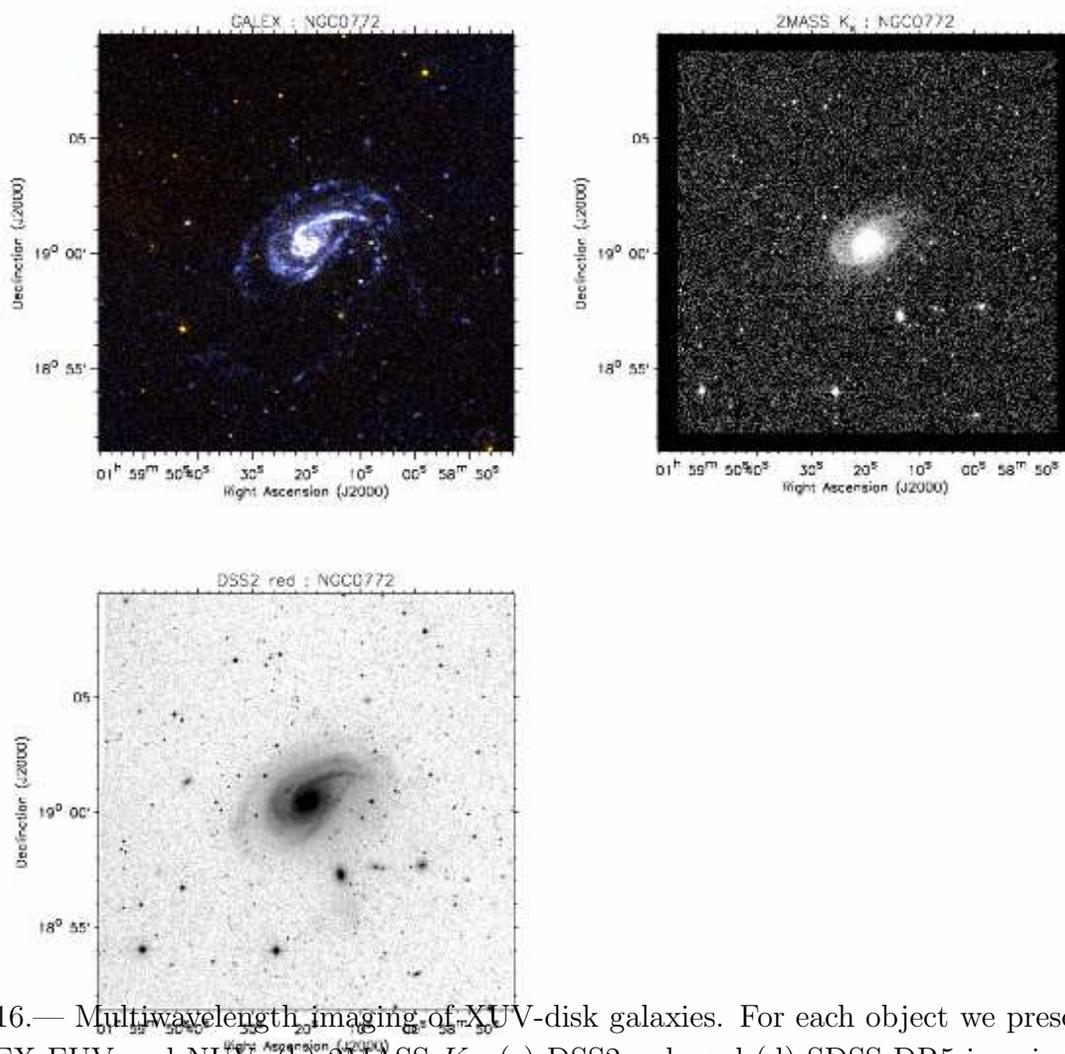}
\vspace*{-20mm}
\caption{Multiwavelength imaging of XUV-disk galaxies.  For each object we present: (a) GALEX FUV and NUV, (b) 2MASS-$K_s$, (c) DSS2-red, and (d) SDSS-DR5 imaging (when available).  In the case of SDSS, we display the {\it gri} bands as blue, green, and red channels. The extent of all images is three times the $D_{25}$ size of the galaxy.   In the print edition, we only present two galaxies of each XUV-disk variety as a guide to content.   NGC~772 and NGC~4254 [M~99] are shown as examples of Type~1, UGC~10445 and NGC7418A for mixed-type objects, and NGC~2090 and NGC~2541 for Type~2 XUV-disks. A complete version of this figure is available online. }
\end{figure}
\clearpage
{\plotone{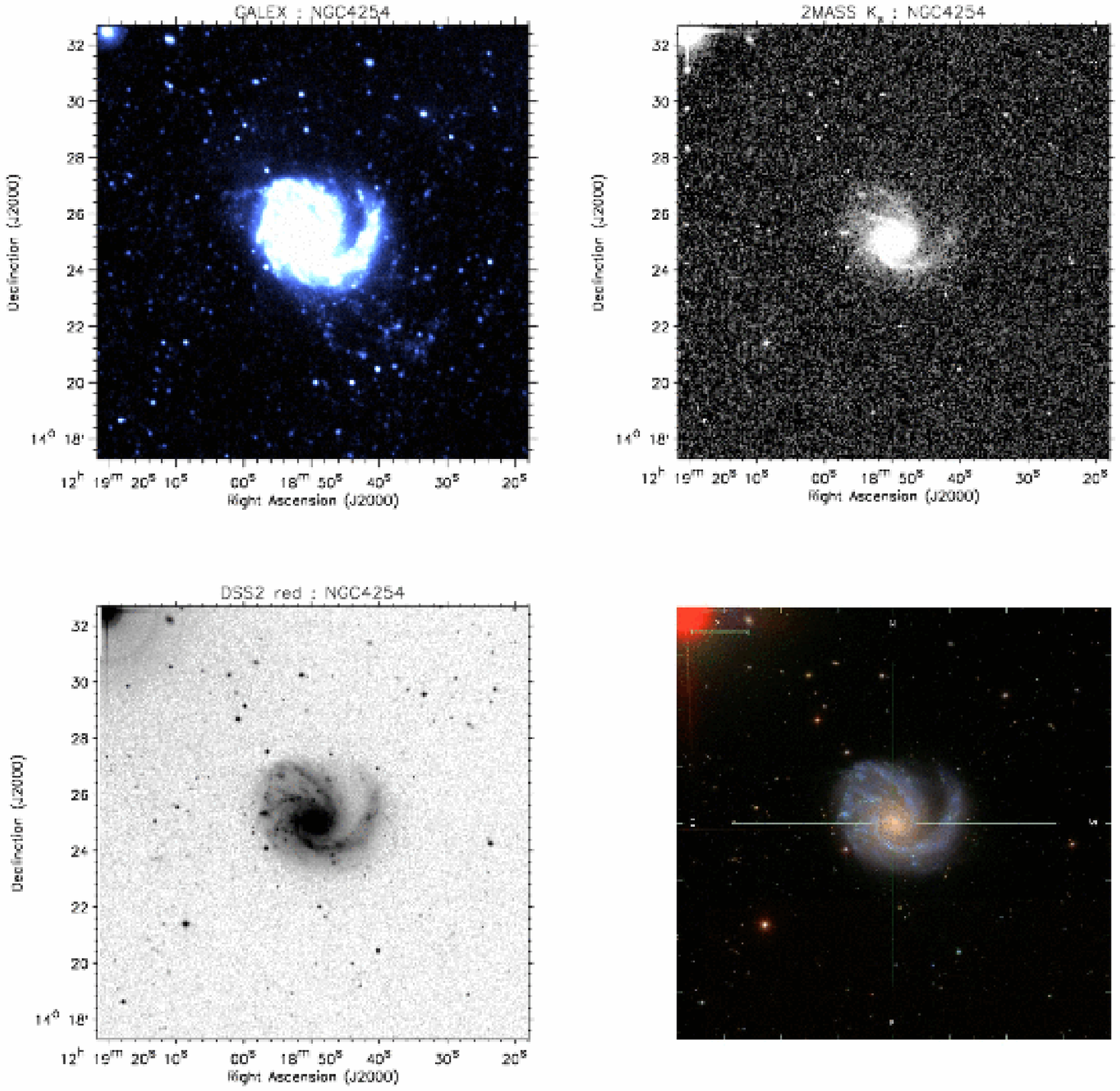}}\\
\centerline{Fig. 16. --- Continued.}
\clearpage
{\plotone{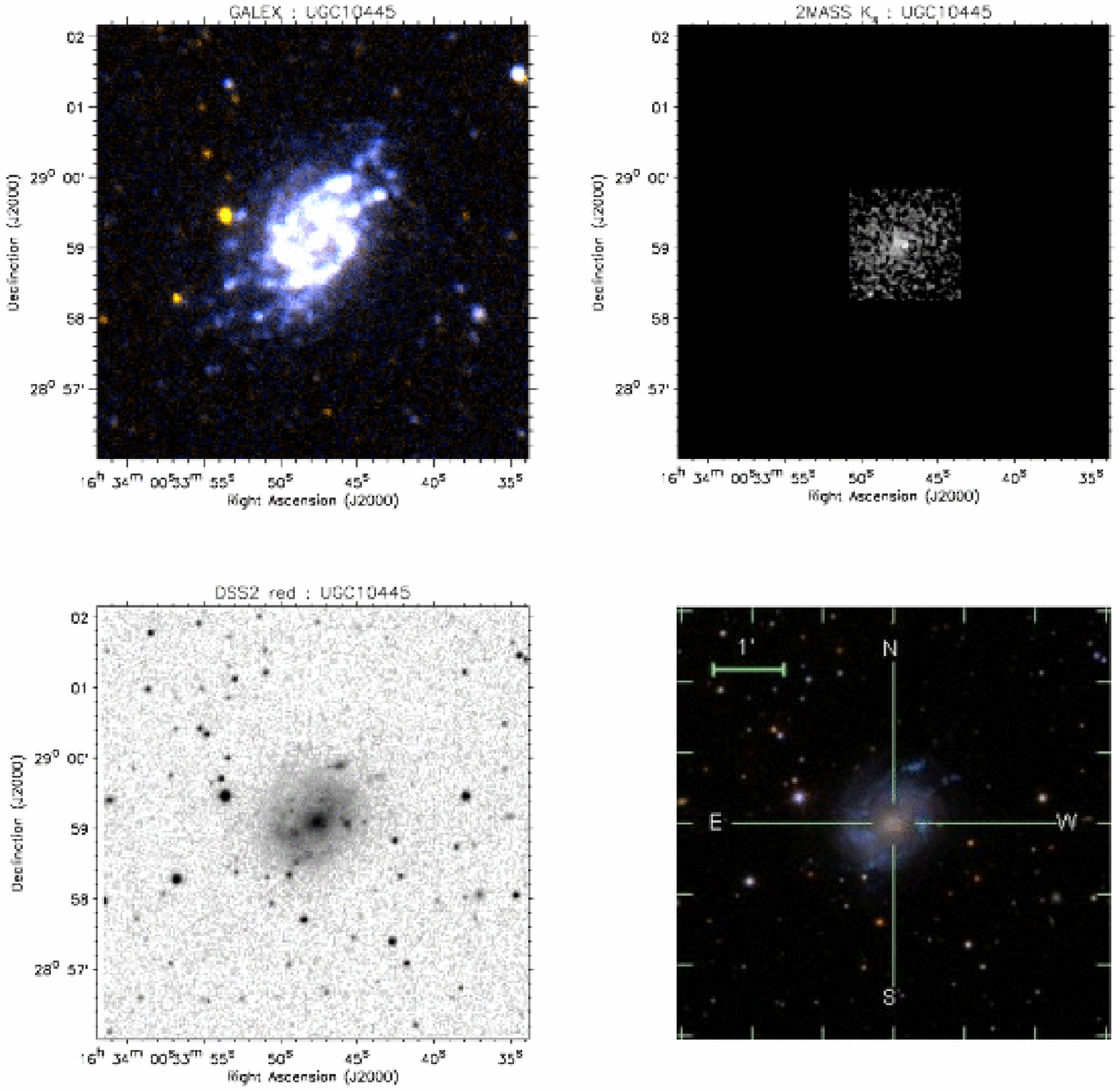}}\\
\centerline{Fig. 16. --- Continued.}
\clearpage
{\plotone{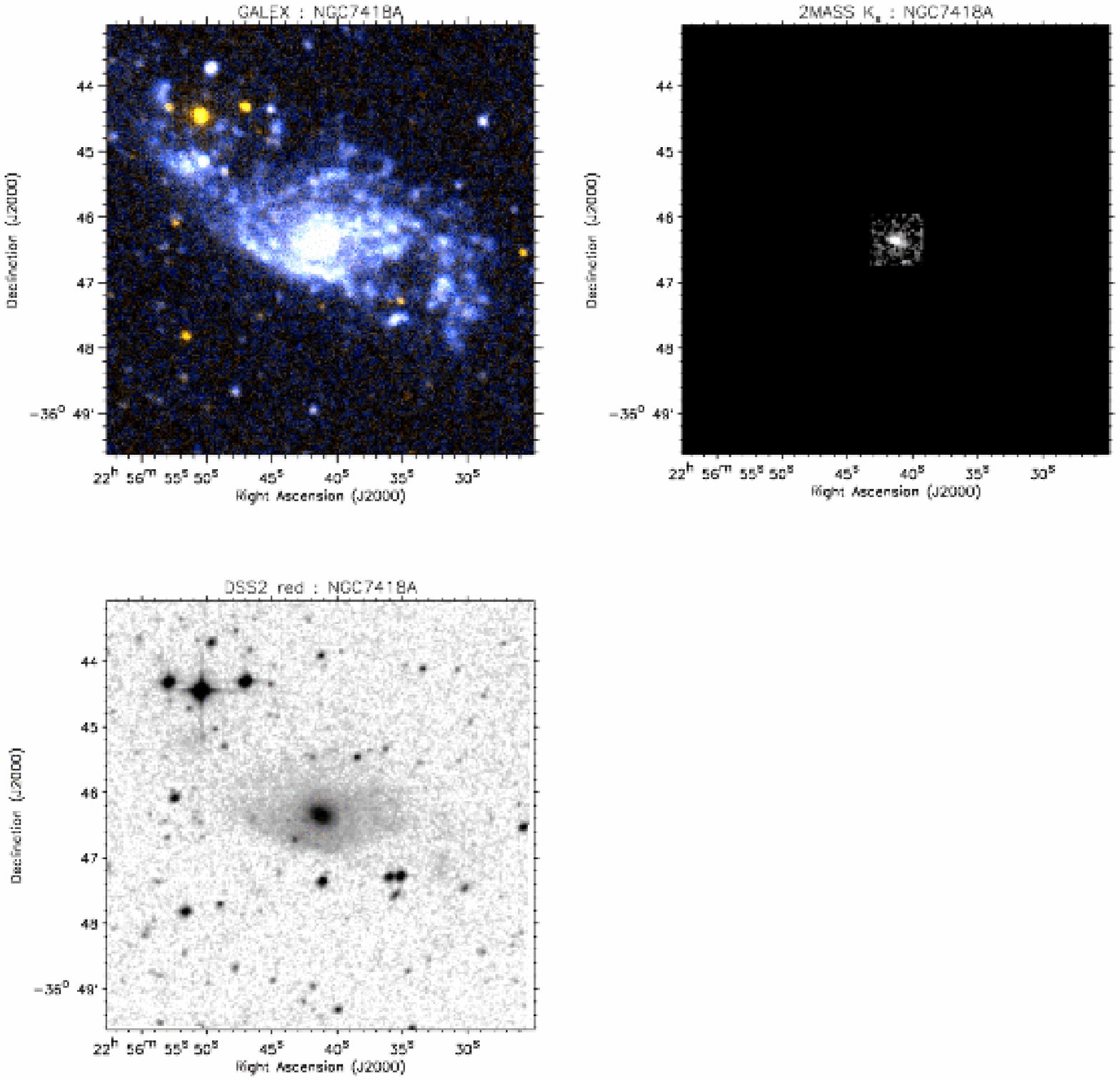}}\\
\centerline{Fig. 16. --- Continued.}
\clearpage
{\plotone{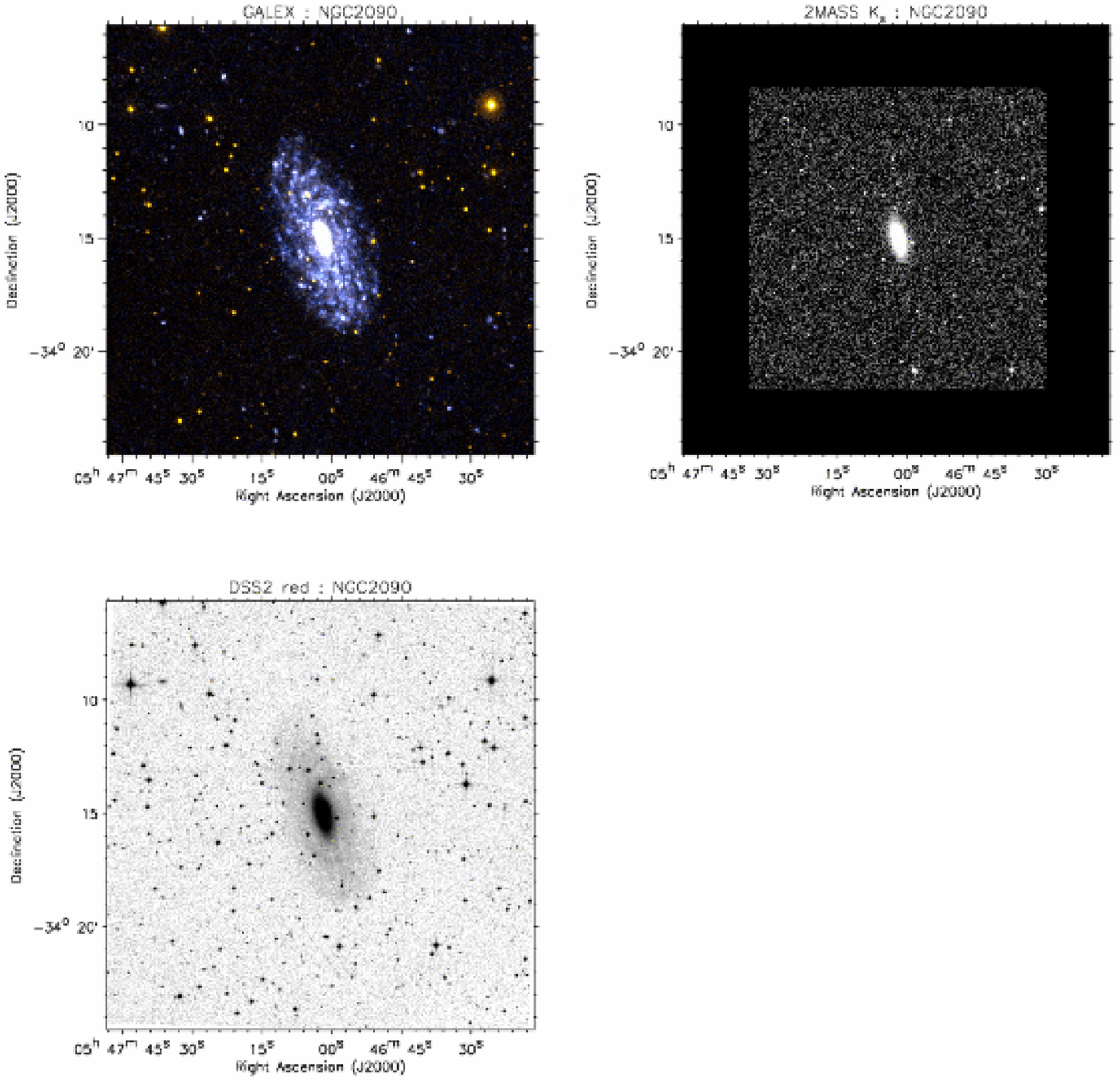}}\\
\centerline{Fig. 16. --- Continued.}
\clearpage
{\plotone{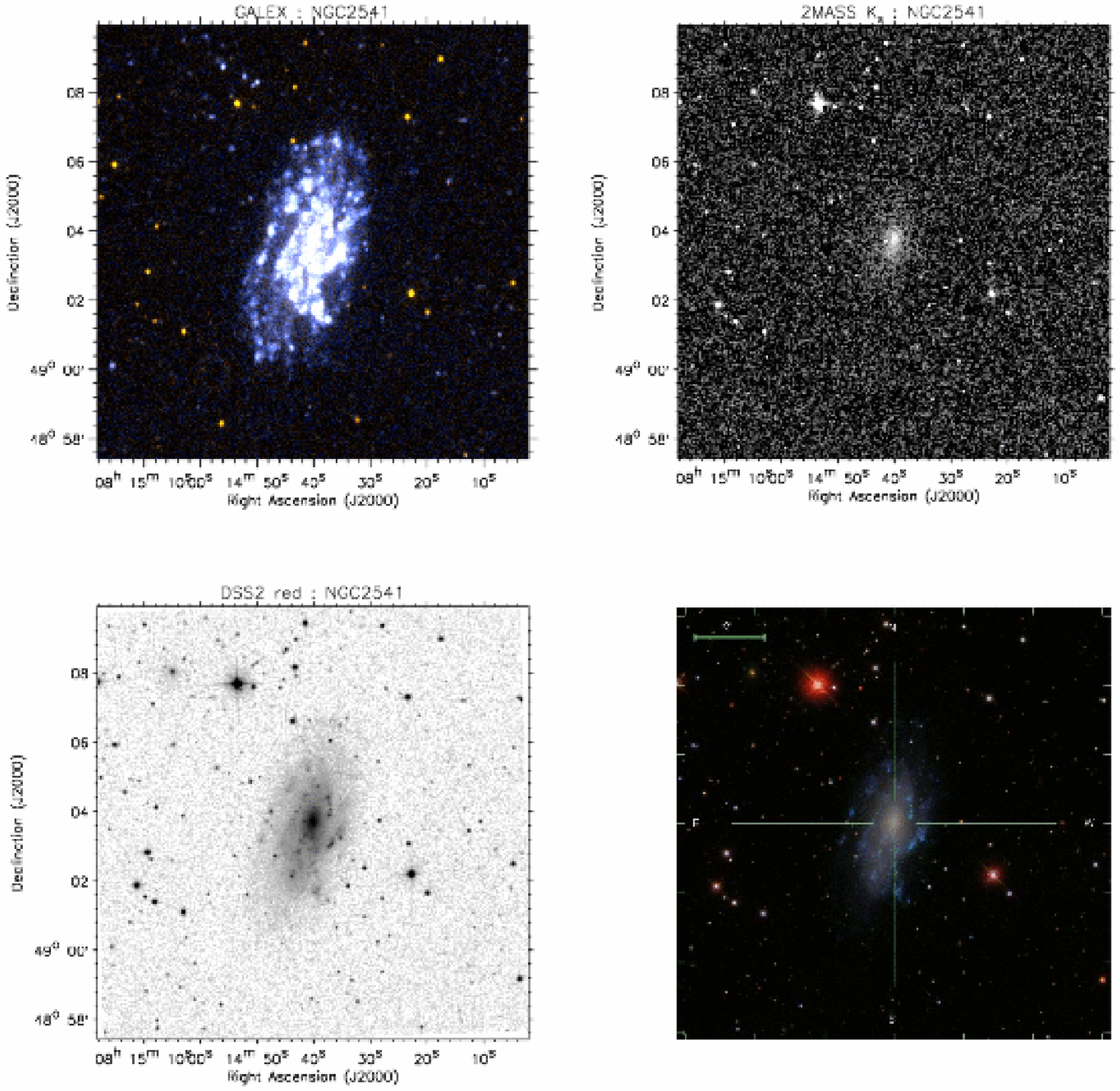}}\\
\centerline{Fig. 16. --- Continued.}

\end{document}